\documentclass[11pt]{article}
\usepackage{helvet}
\usepackage{graphicx}
\usepackage{color}
\usepackage{rotating}
\usepackage{eurosym}
\usepackage{wrapfig}
\usepackage{setspace}
\usepackage{eso-pic}
\usepackage{subfig}
\usepackage{afterpage}
\usepackage{sidecap}

\hoffset= - 2.7 truecm 
\def\lsim{\raise0.3ex\hbox{$<$\kern-0.75em\raise-1.1ex\hbox{$\sim$}}}
\def\gsim{\raise0.3ex\hbox{$>$\kern-0.75em\raise-1.1ex\hbox{$\sim$}}}

\def\beq{\begin{equation}} \def\eeq{\end{equation}} \def\bea{\begin{eqnarray}}
\def\eea{\end{eqnarray}}  
\def\be{\begin{equation}} \def\ee{\end{equation}} \def\ba{\begin{eqnarray}}
\def\ea{\end{eqnarray}} 
\def\dalemb#1#2{{\vbox{\hrule height.#2pt \hbox{\vrule width.#2pt height#1pt
\kern#1pt \vrule width.#2pt} \hrule height.#2pt}}}

\def\ba{\begin{eqnarray}} \def\ea{\end{eqnarray}} \def\be{\begin{equation}}
\def\ee{\end{equation}} 
\def\gtorder{\mathrel{\raise.3ex\hbox{$>$}\mkern-14mu
\lower0.6ex\hbox{$\sim$}}}
\def\ltorder{\mathrel{\raise.3ex\hbox{$<$}\mkern-14mu
\lower0.6ex\hbox{$\sim$}}}

\newcommand{\muK}{\mu  {\rm K}} \newcommand{\muKarcmin}{\mu  {\rm K\cdot  arcmin}}
\newcommand{\kmbysbyMpc}{\,{\rm km~s^{-1}~Mpc^{-1}}}

\usepackage{amssymb,amsbsy,amsmath,amsfonts,amssymb,amscd}
\newcommand{\core}{\textit{COrE\/}}
 
\newcommand{\Herschel}{\negthinspace \textit{Herschel\/}}

\def\m{\ifmmode $m$\else \,m\fi}

\newcommand{\dg}{\nobreak^\circ}

\def\st{\ifmmode ^{\mathrm{st}} \else $^{\mathrm{st}}$\fi}
\def\nd{\ifmmode ^{\mathrm{nd}} \else $^{\mathrm{nd}}$\fi}
\def\rd{\ifmmode ^{\mathrm{rd}} \else $^{\mathrm{rd}}$\fi}
\def\th{\ifmmode ^{\mathrm{th}} \else $^{\mathrm{th}}$\fi}

\newcommand\ltsima{$\; \buildrel < \over \sim \;$}
\newcommand\simlt{\lower.5ex\hbox{\ltsima}}
\newcommand\gtsima{$\; \buildrel > \over \sim \;$}
\newcommand\simgt{\lower.5ex\hbox{\gtsima}}
\newcommand\simprop{\lower.5ex\hbox{$\; \buildrel \propto \over \sim \;$}}

\begin{document}


\AddToShipoutPicture*{\put(20,-160){\includegraphics[width=255mm,height=332mm]%
{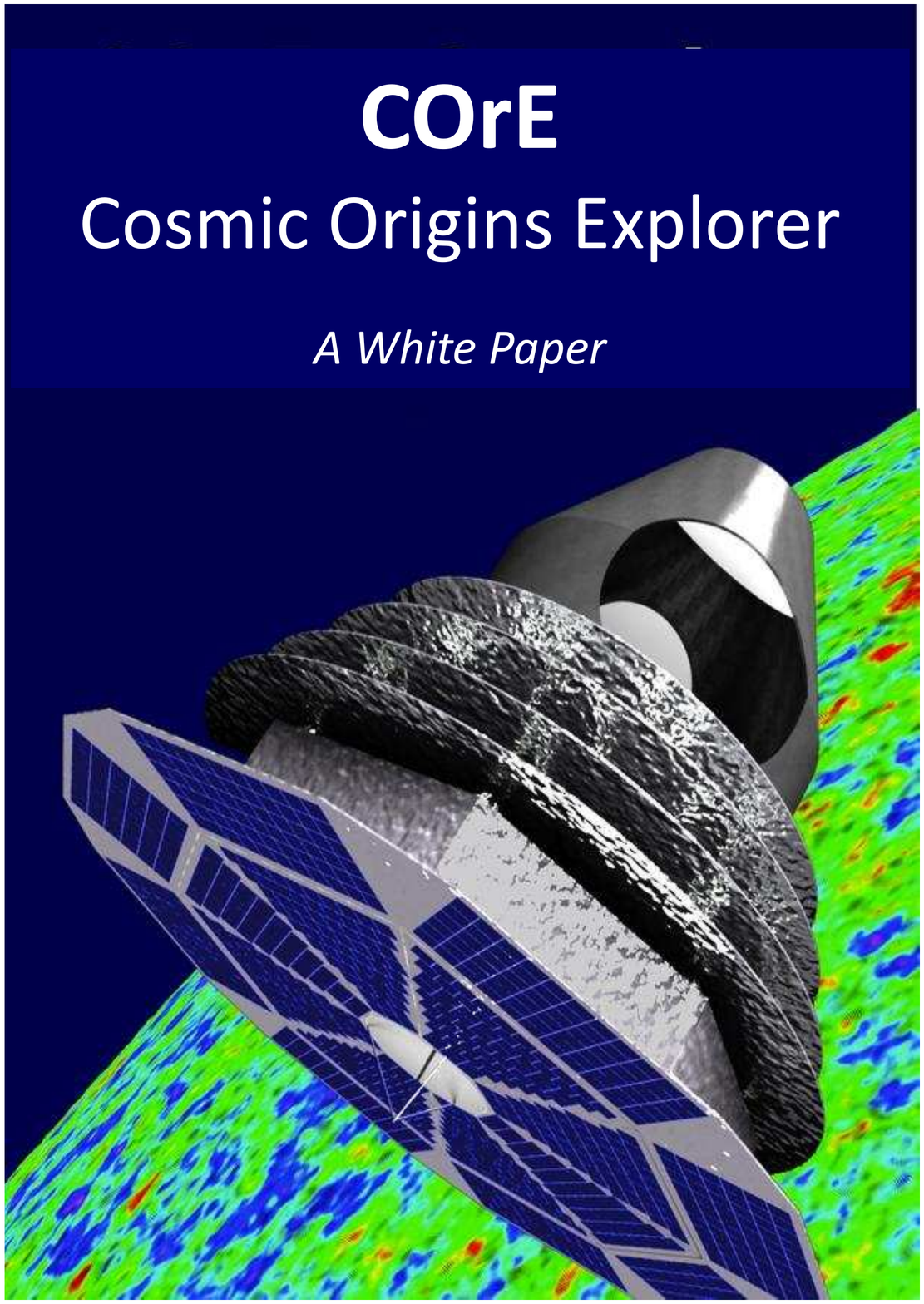}}}%
\phantom{Invisible, but important}
\newpage 

\pagestyle{empty}
\newpage

\vfill\eject
\pagestyle{plain}

\begin{center}
{\fontfamily{helvet}\selectfont\sffamily{\bf \Huge \sffamily COrE}}\\
\vskip 8pt
{\fontfamily{helvet}\selectfont\sffamily{\bf \huge \sffamily Cosmic Origins Explorer}}\\
\vskip 5pt
{\fontfamily{helvet}\selectfont\sffamily{\rm \huge \sffamily A White Paper}}\\
\vskip 13pt
\end{center}

\vskip 15pt

\noindent \textbf{Mission and programmatics working group}\newline
F. R. Bouchet, P. de Bernardis, B. Maffei, P. Natoli, M. Piat, N. Ponthieu,
R. Stompor  \newline

\noindent \textbf{Instrument working group}
\newline B. Maffei, M.
Bersanelli, P. Bielewicz, P. Camus, P. de Bernardis, M. De Petris, 
P. Mauskopf, S. Masi, F. Nati, 
T. Peacocke,
F.  Piacentini, L. Piccirillo, M. Piat, G. Pisano, M. Salatino, R. Stompor, S.
Withington, 
\newline

\noindent \textbf{Science working group}
\newline M. Bucher, M. Avilles, D.
Barbosa, N. Bartolo, R. Battye, J.-P. Bernard, F. Boulanger, A. Challinor, S.
Chongchitnan, S. Colafrancesco, T. Ensslin, J. Fergusson, P.
Ferreira, K. Ferriere, F. Finelli, J. Garc\'ia-Bellido, S. Galli, 
C. Gauthier, M.  Haverkorn, M. Hindmarsh, A. Jaffe, M. Kunz, J. Lesgourgues, A.
Liddle, M. Liguori, P. Marchegiani, S.~Matarrese, A. Melchiorri, P. Mukherjee, 
L. Pagano, D. Paoletti, H. Peiris, L. Perroto, C. Rath, J. Rubi\~no Martin, 
C. Rath, P. Shellard, J.  Urrestilla, B. Van Tent, L. Verde, B. Wandelt
\newline

\noindent
\textbf{Foregrounds working group}\newline
C. Burigana, J. Delabrouille, C. Armitage-Caplan, A. Banday, S. Basak, A. Bonaldi, D. Clements,
G. De Zotti, C. Dickinson, J. Dunkley, M. Lopez-Caniego, E. Mart\'inez-Gonzalez, M. Negrello,
S. Ricciardi, L.~Toffolatti
\newline

\noindent 
This White Paper is an extended version of a 
proposal document that was submitted to ESA in December 2010 in response to a Call for Proposals within the framework of 
ESA's Cosmic Vision 2015-25.
The proposal is the product of the cosmic microwave background observation
communities from France, Italy, Spain, and the United Kingdom, as
well as several other European countries (presently Denmark,
Germany, Ireland, the Netherlands, Norway, Portugal, 
Sweden, Switzerland). The
US community as represented by the PPPT group and its chairman
(S. Hanany) has expressed strong interest in an extensive
collaboration if this proposal is successful. 
A full list of the people currently  involved with \core\
may be found at our website: {\bf http://www.core-mission.org}.
Above we have listed those members of the \core\ community who have contributed
most actively in preparing the proposal and white paper under the working group
to which they have contributed the most.\newline

\vskip 1in
\centerline{Revised version: 28 April 2011}

\newpage

\phantom{BOO}

\vskip 7cm

\centerline{\bf ABSTRACT}

\vskip 1in 

{\large

\noindent
COrE (Cosmic Origins Explorer) is a fourth-generation full-sky, microwave-band satellite 
recently proposed to ESA within Cosmic Vision 2015-2025. 
COrE will map the polarization of the microwave sky with such a high precision that tensor modes 
produced by the inflationary expansion are detected at more than 3$\sigma$ even if they are 0.1\% of 
the scalar modes. This is a factor about 50 times better than what PLANCK can achieve. 
COrE will provide maps of the microwave sky in 15 frequency bands, ranging from 45 
GHz to 795 GHz, with an angular resolution ranging from 23 arcmin (45 GHz) to 
1.3 arcmin (795 GHz) and sensitivities roughly 
10--30 times better than PLANCK (depending on the frequency channel). The COrE mission will lead to 
breakthrough science in a wide range of areas, from primordial cosmology to galactic 
and extragalactic science. COrE is designed to detect the primordial gravitational waves 
generated during the epoch of cosmic inflation at more than $3\sigma $ for $r=(T/S)\ge 
10^{-3}$. It will also measure the CMB gravitational lensing deflection power spectrum to the cosmic
variance limit on all linear scales, allowing us to 
probe absolute neutrino masses better than laboratory experiments and down to plausible 
values suggested by the neutrino oscillation data. COrE will also search for primordial 
non-Gaussianity with significant improvements over PLANCK in its
ability to constrain the shape and amplitude of non-Gaussianity. 
In the areas of galactic and extragalactic science, in its highest frequency 
channels COrE will provide maps of the galactic polarized dust emission allowing us to map 
the galactic magnetic field in areas of diffuse emission not otherwise accessible to probe 
the initial conditions for star formation. COrE will also map the galactic synchrotron 
emission thirty times better than PLANCK. This White Paper reviews the COrE science program, 
our simulations on foreground subtraction, and the proposed instrumental configuration.

}

\vfill
\newpage

\newpage

\vfill\eject

\setcounter{tocdepth}{3}
\begin{spacing}{0.8}
\tableofcontents
\end{spacing}

\vfill\eject
\setcounter{page}{1}

\section{Overview of polarized microwave sky}

\textbf{
COrE is a fourth-generation full-sky, microwave-band satellite 
that has been proposed to ESA within the context of Cosmic Vision 2015-2025
to follow on the successes of the COBE, 
WMAP, and PLANCK space missions. COrE will map the polarization
of the microwave sky with a relative precision comparable to that of the temperature
maps that PLANCK is now in the process of delivering.
COrE will provide maps of the microwave sky in 15 frequency bands,
ranging from 45 GHz to 795 GHz with an angular resolution roughly comparable to PLANCK and a
sensitivity 10--30 times better (depending on the frequency channel).
The sensitivity of COrE, which
nominally corresponds to over 250 years of PLANCK integration time,
matches the need to observe the polarized 
signal whose level is only a few percent of the temperature 
anisotropy.} 

\textbf{
The COrE mission will lead to breakthrough science in a wide variety of  
areas, ranging from primordial cosmology to galactic and extragalactic science. COrE
is designed to detect the primordial gravitational waves generated
during the epoch of cosmic inflation at more than $3\sigma $ for $r=(T/S)\ge 10^{-3}$.
COrE will also measure the CMB gravitational lensing power
spectrum with unprecedented precision, allowing us to probe 
absolute neutrino masses better than is
possible in laboratory experiments and down to plausible values
suggested by the neutrino oscillation data. 
COrE will search for primordial non-Gaussianity, essentially at the 
theoretical limit of what is possible with the CMB. While PLANCK will
measure the E-mode polarization with a signal-to-noise oscillating in
the neighborhood of unity for $\ell \ltorder 800,$ COrE will provide
$(S/N) \gg 1$ maps up to $\ell \approx 2000$---a major step forward since it is the number
of independent resolution elements that determines the lensing science
and non-Gaussianity search capabilities of a survey. On the front
of galactic and extragalactic science, in its highest frequency channels
COrE will provide maps of the galactic polarized dust emission with 
a precision not possible from the ground.
This data will allow us to map the galactic magnetic field in 
areas of diffuse emission not otherwise accessible, thus probing
the initial conditions for star formation. COrE will 
also map the galactic synchrotron emission ten times better
than PLANCK and WMAP at 45 GHz where the Faraday rotation is
small. In the highest frequency channels CORE's angular resolution
will be diffraction limited with a beam size of $1.3'$ full width
at half maximum (fwhm). 
This enhanced resolution combined with COrE's exquisite
sensitivity will lead to the discovery of numerous compact
sources across the whole sky with well-defined selection
criteria. 
}

We review the diverse science objectives motivating the mission,
formulate the instrumental requirements for achieving these
objectives, and discuss the procedure for separating all the many
components contributing to the observed microwave sky in this frequency 
range. An adequate and reliable component separation is a prerequisite for
achieving the targeted science objectives. 
Finally we describe the proposed instrument and the proposed technology.

Observations of the microwave and far infrared sky have 
revolutionized our understanding of the origins of the
Universe and the processes at play within our galaxy and 
beyond. Before proceeding to a detailed analysis of 
some of the highlights of the new science that will be made
possible with COrE, it is appropriate to review what is
known, mainly from temperature maps but also from the presently
available polarization data, concerning the several
components emitting in the microwave bands and then 
quickly review the physical mechanisms causing this emission
to be partially polarized, with a view toward understanding
qualitatively the new science that can be extracted. 

We begin our tour in the center of COrE's broad frequency coverage.
For the temperature
and polarization anisotropies, the central channels
from about 70--220 GHz offer a relatively clean window
through which the primordial CMB anisotropies 
dominate. At lower frequencies, a non-thermal galactic synchrotron
component becomes an increasingly dominant contribution to
the anisotropic signal. The synchrotron component
arises from the interaction of cosmic rays with the galactic magnetic
field and provides invaluable information to constrain the
galactic magnetic field. This component
has a spectral index redder than a blackbody by
approximately three inverse powers in frequency, allowing it
to be removed through this differing frequency dependence.
The WMAP data has provided a wealth of information
concerning the degree of polarization of 
this synchrotron emission, providing one of the inputs
for the COrE polarization cleaning forecasts. Nevertheless
a more precise template will be needed, hence the presence
of 45 GHz channel. At low frequencies, there is also
a free-free component, arising from bremsstrahlung of
the hot HII regions, tightly correlated with H$\alpha $
emission, which serves as a useful tracer. This emission
however is not intrinsically polarized. At the low end of the spectrum,
mostly below the frequency coverage of COrE,
one also has evidence for spinning dust emission. The 
details and degree of polarization of this emission are not
presently well understood. 

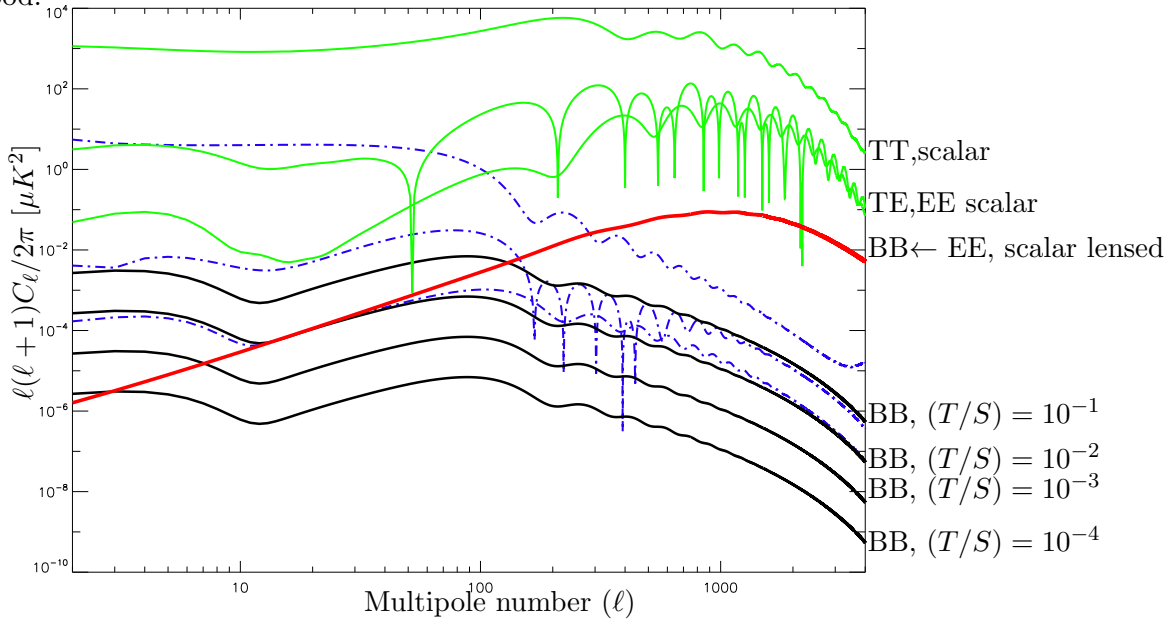
\begin{figure}[ht]
\begin{center}
\vskip 1.9cm
{
\setlength{\unitlength}{1cm}
\begin{picture}(14, 6)(0,0)
\put(11.7,6.2){TT,scalar}
\put(11.7,5.5){TE,EE scalar}
\put(11.7,4.9){BB$\leftarrow $ EE, scalar lensed}
\put(11.7,2.7){BB, $(T/S)=10^{-1}$}
\put(11.7,2.1){BB, $(T/S)=10^{-2}$}
\put(11.7,1.7){BB, $(T/S)=10^{-3}$}
\put(11.7,1){BB, $(T/S)=10^{-4}$}
\put(5.,0.2){Multipole number ($\ell $)}
\put(0.6,3){\begin{rotate}{90} $\ell (\ell +1)C_{\ell }/2\pi ~[\mu K^2]$ \end{rotate}}
\includegraphics[width=12cm]{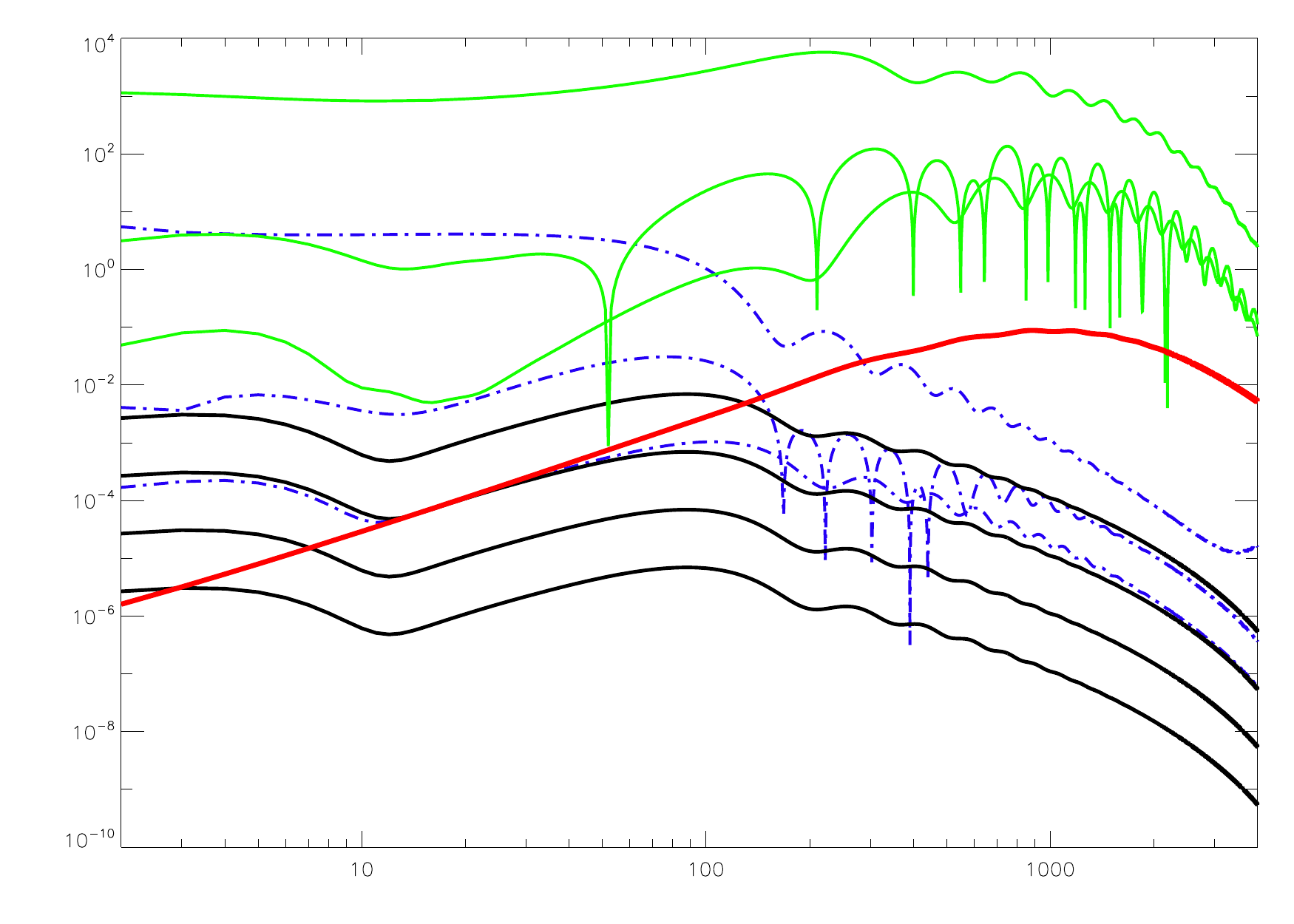}
\end{picture}
}
\end{center}
\vskip -0.5cm
\caption{\small \baselineskip=8pt { {\bf Inflationary prediction
for the CMB temperature and polarization anisotropies for the
scalar and tensor modes.} The horizontal axis indicates the
multipole number $\ell$ and the vertical axis indicates $\ell
(\ell +1)C_\ell ^{AB}/(2\pi )$ in units of $(\mu  K)^2$, which is
roughly equivalent to the derivative of the power spectrum with
respect to $\ln\ell$. The green curves indicate the TT, TE, and EE
power spectra (from top to bottom) generated by the {\it scalar}
mode assuming the parameters from the best-fit model from WMAP
seven-year data ($H_0=71.4\kmbysbyMpc$, $\Omega_b=0.045$, 
$\Omega_{cdm}=0.220$, $\Omega_{\Lambda }=0.73$, $\tau =0.086$, and
$n_s=0.969$). The BB scalar component (indicated by the heavy red
curve) results from the gravitational lensing of the EE polarized
CMB anisotropy at the last scattering surface $z\approx 1100$ by
structures situated mainly around redshift $z\approx 2$. The top
three blue curves (from top to bottom on the left) indicate the TT,
TE, BB, and EE spectra resulting from the
{\it tensor} mode assuming a scale-invariant ($n_T=0$) primordial
spectrum and a tensor-to-scalar ratio $(T/S)$ of $0.1$,
and the solid black curves indicate the BB spectra for the descending
values of $(T/S)=r=0.1, 0.01, 0.001$ and $0.0001.$ 
For the TE cross-correlations 
we have plotted the log of the absolute value, hence the downward spikes
which correspond to sign changes. 
} }
\label{Fig:UnderlyingAnisotropies} 
\end{figure}

Above the so-called CMB channels, above around 250 GHz,
a component of cold ($T\approx 20K$) dust starts to become the 
dominant contribution to the anisotropies as one pushes
into the exponentially falling Wien regime of the CMB 
blackbody spectrum. The physics of the dust grains
of the interstellar medium is complex and only partially
understood. Our understanding is summarized by empirical
models of the grain populations drawing from a variety
of observational inputs (frequency dependence 
of reddening including the presence
of absorption features, polarization of starlight, 
infrared emission,...) as well as more physical 
arguments (bottom-up modelling informed by cosmic element 
abundances). The infrared emission can be described approximately
with 
a blackbody at the dust temperature and 
an emissivity index in the neighborhood of $\alpha =1.7.$
Presently the most widely used dust model has
two dust components with separate 
temperatures and emissivity indices.

{\small

\begin{SCtable}
\begin{minipage}{0.6\textwidth}%
\begin{center}
\begin{tabular}{|c|c|c|c|c|c|c|}
\hline 
\hline 
$\nu $ & $\theta _{fwhm}$ & $n_{det}$ & \multicolumn{2}{|c|}{Temp (I)} &   \multicolumn{2}{|c|}{Pol (Q,U)} \\       
    & &           & \multicolumn{2}{|c|}{$\mu K\cdot $arcmin } &   \multicolumn{2}{|c|}{$\mu K\cdot $arcmin } \\       
\cline{4-7}
GHz & arcmin &   & RJ    & CMB   & RJ    & CMB   \\
\hline 
   23  & 52.8 &  2 &  413  &     418  &     584  &     592 \\
   33  & 39.6 &  2 &  413  &     424  &     584  &     600 \\
   41  & 30.6 &  4 &  365  &     381  &     516  &     539 \\
   61  & 21.0 &  4 &  438  &     481  &     619  &     681 \\
   94  & 13.2 &  8 &  413  &     516  &     584  &     729 \\
\hline 
\end{tabular}
\\
\vskip 12pt
{\bf WMAP (9 year mission)}\\
\vskip 0.5in
\begin{tabular}{|c|c|c|c|c|c|c|c|}
\hline 
\hline 
$\nu $ & $n_{unpol}$ & $n_{pol}$ & $\theta _{fwhm}$ & \multicolumn{2}{|c|}{Temp (I)} &  \multicolumn{2}{|c|}{Pol (Q,U)} \\       
 &             &           & & \multicolumn{2}{|c|}{$\mu K\cdot $arcmin} &  \multicolumn{2}{|c|}{$\mu K\cdot $arcmin} \\       
\cline{5-8}
GHz & & &  arcmin   & RJ    & CMB   & RJ    & CMB   \\
\hline 
   30 &   4   & 4 &  32.7& 198.5 & 203.2 & 280.7 & 287.4 \\
   44 &   6   & 6 &  27.9& 228.0 & 239.6 & 322.4 & 338.9 \\
   70 &  12  & 12 &  13.0& 186.5 & 211.2 & 263.7 & 298.7 \\
  100 &   8   & 8 &   9.9&  23.9 &  31.3 &  33.9 &  44.2 \\
  143 &  11   & 8 &   7.2&  11.9 &  20.1 &  19.7 &  33.3 \\
  217 &  12   & 8 &   4.9&   9.4 &  28.5 &  16.3 &  49.4 \\
  353 &  12   & 8 &   4.7&   7.6 & 107.0 &  13.2 & 185.3 \\
  545 &   3   & 0 &   4.7&   6.8 & $1.1\times 10^{3}$  &  ---  &   --- \\
  857 &   3   & 0 &   4.4&   2.9 & $8.3\times 10^{4}$  &  ---  &   --- \\
\hline 
\end{tabular}
\\
\vskip 12 pt
{\bf PLANCK (30 month mission)} \\
\vskip 0.5in
\begin{tabular}{|c|c|c|c|c|c|c|c|}
\hline 
\hline 
$\nu  $ & $(\Delta \nu )$ & $n_{det}$ & $\theta _{fwhm}$ & \multicolumn{2}{|c|}{Temp (I)}            &       \multicolumn{2}{|c|}{Pol (Q,U)} \\       
 &          &           &          & \multicolumn{2}{|c|}{$\mu K\cdot $arcmin} &  \multicolumn{2}{|c|}{$\mu K\cdot $arcmin} \\       
\cline{5-8}
$GHz$ & $GHz$&  & arcmin   & RJ    & CMB   & RJ    & CMB   \\
\hline 
      45   &   15 &    64    & 23.3& 4.98  &  5.25 &    8.61 & 9.07       \\   
      75   &   15 &   300    & 14.0& 2.36  &  2.73 &    4.09 & 4.72       \\   
     105   &   15 &   400    & 10.0& 2.03  &  2.68 &    3.50 & 4.63       \\   
     135   &   15 &   550    &  7.8& 1.68  &  2.63 &    2.90 & 4.55       \\   
     165   &   15 &   750    &  6.4& 1.38  &  2.67 &    2.38 & 4.61       \\   
     195   &   15 &  1150    &  5.4& 1.07  &  2.63 &    1.84 & 4.54       \\   
     225   &   15 &  1800    &  4.7& 0.82  &  2.64 &    1.42 & 4.57       \\   
     255   &   15 &   575    &  4.1& 1.40  &  6.08 &    2.43 & 10.5       \\   
     285   &   15 &   375    &  3.7& 1.70  &  10.1 &    2.94 & 17.4       \\   
     315   &   15 &   100    &  3.3& 3.25  &  26.9 &    5.62 & 46.6       \\   
     375   &   15 &    64    &  2.8& 4.05  &  68.6 &    7.01 & 119        \\   
     435   &   15 &    64    &  2.4& 4.12  &  149  &    7.12 & 258        \\   
     555   &  195 &    64    &  1.9& 1.23  &  227  &    3.39 & 626        \\   
     675   &  195 &    64    &  1.6& 1.28  & 1320  &    3.52 & 3640       \\   
     795   &  195 &    64    &  1.3& 1.31  & 8070  &    3.60 & 22200      \\   
\hline 
\end{tabular}
\\
\vskip 12pt
{\bf COrE summary (4 year mission)}
\\
\end{center}
\end{minipage}
\caption{
{\bf COrE performance summary compared to WMAP and PLANCK.} 
The Planck numbers reported here are based on instrument performance as
measured in flight and projected to the best current estimate of the
mission lifetime (i.e., 30 months). In the low frequency channels the
superior performance of COrE derives from a combination of the increased
number of detectors and the more sensitive bolometer technology, which
allows measurements to be made at nearly the quantum shot noise limit. For
example at 45 GHz, the lowest frequency channel of COrE, the Planck-LFI
integration time would have to be increased by approximately a factor of a
thousand to match the COrE sensitivity. In the central channels this
factor is reduced to about 100, because Planck-HFI, like COrE, already
uses bolometers cooled to 0.1K. The COrE bolometers, however, are much
more numerous than those of Planck and their individual perforance is
closer to the quantum limit. For galactic and extragalactic science using
polarization data, Planck is severely limited by the fact that the
Planck-HFI channels at 545 GHz and 857 GHz lack polarization sensitive
bolometers. Moreover, the angular resolution of COrE in the highest
frequencies is diffraction limited whereas for PLANCK it is not.
This is a major limitation because interstellar dust and many
other sources of interest are best studied at these high frequencies. COrE
will thus open a new window in mapping far infrared polarized emission.
These high frequency channels require only a small fraction of the COrE
resources, as they occupy a small portion of the focal plane area and
contribute a small fraction to the total data rate transmitted to Earth.
In Sec.~\ref{sect:Sens} we propose a modification 
of the baseline, where the
sensitivity at 795 GHz is increased by a factor of five, at negligible
cost to the primordial science, but greatly enhancing the COrE
capabilities for Galactic science.
}
\label{tab:CorePerf}
\end{SCtable}

}

\afterpage{\clearpage}

\begin{figure}[htb]
\begin{center}
\includegraphics[width=15cm]{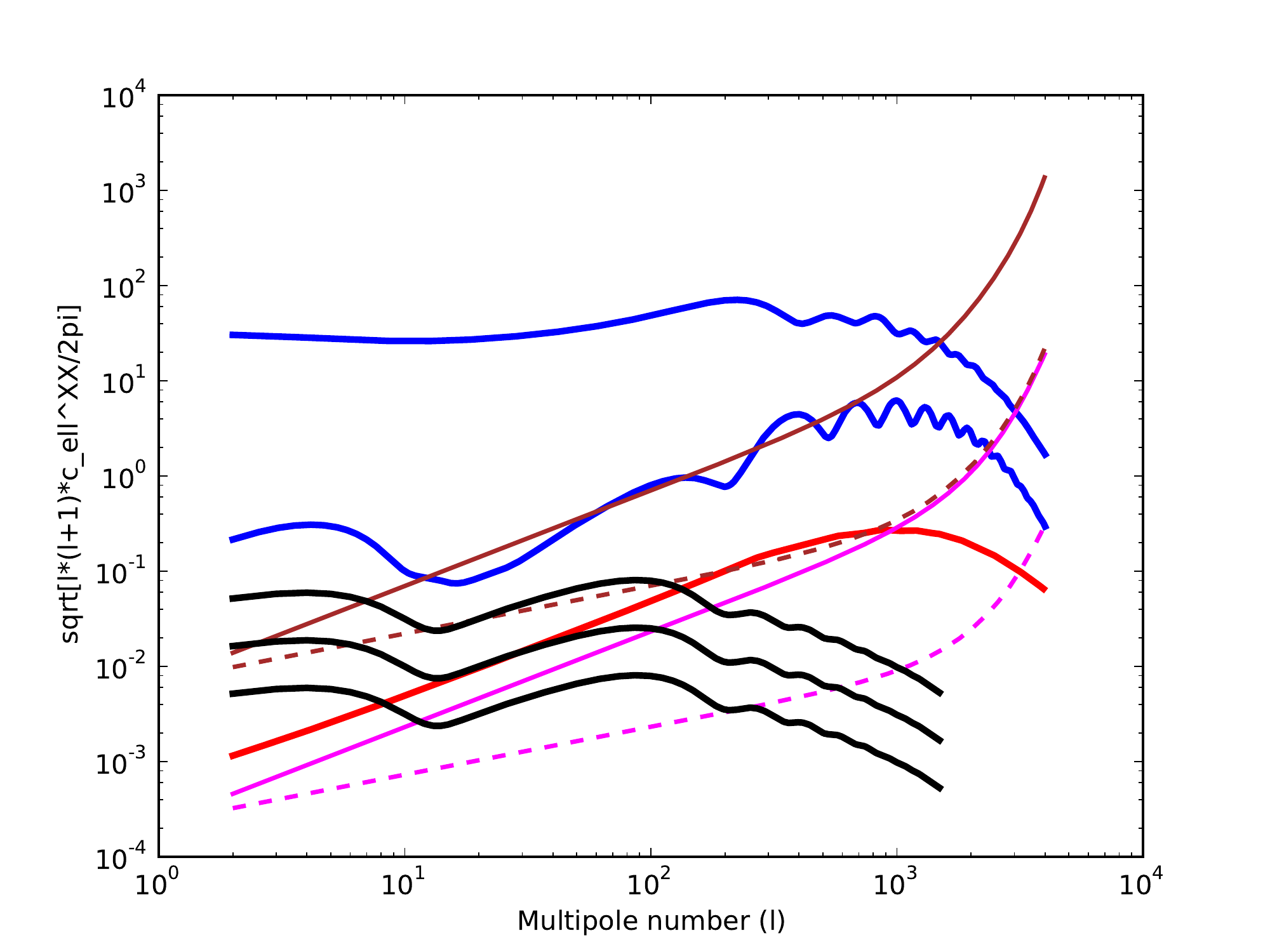}
\end{center}
\caption{{\bf CMB scalar and tensor anisotropies relative to PLANCK and COrE
sensitivities}
We show again the scalar (blue and red) and tensor (black) anisotropies 
but now together with the PLANCK and CORE instrument noise in brown and 
magenta, respectively. The scalar anisotropies are the TT (top) and EE (bottom)
spectra in blue and a BB lensing contribution (in red). These are relatively
certain. The BB is shown for $r$ of $0.1,$ $0.01$ and $0.001$ in black
(top to bottom). The solid brown and magenta curves show the instrument noise power
spectra for PLANCK and COrE, and the dashed counterparts below indicate what
sensitivity can be obtained by wide binning $(\Delta \ell )/\ell \approx 1.$
Modes of the CMB spectra lying above the solid noise curves are detected
with $(S/N)\gtorder 1,$ implying that measurements at higher instrumental 
sensitivity would result in marginal additional science value. On the other
hand, CMB spectra lying between the solid and dashed sensitivity give 
a signal detectable when many modes are combined in the analysis. 
We see that for $\ell \ltorder 800 $ PLANCK the $C_{EE}$ measurements
oscillate in and out of the $(S/N)=1$ threshold while for COrE 
modes up to $\ell \approx 2000$ are resolved with $(S/N)\gg 1.$
This means that for studies of non-Gaussianity one has
about ten times as many resolution elements. 
}
\end{figure}

All these emission processes (with the exception of free-free)
are partially polarized. The degree of polarization of the 
synchrotron emission is around 10\% and that of the dust around 
5--7\% , although large uncertainties persist. 
For the primordial component,
polarization can heuristically be understood as a measure
of the CMB quadrupole moment as seen by a typical electron
of last scattering, roughly proportional to the 
double derivative of the `temperature' and the square of the
mean distance between last and next-to-last scattering. This proportionality
explains why the electric-type (E) polarization 
from the scalar perturbations receives an appreciable
contribution from late-time reionization despite the fact that
the reionization optical depth is quite small 
$(\tau \approx 0.087).$ This is the mechanism 
that allowed the WMAP polarization data to break
the degeneracy between the amplitude of the primordial scalar
perturbations $A$ and the reionization optical depth $\tau .$ Current observations
of the polarization of the primordial CMB component are
in good agreement with the scalar perturbations and 
cosmological parameters implicit from the temperature
perturbations. Polarization provides a powerful cross
check because it probes the other of the two quadratures of the 
cosmological perturbations. 

The frontier of CMB polarization observations
lies in searching for the B mode.
The polarization field on the celestial sphere may be
divided into two components: an E mode, which may be 
expressed by means of second derivatives acting on 
a potential, and a B mode where this pattern is rotated
by $45^\circ .$ For scalar perturbations, which are the only ones 
that contribute to the matter power spectrum, only the 
E mode polarization is possible within linearized cosmological 
perturbation theory.
Nonlinear corrections, which may be calculated reliably, 
contribute a gravitational lensing background (shown in red in Fig.~\ref{Fig:UnderlyingAnisotropies})
having a white noise spectrum at low multipole number $\ell .$
This `white noise' at low $\ell $ has 
a magnitude of around $5 \muK\cdot {\rm arcmin},$
the precise value depending slightly on the cosmological
parameters. Any B mode with a black body spectrum 
beyond the level expected from lensing is the tell-tale
sign of primordial gravitational waves from inflation,
whose multipole spectral shape predictions
are shown in Fig.~\ref{Fig:UnderlyingAnisotropies}
alongside the predictions for the scalar anisotropies,
shown in green. The expected B-mode anisotropies from
inflation is parameterized by 
the ratio
of the primordial tensor perturbations relative to
the scalar perturbations 
$(T/S)$ or $r.$

The current COrE concept does not attempt to `clean' the
gravitational lensing B mode, but rather accepts it as
a background that can be well characterized and included
in the analysis in the same way as one typically deals with 
instrument noise in CMB experiments. Hence the science requirement
is to deliver a foreground cleaned map with an accuracy
in the neighborhood of or slightly better than  $5\mu K\cdot {\rm arcmin}$
after foreground subtraction. This requires a superior raw
sensitivity in order to clean out the foregrounds.

The so-called ``foregrounds'' may, on the one hand, 
be considered a nuisance for primordial cosmology.
They are the `dirt' that must be
removed to gain a glimpse at the pristine state of the 
primordial universe. But on the other hand, these foregrounds, which will be
characterized with exquisite precision, constitute
a gold mine for galactic and extragalactic science. In particular, the dust
polarization maps produced, when combined with 21 cm maps to
provide depth information, can be used to gain a better
understanding of the galactic magnetic field, which according
to equipartition arguments plays a key role in the dynamics
of the interstellar medium. Numerous extragalactic polarized
point sources will be discovered as well.

COrE will be competing with suborbital experiments likewise
aiming to detect non-zero $(T/S)$ and carry out other components
of the science program presented here. Suborbital experiments have
indeed played and will continue to play an important role in developing
and demonstrating new technology for space-based CMB observation. 
Nevertheless, suborbital experiments are substantially handicapped
in a number of ways that are analyzed in detail in Sec.~\ref{sec:why:space}.

\section{Science with COrE}
\label{sec:science}

\subsection{Cosmic inflation}

{\bf 

Inflation represents our current best understanding of the physics at play
in the primordial Universe. Originally inflation was proposed to solve
several cosmological conundrums---the famous horizon, flatness, smoothness,
and monopole problems---using ideas from high-energy physics near the Planck
scale, far beyond the reach of accelerator experiment. The successful
resolution of these problems is a result of analyzing inflation using
classical physics. However it was soon realized that when quantum
effects are taken into account, inflation also provides a mechanism
for explaining the nearly Gaussian nature of the primordial cosmological perturbations
and explains the empirically deduced Harrison-Peebles-Zeldovich scale
invariant spectrum of density perturbations. The scalar perturbations,
which seed the large-scale structure seen today in the universe, has been
extensively probed by WMAP, PLANCK, and other CMB experiments as well as
galaxy surveys. Inflation however makes an additional prediction that
has not yet been verified observationally---namely, that these scalar perturbations
should be accompanied by primordial gravitational waves having a very
red, scale-invariant spectrum. The confirmation of this prediction would
provide a remarkable qualitatively new test of inflation. It would
also establish the energy scale of inflation, providing invaluable
data for physics near the Planck scale. The amplitude of these gravitational
waves is parameterized by $r,$ the tensor-to-scalar ratio. PLANCK will be
able to detect around $r\ge 0.1.$ However, COrE can detect $r\ge 0.001$ at $3\sigma $---that is,
almost two orders better than PLANCK. Moreover, if PLANCK makes a marginal detection,
say somewhere around $0.1,$ the follow-up by COrE would provide a measurement at the cosmic variance 
limit.
}

There is compelling evidence that the early Universe underwent a period of very rapid expansion, 
driven by an approximately constant vacuum energy. As a result of this {\it cosmic inflation}, 
the Universe ended up in a very special state, almost perfectly homogeneous and empty, with a 
geometry that is almost exactly Euclidean. After inflation, the vacuum energy was converted into 
radiation and matter, which filled the Universe. It is satisfying that present cosmological 
observations can be elegantly explained using simple inflationary models. 

A key prediction from inflation is that the observed large scale structure was seeded by quantum 
fluctuations in the fabric of spacetime, stretched to cosmological distances by the expansion. 
Those seeds can be described by almost perfect Gaussian random fields, with an almost 
scale-invariant spectrum for both scalar and tensor fluctuations. The scalar fluctuations are the 
seeds for adiabatic density perturbations that lead to the formation of the cosmic web. The 
tensor fluctuations are the gravitational waves that may soon be detected in the large scale 
polarization of the cosmic microwave background (CMB).

Inflationary theory has been extraordinarily successful when confronted with the new spate of 
high quality cosmological data. Most notably, the WMAP data have confirmed that large scale 
density fluctuations can be characterized in terms of an almost scale-invariant, adiabatic, Gaussian 
random field. Furthermore, PLANCK will soon characterize the fine details of this random field, 
in particular by probing expected small deviations from exact scale invariance and constraining
(or possibly discovering) deviations from Gaussianity and adiabaticity. These precise measurements 
can then be exploited to uncover the details of cosmological inflation itself.
Some of the questions we would like to address include:  What type of fields were responsible 
for inflation? What are their properties? How do they interact with the remaining fields of the 
standard model of physics and cosmology? Our present understanding is very incomplete. 

We can begin to understand the workings of the early Universe by targeting observations that 
directly probe the energy scale and the dynamics of inflation, through measurements of the 
running of the spectral index (defined below) and the amplitude of gravitational wave 
fluctuations relative to density fluctuations, as well as through deviations from Gaussianity 
and adiabaticity. The temperature and polarization anisotropies of the CMB offer the best 
observables for making progress on this front. 

\subsubsection{Physics of inflation}

We now describe the simplest model of inflation based on a single scalar field $\phi 
$ having a potential $V(\phi)$. The description presented here can be easily generalized to a 
broad class of models involving more fields. If we focus on the overall dynamics of the 
Universe, the energy density residing in a homogeneous scalar field is the sum of the kinetic and potential 
terms $\rho=\frac{1}{2}{\dot \phi}^2+V(\phi)$, while its pressure is given by $p= 
\frac{1}{2}{\dot \phi}^2-V(\phi)$. The energy and pressure of the scalar field drive the 
expansion of the Universe according to the Einstein equations
\begin{eqnarray}
H^2\equiv\left ( \frac{\dot a}{a} \right )^2 = \frac{8\pi G}{3}\rho- \frac{K}{a^2} \, ,
\hspace{2cm}
\frac{\ddot a}{a} = -\frac{4\pi G}{3}(\rho+3p) 
\label{FRW}
\end{eqnarray}
where $a(t)$ is the scale factor of the Universe.
To obtain an epoch of inflationary expansion, 
we consider a regime where as a result of the rapid expansion of the Universe, the scalar field 
is moving sufficiently slowly so that the kinetic term is negligible compared to the potential 
term. In this {\it slow roll} regime, $p\approx -\rho\approx -V(\phi),$ and the scalar field 
plays the role of a cosmological constant, albeit slowly decaying. From eqn.~(\ref{FRW}) we 
observe (for a spatially flat universe with $K=0$) that the expansion accelerates with the scale factor evolving as 
$a(t)\propto e^{Ht}$ where $H^2=(8\pi G/3)V(\phi )$. The evolution equation 
$\ddot{\phi}+3H\dot{\phi}=-V'$ in the slow-roll approximation becomes 
`friction-dominated,' with $3H\dot{\phi}\approx -V'$.

From this simple one-field model we can extract some key consequences. The geometry of the 
Universe is closely tied to the fractional energy density of the Universe, $\Omega\equiv 
\rho/\rho_C$, where $\rho_C=(3/8\pi G)H_0^2$ is the critical energy density and $H_0$ is the 
Hubble constant today.  During an inflationary period we have that $|1-\Omega|\propto 
\exp(-2Ht)$, which implies that cosmic evolution will drive $\Omega\rightarrow 1$, (i.e., to 
a spatially flat, Euclidean geometry). This is a generic prediction of 
inflation and has been borne out through 
observations of the CMB, such as those from the WMAP satellite, which constrain $\Omega$ to be 
close to $1$ within a few per thousand. The PLANCK constraints promise to be even more stringent.
A period of inflation also resets the initial state of the 
observable Universe, since a patch of space that undergoes inflation becomes exponentially 
stretched and smoothed. Again, observations of the CMB show that the Universe is smooth to one 
part in $10^5$ on large scales, up to several gigaparsecs.

\subsubsection{Observing inflation}

The statistical properties of the large-scale structure imprinted during inflation depend on the 
form of the potential $V(\phi)$ for the scalar field $\phi $ driving inflation. It is convenient 
to define the following dimensionless {\it slow-roll} parameters, characterizing the shape of the 
inflationary potential
\begin{eqnarray}
\epsilon \equiv \frac{M^2_\mathrm{Pl}}{2}\left( \frac{V'}{V}\right) ^2
=\frac{1}{2}\left( \frac{d \ln V(\phi)}{d(\phi /M_\mathrm{Pl})} \right)^2,
\hspace{2cm}
\eta \equiv M^2_\mathrm{Pl}\frac{V''}{V}= \frac{d^2\ln V(\phi)}{d(\phi /M_\mathrm{Pl})^2}-2\epsilon 
\end{eqnarray}
where $M_\mathrm{Pl}=(8\pi G)^{-1/2}=2.4\times 10^{18}$ GeV is the reduced Planck mass. Both 
parameters must be small during inflation and we shall assume the slow-roll approximation 
$\vert \epsilon \vert, \vert eta \vert \ll 1$.

Scalar perturbations arising from inflation imprint inhomogeneities in the energy density of the 
Universe, which can be described as a Gaussian random field with amplitude $A_S$ (i.e., 
$\left<\delta_k^2\right>=(2\pi^2/k^3)A_S^2$), that depends on the wavenumber $k$ with spectral 
index $(n_S-1)$, so that
\begin{eqnarray}
A_S(k)=\frac{2}{5}{\mathcal P^{1/2}_{R}} (k)  \approx 
\frac{\epsilon^{-1/2}}{5\pi\sqrt{3}}\left.\frac{V^{1/2}(\phi)}{M_\mathrm{Pl}^2}\right|_{k=a(\phi )H(\phi )},
\hspace{1cm}
n_S\equiv1+\frac{d\ln A_S^2(k)}{d\ln k}\approx 1+2\eta-6\epsilon .
\end{eqnarray}
Here $k$ is the comoving wavenumber, and the fluctuations on a given scale 
are imprinted as that scale ``crosses 
the horizon.''  ${\mathcal P_{R}}^{1/2}$ is the rms amplitude of the scalar metric perturbations. 
In the extreme slow-roll limit, the spectrum of density perturbations is 
exactly scale invariant---in other words, $n_S= 1.$

Inflation also generates tensor perturbations or primordial gravitational waves. Tensor perturbations
are transverse traceless perturbations of the spacetime metric $g_{ij}=a^2(t)(\delta_{ij}+2h_{ij})$. 
They include two spin-2 polarization states $h_+$ and $h_\times,$ which obey the evolution
equations of a massless field.
Their amplitude and spectral index are given by
\begin{eqnarray}
A_T(k)\equiv \frac{1}{5\sqrt2}{\mathcal P^{1/2}_{gw}}
\approx\frac{1}{5\pi\sqrt{3}}\frac{V^{1/2}(\phi)}{M_\mathrm{Pl}^2}\Biggl| _{k=a(\phi )H(\phi )},
\hspace{1cm}
n_T\equiv\frac{d\ln A_T^2(k)}{d\ln k} \approx -2\epsilon , 
\end{eqnarray}
where $A_T$ is defined so that $A^2_T/A^2_S=\epsilon$ to lowest order. The relative contribution 
of gravity waves to curvature perturbations is given by the {\it tensor-to-scalar} ratio
\begin{eqnarray}
r =T/S \equiv \frac{{\mathcal P_{gw}}}{{\mathcal P_{R}}} =
16\frac{A^2_T}{A^2_S}\approx 16\,\epsilon\label{ratio} = -8\,n_T\,,
\end{eqnarray}
giving the so-called {\it consistency condition}.
Using the COBE normalization $A_S=1.91\times10^{-5}$, we find that
\begin{eqnarray}
M_{\rm inf}\equiv V^{1/4}=3.3\times 10^{16} \: r^{1/4} \:{\rm GeV} \label{EnergyInf}
\end{eqnarray}
relating $r$ to the energy scale of inflation, $M_{\rm inf}$.

\subsubsection{Models of inflation}

Inflation to date is still a paradigm more than a well defined model. Many different inflationary 
models and implementations of the inflationary mechanisms have been proposed, but these models by 
no means cover all the possibilities. Since inflation happened in the very early Universe, 
observational tests of inflation offer a window into extremely high energy scales. The physics of 
inflation lies far outside the reach of terrestrial experiments. Thus cosmological tests of inflation 
offer a unique opportunity to use the early Universe as a laboratory to learn about ultra-high 
energy physics.  To go beyond the generic picture of a scalar field in a flat potential, we need 
to answer questions about the underlying physics, such as the following: What principles fix 
the shape of the potential? How do the symmetries that protect its flatness arise? Was there more 
than one dynamically active field? How does the Universe reheat after inflation? What sets the 
initial conditions for inflation itself? 
To answer such questions and directly test the physics of inflation requires new clues. 
The stochastic background of gravity waves is a new and unique prediction of inflation, unlike the
scale invariant density fluctuations,  which were postulated prior to the development of inflation
and then beautifully explained by inflation. 
One might say that the tensor modes with the predicted scale-invariant spectrum are the `smoking gun' of
inflation. More conventional sources of gravitational waves peak at much higher frequencies. Consequently
the detection of gravitational waves on cosmological scales would constitute a truly revolutionary discovery. 
By measuring or constraining primordial gravitational waves, it is 
possible to start quantitatively ruling in or out specific models---that is, physical implementations 
of the inflationary paradigm, and thus shed light on the specific connections with fundamental 
physics at the highest energies.

\vskip 4pt
\centerline{\bf Single-field inflation} 
\vskip 4pt

The simplest class of inflation models is that of single-field inflation. Such models have been 
shown to satisfy the following relation, originally due to Lyth~\cite{lyth:1997},
\begin{eqnarray}
{(\Delta \phi )\over M_{\rm Pl}}\approx \left({r\over 0.01}\right)^{1/2},\label{lyth}
\end{eqnarray}
where $(\Delta \phi )$ is the variation in the inflaton field between the end of inflation and the 
time at which CMB-scale perturbations were generated. The Lyth relation implies that tensor modes 
may be detectable if inflation involves a large field variation. It is therefore convenient to 
classify single-field models into two broad categories, namely `large field' 
($\Delta\phi\gtrsim M_{\rm Pl}$) and `small field' ($\Delta\phi< M_{\rm Pl}$) inflation.

%

Inflationary models with a monomial potential $V(\phi)\propto \phi^{\alpha}$ are perhaps the most widely 
studied class of large-field models. The case when $\alpha$ is a positive integer is typical of 
the `chaotic' inflation scenario \cite{staro,linde83} in which inflation begins with a chaotic 
initial condition. In particular, the model with $\alpha=2$ is generally regarded as the simplest 
and best motivated model still consistent with observation. Other models in this class include 
those in which $\alpha$ is a negative integer \cite{barrow90} or a fraction \cite{silverstein08}, 
as in the case of some string-inspired models.  Another theoretically compelling large-field model 
is `natural' inflation \cite{freese} with $V(\phi)\propto 1+ \cos({\phi/f})$, where $\phi$ may be 
identified with the axion and $f$ is the energy scale at which a global $U(1)$ symmetry is broken 
in the early Universe. In general, large-field models will be severely constrained if no tensor 
modes are detected at the level of $r\sim0.1$.

Small-field inflation models are much more difficult to rule out via the tensor amplitude, 
with most models generically predicting $r<0.01$. These models include 
`new' \cite{lindenew, albrechtnew} inflation associated with spontaneous symmetry breaking in the early Universe. 
Many small-field models can be represented by a hill-top potential, $V(\phi)\propto 1-(\phi/\mu)^p$ ($\mu,p$ are constants), 
upon which the inflaton rolls down towards a displaced vacuum. Because the tensor amplitude 
generated is undetectably small, small-field models may be constrained more effectively by 
measurements of the spectral index and its running.

There is, of course, still a plethora of single-field models in the literature, with some 
straddling the division between large field and small field. A measurement of the tensor amplitude using 
the CMB will be invaluable in discriminating and ruling out a large class of single-field models.

\vskip 4pt
\centerline{\bf Multi-field inflation} 
\vskip 4pt

Unified field theories (SUSY, SUGRA, GUTs, etc.) contain an abundance of light scalar fields. 
Therefore it is conceivable that inflation may involve interactions between a number of scalar 
fields. A simple example is `hybrid' inflation \cite{lindehybrid}, in which the inflaton field 
$\phi$ rolls slowly until it reaches a critical value set by a `waterfall' field $\chi$, which 
breaks some as yet unidentified symmetry and falls to the true vacuum, thus ending inflation suddenly. Hybrid models 
can produce both red ($n_S <1$) and blue spectra ($n_S >1$), and yields a negligible tensor amplitude. 
More complicated multi-field models can be constructed with a large number of fields (possibly of 
order $10^3$ as in the case of ``assisted" or $\mathcal{N}$-flation \cite{liddle1, dimopoulos}), 
all of which may evolve along separate potentials. These models typically predict large 
isocurvature perturbations, which seem to be in conflict with measurements of CMB anisotropies. 
Moreover, these additional degrees of freedom inevitably lead to a broad spectrum of predictions 
for the amplitude of tensor modes. The prospects for constraining multi-field inflation therefore 
appear extremely challenging.

\vskip 4pt
\centerline{\bf String models} 
\vskip 4pt

String theory at present offers the most compelling theory proposed to unify all the fundamental 
forces. If the energy scale of inflation is close to that at which string theory operates, it may 
be possible to detect `stringy' signatures in cosmological observations, especially in the 
amplitude of tensor modes. In the context of string theory, one possibility is that the 
observable Universe is part of a 3-dimensional `brane,' embedded within a higher dimensional 
`bulk' \cite{dvalitye, kklmmt} . Inflation occurs as the brane moves along a region of the bulk, 
perhaps towards an anti-brane of the opposite charge, with the inter-brane distance playing the 
role of the inflaton. Inflation ends when the brane separation reaches a critical value and the 
standard hot Big Bang subsequently ensues when the branes annihilate. Typically extra `flux' 
fields are also required to stabilize the very light fields associated with very flat potentials 
needed for successful inflation. Generally, brane inflation predicts a very low tensor 
amplitude ($r \ll10^{-3}$), and therefore a measurement of $r\sim\mathcal{O}(10^{-2})$ would 
severely constrain such string inflation models.

Brane inflation is typically accompanied by copious production of cosmic strings. These strings 
subsequently decay into gravitational waves with amplitude depending on the tension, $G\mu$, 
ranging from $10^{-6}$ (GUT-scale strings) down to $10^{-13}$ (cosmic \textit{super}strings) 
or smaller. A tensor amplitude of $T/S\sim10^{-3}$ roughly corresponds to a string tension of
$G\mu\sim10^{-7}$, and 
thus there is a good prospect for ruling out cosmic strings with large tension. In addition, 
cosmic superstrings with cusps and kinks can produce intense bursts of gravitational waves with a 
distinctive spectrum. A measurement of tensor modes will therefore improve our understanding of the 
interactions of cosmic strings and the dynamics of brane inflation.

\begin{figure}[htbp]
\center{
\includegraphics[width=6.5in]{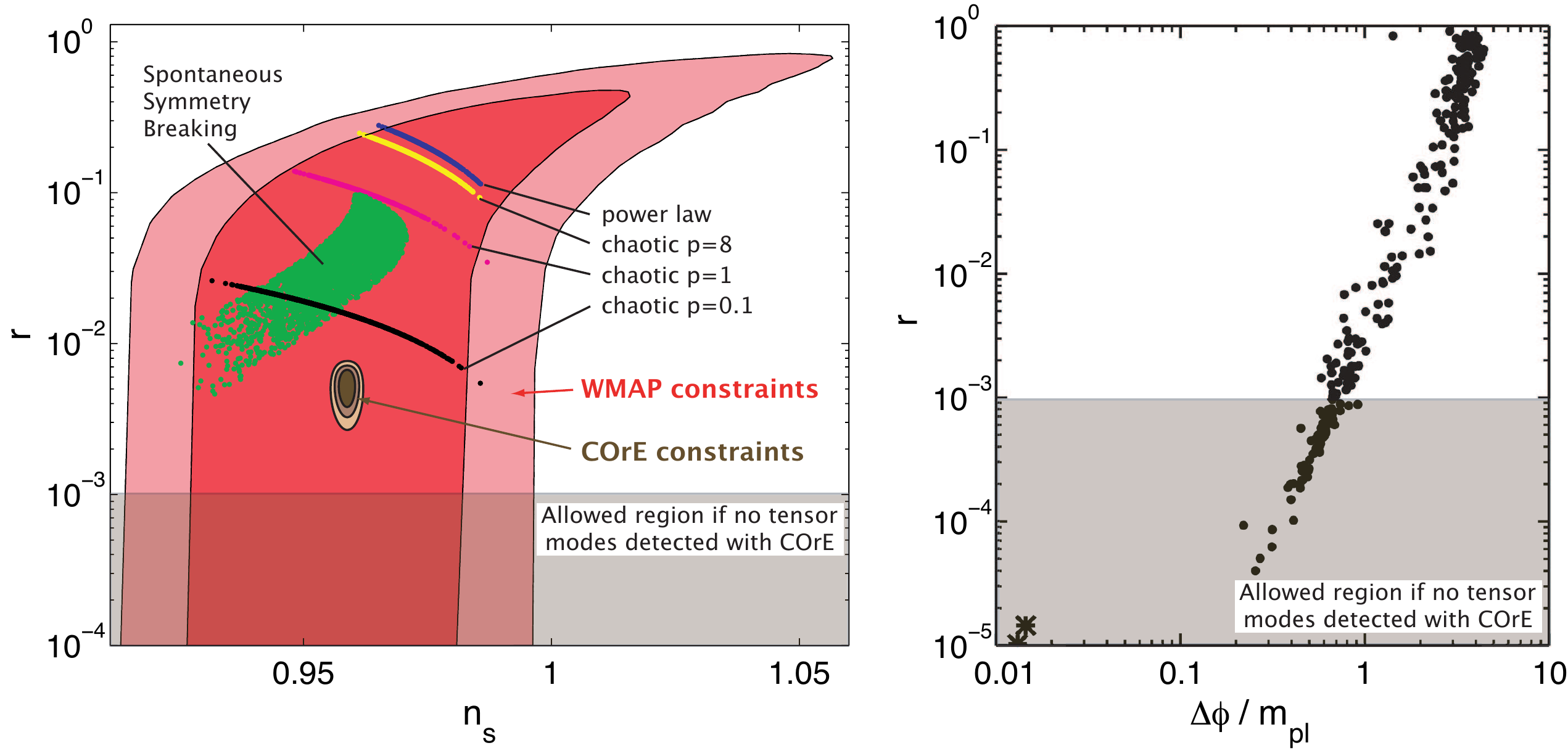}
}
\caption{{\bf Constraints on inflation from COrE}. For a broad range of inflationary models 
COrE can be expected to detect primordial gravitational waves from inflation. The 
large contours on the left panel show the present constraints from WMAP seven-year data 
in the $r$-$n_S$ plane. A few parameterized families of inflationary models give an idea of 
representative model predictions. The small contours illustrate what a COrE detection would 
look like if $r >5\times 10^{-3}$. The part of parameter space still allowed at 2$\sigma$ in the case of 
a non-detection is shown in grey. The right panel shows the `main sequence' of inflationary models 
generated using a model independent approach.}
\label{figure1}
\end{figure}

\subsubsection{Model independent analysis} 
\label{sec:modelindependent}

One can reformulate the exact dynamical equations for inflation as an infinite hierarchy of flow 
equations described by the generalized `Hubble slow roll' (HSR) parameters 
\cite{hoffman/turner:2001,kinney:2002,Liddle2003,easther/kinney:2003, peiris/etal:2003, 
kinney/etal:2004,VPJ06}. In the Hamilton-Jacobi formulation of inflationary dynamics, one 
expresses the Hubble parameter directly as a function of the field $\phi$ rather than a function 
of time, $H \equiv H(\phi)$ under the assumption that $\phi$ is monotonic in time. Then the 
equations of motion for the field and background are given by 
$$\dot{\phi} =-2 M_\mathrm{Pl}^2 
H'(\phi),\qquad V(\phi) = 3M_{\rm Pl}^2H(\phi)^2 - 2M_{\rm Pl}^4H'(\phi)^2$$
where the primes denote 
derivatives with respect to the field $\phi$. The second equation allows us to consider inflation 
in terms of $H(\phi)$ rather than $V(\phi)$. This approach has the advantage that it allows us to 
remove the field from the dynamical picture altogether, and study the generic behavior of slow 
roll inflation without making assumptions about the underlying particle physics (although the 
underlying assumption of a single order parameter is still present). In terms of the HSR 
parameters, $\epsilon$ and $^{\ell}\lambda_H$, the dynamics of inflation is described by the 
infinite hierarchy
\begin{equation}
\epsilon(\phi) = 2M_{\rm Pl}^2\left(\frac{H'(\phi)}{H(\phi)}\right)^2\,,
\hspace{1cm}
^{\ell}\lambda_H \equiv \left(2M_{\rm Pl}^2\right)^\ell
  \frac{(H')^{\ell-1}}{H^\ell} \frac{d^{(\ell+1)} H}{d\phi^{(\ell+1)}}, \qquad {\rm for} ~~\ell\geq 1 .
  \label{eq:hier}
\end{equation}

The flow equations allow us to consider the model space spanned by inflation using Monte Carlo 
techniques. Since the dynamics is governed by a set of first-order differential equations, one 
can specify the entire cosmological evolution by choosing values for the slow-roll parameters 
$^{\ell}\lambda_H$, which completely specifies the inflationary model. However, in 
practice one has to truncate the infinite hierarchy at some finite order. We retain terms up to 
tenth order having checked robustness against the choice of truncation order.  We will 
discuss the meaning and the implications of this truncation below. Moreover, the choice of 
slow-roll parameters for the Monte Carlo process requires assumptions about priors for the 
ranges of values taken by the $^{\ell}\lambda_H$. In the absence of {\it a priori} 
theoretical knowledge about these priors, one can assume flat priors with some ranges dictated by 
current observational limits, and the requirement that the potential satisfies the slow-roll 
conditions. Changing this `initial metric' of slow-roll parameters changes the clustering 
of phase points on the resulting observational plane of a given Monte Carlo simulation. Therefore, 
the results from these simulations cannot be interpreted in a statistical way. 
Nevertheless much intuition can be gained from the results of such simulations (e.g.,
\cite{kinney:2002, peiris/etal:2003,chongchitnan}). The simulations show that models do not cover the 
observable parameter space uniformly but instead cluster around certain attractor regions.
  
Inflation requires a form of stress-energy which sources a nearly constant Hubble parameter.  
This can arise via a truly diverse set of mechanisms with disparate phenomenology and varied 
theoretical motivations. As mentioned above, in the Hubble slow-roll approach it is possible to 
characterize, in a model independent way, single field models of inflation, including `hybrid' 
models where, although more than one field is involved, the generation of primordial 
perturbations is still governed by a single scalar field. This approach has shown 
\cite{VPJ06,efstathiou/mack:2005} that there is a model independent relation between the 
excursion of the field during inflation ($\Delta \phi$) and the amplitude of tensor modes, a 
generalization of the well known Lyth bound \cite{lyth:1997},
\begin{eqnarray}
&& \frac{\Delta \phi}{M_\mathrm{Pl}}\sim \left(\frac{r}{0.01}\right)^{1/2} \hspace{1.6cm}  {\rm for}  \,\,\,\,\, \frac{\Delta \phi}{M_\mathrm{Pl}} \ll 1, \\
&& \frac{\Delta \phi}{M_\mathrm{Pl}}\sim  6.3\, \left(\frac{r}{0.01}\right)^{1/4} \hspace{1cm} {\rm for}\,\,\,\,\,    \frac{\Delta \phi}{M_\mathrm{Pl}}\sim 1.
\end{eqnarray}
The second line is a fit to the relation shown in Fig.~\ref{figure1}. 
 
The right panel in Fig.~\ref{figure1} shows extracts from a 2 million point Monte Carlo 
simulation of the inflationary flow equations (adapted from ref.~\cite{VPJ06}).

While the choice of slow-roll parameters for the Monte-Carlo process requires the assumption of 
some prior ranges, the results of the simulations do not depend strongly on these choices once 
known observational constraints (on $n_S$ and $dn_S/d\ln k$) are imposed.  This observation is 
what makes the conclusions of this section model independent.  Note that the simulations show 
significant concentrations of points with a significant tensor-to-scalar ratio.

From these considerations, it is clear that a value of $r > 10^{-3}$ would imply that inflation occurred 
at energy scales $\gtrsim 2 \times 10^{15}$~GeV and that there was a super-Planckian field variation. 
Therefore $r \sim 10^{-3}$ is a natural target for a CMB polarization experiment. A detection of 
primordial tensor perturbations $r > 10^{-3}$ would probe physics at an energy that is a staggering twelve 
orders of magnitude larger than the center-of-mass energy at the Large Hadron Collider. Of equal 
importance is the fact that a detection or constraint on the tensor-to-scalar ratio $r$ at this level will 
answer a fundamental question about the range $\Delta \phi$ of the scalar field excursion during inflation 
as compared to the Planck mass scale. This would yield vital clues about physical symmetries at these unexplored 
energy scales including the ultraviolet completion of gravity.

\subsubsection{Forecasts for $r=T/S$}

As the foreground signal is expected to dominate the cosmological signal at low $\ell$ at all frequencies, 
it is of crucial importance to propagate uncertainties connected to foreground contamination into the 
parameter error forecasts. 

An estimate of the error on cosmological parameters in the presence of residual foregrounds is obtained here following the approach proposed in 
ref.~\cite{VPJ06} and further developed in ref.~\cite{Baumannetal}.
We assume that the  knowledge  about foreground residuals and their uncertainty is available  from 
estimates obtained by component separation methods as discussed in section \ref{sec:fgs}. We assume the same level 
of foreground residuals for both E and B CMB modes. Component separation residuals in $T$, assumed to 
be much smaller than the CMB temperature cosmic variance, are neglected here.

\vskip 4pt
\centerline{\bf Forecasts: method}
\vskip 4pt

In refs.~\cite{VPJ06} and 
\cite{Baumannetal} it was assumed that in each frequency band foregrounds could be subtracted down to a 
certain percentage of the original signal. As this supposes that the level of the foreground contamination
after component separation is known \emph{a priori}, we use here instead the residual foregrounds, uncertainties,
 and final noise level obtained from cleaning simulated maps 
of CMB plus Galactic and point source emission, as described in \S \ref{sec:fgs}.  

We find that the foreground cleaning approach based on the Needlet ILC (NILC) yields foreground residuals in agreement with the prediction made using the pixel-based linear component separation method (pix-LCS) as
described in \cite{2011arXiv1101.4876B}. This gives us confidence in the validity of the residual foreground estimates.

In the case of a realistic experiment (with partial sky coverage, and noisy data), assuming that CMB 
multipoles are Gaussian-distributed and that the noise is Gaussian, the likelihood $\cal L$ is given by

\begin{equation}
\frac{-2\ln{\cal L}}{f_\mathrm{sky}} = \sum_{\ell}(2\ell+1) \left\{ {\rm tr}[\hat R_\ell R_\ell^{-1}] - {\rm ln}\, {\rm det} \, [\hat R_\ell R_\ell^{-1}] -3 \right \},
 \label{eq:likelihood}
 \end{equation}
where $R_\ell$ is the theoretical multivariate covariance matrix of the $T,E,B$ fields, and $\hat R_\ell$ its empirical estimate on the observed data.
%
%
%
Here $R=C_{th}+N$ where
$$
C_{th} =\begin{pmatrix}
C_{th}^{TT} & C_{th}^{TE} & 0           \cr
C_{th}^{TE} & C_{th}^{EE} & 0           \cr
0           & 0           & C_{th}^{BB} \cr
\end{pmatrix},
\qquad
N =\begin{pmatrix}
N^{TT} & 0           & 0           \cr
0      & N^{PP}      & 0           \cr
0      & 0           & N^{PP}      \cr
\end{pmatrix}
$$
and $\hat R$ is defined analogously but with $C_{th}$ replaced by $C_{obs}.$
Here $f_\mathrm{sky}$ denotes the fraction of sky observed.
Note that this accounts for the 
effective number of modes accessible with partial sky coverage, but does not account for mode correlations 
introduced by the sky cut, which may smooth power spectrum features. The details of the mode-correlation depend on 
the specific details of the mask, but for $\ell$ greater than the characteristic size of the survey,
our approximation should be valid.

The expression for the likelihood in eqn.~(\ref{eq:likelihood}) with the specified values of $R$ and $\hat R$ is valid only for a single frequency experiment. 
However, cleaning the simulated 
multi-frequency maps yields (fore each of $T$, $E$ and $B$) a single map of one effective frequency, with an effective noise power spectrum and 
effective foreground residuals. We express the effect on the power spectrum of the residual Galactic contamination as 
an additional ``noise-like'' component, with a dependence on $\ell$ given by the spectral energy distribution
of the residual foregrounds. The effective noise also includes a term which is a function of the noise levels of the individual frequency channels.
The forecasts are then obtained from the Fisher information matrix (FIM)
\begin{equation}
F_{ij}= - \left. \left\langle \frac{\partial^2 \ln {\cal L}}{\partial \theta_i\partial\theta_j}
\right\rangle \right|_{\theta=\bar{\theta}}
\end{equation}
where the $\theta_i$ denote the various model parameters. We then have
$\sigma_{\theta_i}\ge (F^{-1})_{ii}^{1/2}$ (Cramer-Rao inequality).
This becomes an equality for the maximum likelihood estimator in the limit of large data sets, which we assume here.
We consider in our analysis the cosmological parameter set $$\bar{{\bf \theta}} = \{
r,n_S,n_T,dn_S/d\ln k, Z ,\Omega_b h^2 \equiv \omega _b, \Omega_c h^2 \equiv \omega _c, h, \Omega_k, A(k_0) \},$$ 
where $Z\equiv\exp(-2 \tau)$.

\vskip 4pt
\centerline{\bf Forecasts: results}
\vskip 4pt

Table~\ref{tab:forecastsr} reports the 1$\sigma$ forecast errors
on $r$ after marginalizing over  the other cosmological parameters, with 
and without imposing the consistency relation (CR) $r=-8n_t$.
The $B$ mode polarization signal comes mainly from two regions of the angular power spectrum. 
At $\ell <20$, the signal comes through the so-called `reionization bump', and there is therefore some sensitivity to 
the adopted fiducial value for $\tau $. At $\ell>20$ the signal comes from the $BB$ peak, but is contaminated by the 
lensing signal which depends on the perturbation amplitude, as well as by the instrumental noise, which becomes 
increasingly important with increasing $\ell $.

\vspace*{5mm}
\begin{table}[htb]\footnotesize
\label{tab:forecastsr}
\begin{center}
\begin{tabular}{|r|c|c|c|}
\hline
parameter & 1$\sigma$ error (no CR) & 1$\sigma$ error (CR) & $f_{\rm sky}$ used \\
\hline
\hline
($r_\mathrm{fid}=5 \times 10^{-3}$)\,\,\,\,\,$r$                    &    $0.0014$  &  $6.2\times 10^{-4}$ & \\
$n_S$              &   $0.0017$ &  $0.0017$ & 70 \%\\
$n_T$              &    $0.12$    &    N/A  & \\
\hline
($r_\mathrm{fid}=5 \times 10^{-3}$)\,\,\,\,\,$r$                    &    $0.0013$  &  $5.3\times 10^{-4}$ & \\
$n_S$              &   $0.0017$ &  $0.0017$ & 65 \%\\
$n_T$              &    $0.096$    &    N/A  & \\
\hline
\hline
($r_\mathrm{fid}=1 \times 10^{-3}$)\,\,\,\,\,$r$                    &    $0.001$  &  $5.0\times 10^{-4}$ & 70 \%\\
\hline
($r_\mathrm{fid}=1 \times 10^{-3}$)\,\,\,\,\,$r$                    &    $7.0\times 10^{-4}$  &  $2.8\times 10^{-4}$ & 65 \%\\
\hline
\end{tabular}
\caption{Forecast errors on $r$ for two fiducial models, assuming foreground residuals as obtained with two independent foreground cleaning methods applied to realistic simulated maps. 
Results for $r$ are given for two different masks, which correspond to observed sky fractions as listed in the last column.
\label{tab:forecastsr}}
\end{center}
\end{table}

\begin{figure}[htbp]
\vspace{-5mm}
\begin{center}
\includegraphics[width=4.5in]{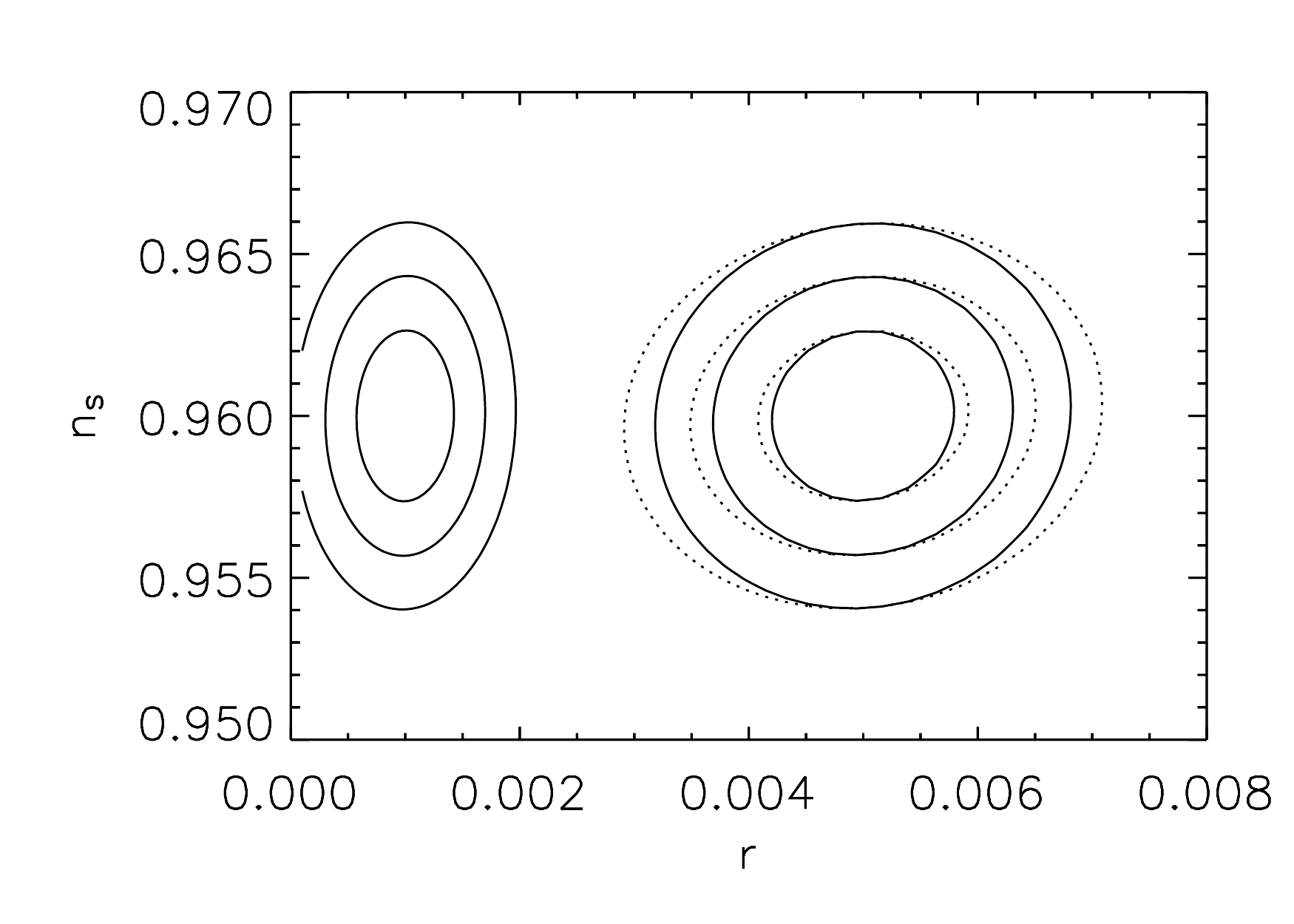}
\end{center}
\caption{
Forecast  $1$, $2$ and $3\sigma$ joint constraints in the $r$-$n_S$ plane including the 
effects of  foreground subtraction errors. The consistency relation has been imposed.
All other cosmological parameters have been marginalized over.  
The ellipses on the left side correspond to component separation 
with a large galactic mask blanking out the galaxy at galactic latitudes $|b| \leq 20^\circ$ ($f_{\rm sky} = 65$\%), for $r=0.001$ and $n_s=0.96$. Ellipses centered around $r=0.005$ and $n_s=0.96$ correspond to
a case with five times larger B modes, analyzed with a smaller galactic mask ($f_{\rm sky} = 70$\%, dotted contours) or with the same mask as in the case $r=0.001$ ($f_{\rm sky} = 65$\%, solid contours).
}
\label{fig:ellipses}
\end{figure}

Fig.~\ref{fig:ellipses} shows the joint 1, 2 and 3$\sigma$ regions in the $r$-$n_S$ plane, which is relevant for 
inflationary models. The single field consistency relation has been imposed, and all other cosmological parameters 
have been marginalized over. Here the result plotted for $r=0.001$ corresponds to the larger mask ($f_{\rm sky} = 65$\%).
Dotted and solid contours for $r=0.005$ show the (small) effect
of changing the mask for larger values of $r$, which indicates that the cosmic variance is becoming the major source of uncertainty.

The conclusion of the present analysis is that the proposed instrumental setup of COrE will enable us to measure at high statistical significance 
the target values for $r$ motivated in Sect.~\ref{sec:modelindependent} of $r\gtrsim 5 \times 10^{-3}$, and to 
detect $r\sim 1\times 10^{-3}$. This means that it will be possible to explore most large-field models and 
reach energies $M_{\rm inf}\sim$ few $\times 10^{15}$~GeV.

\smallskip
This analysis can be generalized in the future by using the multi-detector multi component spectral matching independent component analysis (SMICA)
\cite{2003MNRAS.346.1089D,2008ISTSP...2..735C,2009A&A...503..691B}.
SMICA indeed is a parameter estimation method that uses a multi-detector, multi component extension of the likelihood of eqn.~\ref{eq:likelihood}, 
in which a parametric model of theoretical covariances $R(\bar{\theta})$ is matched to measured covariances $\hat R$. 
The approach discussed here is equivalent to the special case of a three-channel SMICA likelihood in which the observations are one map for each of $T,E,B$, the parameters
$\bar{\theta}$ are cosmological parameters, and measured covariances $\hat R$ comprise a cosmological term as well as a noise term and a foreground residual term (the latter two being assumed to be known). In SMICA, just as here, the final covariance matrix of the parameters of the fit is obtained from the Fischer information matrix.
In addition to forecasting the errors (i.e. computing the FIM), SMICA also maximizes the likelihood,
using as parameters of the spectral fit not only the set of cosmological parameters, but a set of covariance matrices for various foregrounds. This makes it possible to estimate cosmological parameters directly from the multi-channel data set, while marginalizing over foreground contamination --- relaxing the assumption that the exact spectral energy distribution of the residual foreground contamination is known. 

This further development will extend on the best of both the method presented here, and the implementation in \cite{2009A&A...503..691B} which assumes cosmological parameters other than $r$ to be known, but unknown foreground residuals.
Note that SMICA has ``built-in'' an estimation of the goodness of fit of the multi-component model of sky emission, is a natural 
way to handle the problem of estimating the error induced by residual foregrounds.

\subsection{Gravitational lensing science}

{\bf

Gravitational lensing of the CMB anisotropies provides a powerful
and clean probe of the mass distribution integrated to high redshift.
Clustered matter lying between the surface of last scatter and
us today distorts the CMB anisotropies by shear and magnification 
distortions. This is a clean probe because it is the mass that is
being probed directly and no complicated astrophysical modelling
is required to interpret the data. Moreover since the surface
of last scatter is more distant than other objects subjected to
gravitational lensing, linear theory (supplemented by small reliable non-linear
corrections) suffices to compare observation to theory. The precise
determination of the CMB lensing power spectrum with
COrE will probe the dark energy sector and measure absolute
neutrino masses with a precision not possible with laboratory
experiments. In particular, COrE will be able to probe the 
two hierarchies (`direct' and `inverted') suggested by neutrino
oscillation experiments.

}

\subsubsection{Physics of CMB lensing}

The observed temperature anisotropies and polarization of the
CMB are at zeroth order an angular projection of the perturbations to the
photons and the spacetime metric around the time of cosmological
recombination ($z\approx 1100$).
The effect of later cosmic history,
and in particular ``dark parameters'' such as those describing the
dark energy and light neutrino masses (much below $1\,\mathrm{eV}$),
is only felt through the
angular diameter distance to recombination plus small corrections
on large angular scales from the late-time decay of gravitational potentials.
The dark parameters are therefore largely degenerate in the primary CMB
anisotropies.

The degeneracies can be broken with external datasets but also with the CMB
itself by exploiting the effect of weak gravitational lensing by large-scale
structure on the propagation of CMB photons from recombination to the
present. The net deflection, with an r.m.s.\ of 2.7~arcmin and degree-scale
correlations, is a sensitive probe of the growth of structure over a range
of redshifts peaking around $z \approx 2$ (see~\cite{2006PhR...429....1L}
for an extensive review).
Lensing has three main effects on the CMB: (i) it blurs out
features in the temperature and polarization
leading to a smoothing of the acoustic peaks in their power spectra
and a transfer of power from large scales into the damping tail on
arcminute scales; (ii) it converts $E$-mode polarization into
$B$-mode polarization corresponding to an additional white-noise spectrum
at the $(5\,\mu\mathrm{K}\,\mathrm{arcmin})^2$ level for multipoles
$l < 300$; and (iii) it generates small levels of non-Gaussianity (in addition
to any primordial non-Gaussianity).
The latter is easily calculable, and exploiting this non-Gaussianity is
key to extracting the lensing information from the observed CMB and
to providing a new window for probing the physics of the dark sector.

In essence, a CMB experiment such as COrE with
$\mu\mathrm{K}\,\mathrm{arcmin}$ sensitivity and resolution of
a few arcmin for the cosmological channels includes a weak-lensing experiment for free.
While lensing studies using the CMB have much in common with present
and planned surveys of the weak lensing of galaxies, there are important
differences. The CMB provides the most distant source plane and
so the deflections are larger and are sourced at higher redshift
than for galaxy lensing. At higher redshift, the amplitude of
the fluctuations is lower and so a wider range of scales can be probed in the
linear regime, removing many of the complications of non-linear evolution
that complicate galaxy lensing. Moreover, there are no astrophysical issues
such as intrinsic alignments of galaxy shapes to worry about. In addition,
the statistics of the CMB source plane are well understood so that
shear and magnification are equally useful observables. On the downside,
the CMB is a single source plane so there is no possibility of 
performing tomographic studies. Note also that instrumental systematic
effects are quite different for the two routes. The different redshift
ranges and systematic effects make CMB and galaxy lensing highly
complementary, and cross-correlations should be able to provide interesting
new results. 

\subsubsection{Measurement and interpretation of CMB lensing signal}

\begin{figure}
\begin{center}
\includegraphics[width=11.0cm]{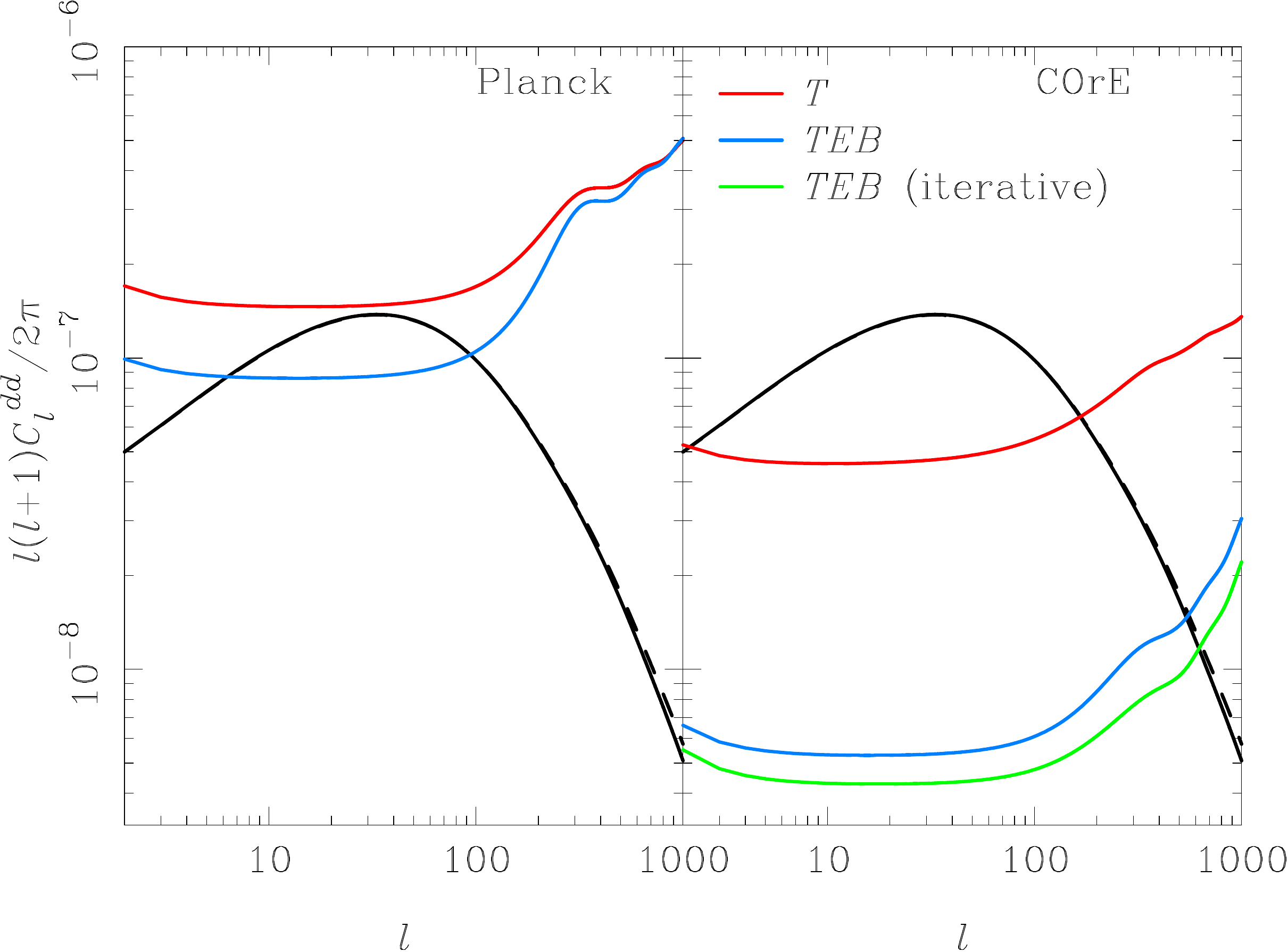}
\end{center}
\caption{{\bf Lensing reconstruction noise} on the deflection power spectrum
for an extended PLANCK mission (24 months; left) and COrE (right) using
temperature
alone (red) and temperature and polarization (blue). For COrE, we also
show the approximate noise level (green) for an improved iterative version of the
reconstruction estimator following Ref.~\cite{lens:cmbpol_stud}. The deflection power
spectrum is also plotted based on the linear matter power spectrum (black
solid) and with non-linear corrections (black dashed). The maximum
multipole used in the reconstruction is $l_{\mathrm{max}}=2500$.
}
\label{fig:lens_Planck_COrE_proposal_recon_noise}
\end{figure}

\begin{figure}
\begin{center}
\includegraphics[width=11.0cm]{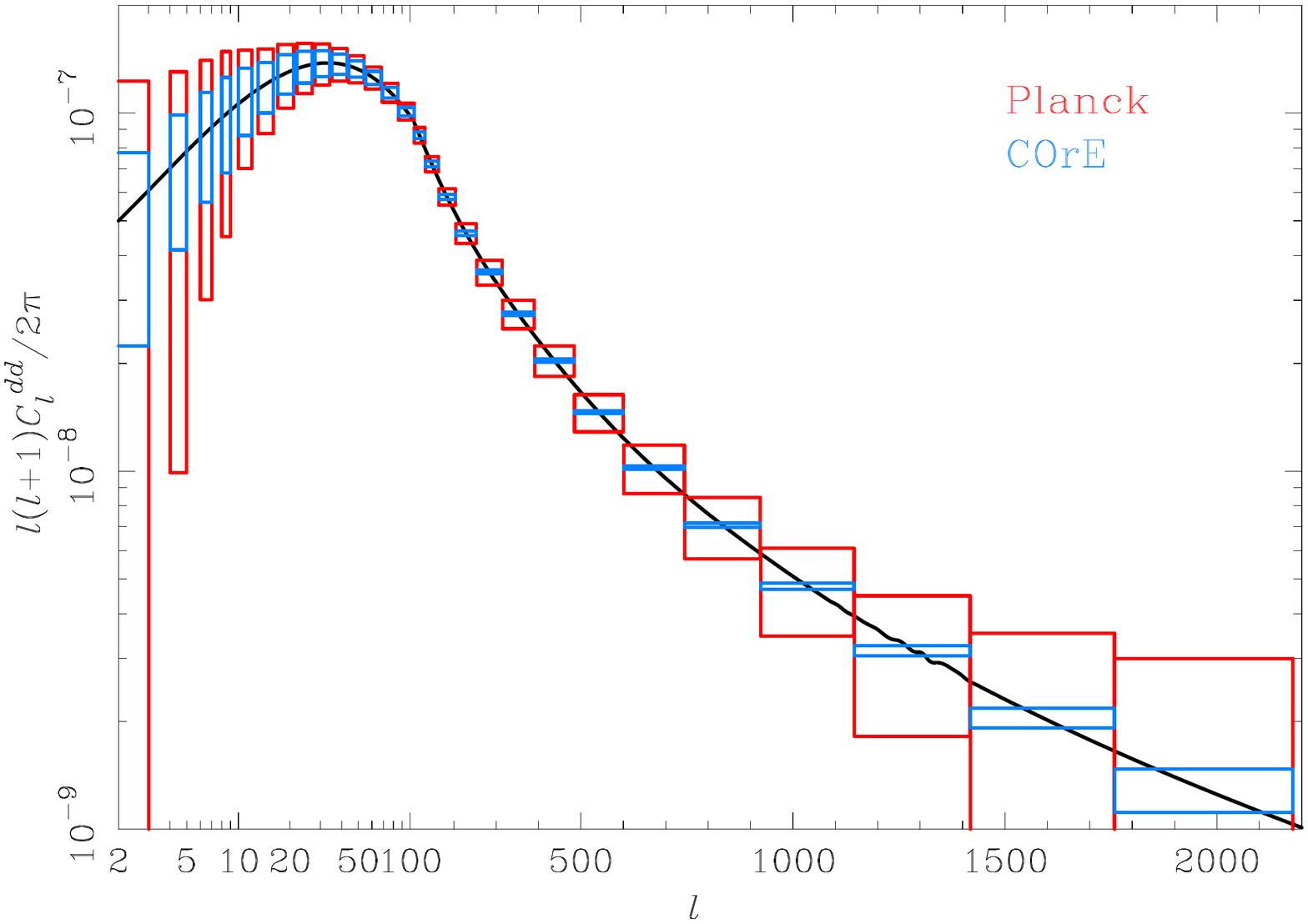}
\end{center}
\caption{%
{\bf COrE vs PLANCK for measuring the lensing power spectrum}.
Bandpower errors are plotted (including cosmic variance) on the deflection
power spectrum from PLANCK (24 months; red) and COrE (blue) using lens reconstruction
with temperature and polarization (no
iteration). With COrE, the power spectrum is cosmic-variance limited to
$l \approx 500$.
}
\label{fig:lens_Planck_COrE_proposal_recon_power}
\end{figure}

Fixed lenses correlate the observed CMB with the gradient of the
unlensed CMB and this property can be used to construct quadratic
estimators for the lensing deflection field~\cite{lens:hu_ok}.
Once reconstructed, the deflection field can be used as another
cosmological observable.
The majority of the information on lensing comes from modes at the resolution limit
of the survey so good angular resolution is essential even to reconstruct
large-scale lenses. At the WMAP resolution and sensitivity, the
reconstruction is very noisy but cross-correlation studies with
galaxy surveys have led to a detection at
the $3\sigma$ level~\cite{lens:smith_detect}. A highly significant detection of the
deflection power spectrum should be possible with PLANCK's temperature
maps (see Fig.~\ref{fig:lens_Planck_COrE_proposal_recon_noise} for forecasts)
and detailed analysis of the PLANCK maps is underway. Detections
are also expected soon from the ground (e.g., with SPT and ACT). However
the statistical noise in temperature reconstructions (due to
cosmic variance of the primary anisotropies) is such that the deflection
power spectrum will only ever be measured to the cosmic variance limit
for multipoles $l < 100$. To do better, and hence increase the lever arm
to constrain dark-sector physics, one must use polarization data.
PLANCK is too noisy for polarization to add much, but with COrE it
improves the reconstruction greatly; see 
Figs.~\ref{fig:lens_Planck_COrE_proposal_recon_noise} 
and~\ref{fig:lens_Planck_COrE_proposal_recon_power}.
Most of the information comes from the $E$-$B$ correlations
and the reconstructed deflection power should be cosmic-variance limited
to $l \approx 500$, roughly a 25-fold increase in the number of
reconstructed modes with $S/N > 1$. Significantly, {\bf COrE can mine all
of the information in the deflection power spectrum where linear
theory is reliable}.

\subsubsection{Neutrino masses as a unique probe of physics beyond the 
standard model}

There are compelling theoretical reasons why the Standard Model
cannot be the last word on particle physics and there are
great expectations that the LHC will provide truly telling clues on how the
Standard Model should be extended. 
One area of particle experiment where there has been significant progress
towards this goal is exploration of the neutrino sector.

While the Standard Model, as first proposed, 
implies that there are three exactly massless chiral
neutrinos, an abundance of evidence for flavor oscillations has now
been accumulated that requires neutrinos to have mass.
The most recent data compilations~\cite{lens:pdg} indicate squared mass differences
between the three mass eigenstates of 
$\Delta m_{12}^2=(7.59\pm 0.20)\times 10^{-5}\,\mathrm{eV}^2$ and $\Delta m_{31}^2 = \pm (2.43\pm 0.13)
\times 10^{-3}\,\mathrm{eV}^2$ ($1\sigma$ errors).
While the evidence for neutrino mass splittings is overwhelming
(the 2002 Nobel Prize in physics was awarded to Davis and Koshiba
for this discovery), oscillation experiments are not able
to probe the absolute mass scale of the neutrinos. Taking the
mass and mixing matrix of charged leptons and quarks as a guide,
we would conclude
that the most probable values for the neutrino masses would fall
into two possible hierarchies: a \emph{normal hierarchy} with
$m_1,m_2\ll m_3$ and a total mass close to $0.058\,\mathrm{eV}$; or
an \emph{inverted hierarchy} with  
$m_3\ll m_1, m_2$ and total mass $0.1\,\mathrm{eV}$.
However, one should be mindful that neutrinos
could well be \emph{degenerate} with $m_1 \sim m_2 \sim m_3 \gg 0.05\,\mathrm{eV}$. 

While laboratory $\beta$-decay experiments can probe absolute masses
(or more correctly, effective masses involving the absolute masses
and the elements of the electron row of the mixing matrix) the target
values for hierarchical masses are well below the detection limit
of current and also future experiments.  Current searches for neutrinoless
double beta decay, which requires massive Majorana neutrinos, limit
the effective electron-neutrino mass $m_{\beta\beta} < 0.27\,\mathrm{eV}$
(90\% confidence) and are complicated by uncertainties in nuclear
matrix elements. The other laboratory route uses the kinematical
effect of neutrino masses on the
extreme tail of the electron energy distribution in ordinary $\beta$ decay.
This is challenging due to the very low number of events there. The current
upper limit on the effective mass is $m_\beta < 2\,\mathrm{eV}$
(95\% confidence); the KATRIN experiment due to start in 2012 should
improve this to $m_\beta < 0.2$~eV.

\begin{figure}
\begin{center}
\includegraphics[width=15.0cm]{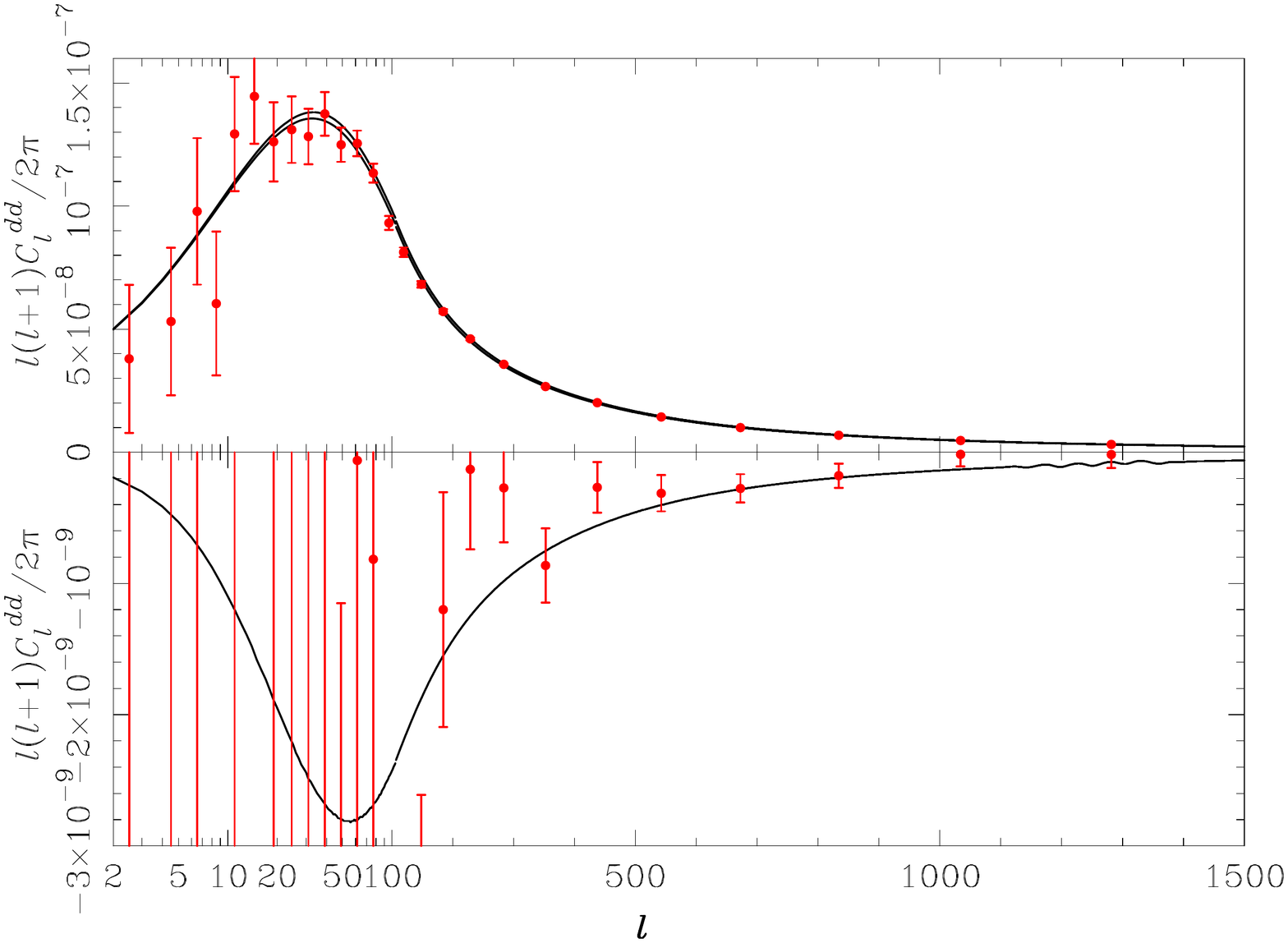}
\end{center}
\caption{%
{\bf Gravitational lensing deflection power spectrum.}
The simulated deflection power spectrum from COrE is shown assuming
an inverted hierarchy of neutrino masses with the minimum total mass
allowed by oscillation data ($m_1 \approx m_2 = 0.05\,\mathrm{eV}$ and
$m_3=0\,\mathrm{eV}$). In the upper panel, the solid lines are the theory
power spectrum for this scenario (lower) and for three massless neutrinos
(upper). The difference between these spectra is plotted in the lower panel
illustrating how COrE can distinguish these scenarios from $C_l^{dd}$ in the
range $l > 200$.
We have assumed 70\% sky coverage after Galactic masking.
}
\label{fig:COrE_proposal_invnu_diffplot_fskyp7}
\end{figure}

Cosmology, however, provides an alternative probe of absolute neutrino
masses. Mass splittings from oscillation data imply that at least two mass
eigenstates are non-relativistic today. Keeping the densities of all
other species (and dark energy) fixed, non-relativistic neutrinos increase
the expansion rate at late times over that for massless neutrinos.
This impedes the growth of density perturbations on scales small
compared to the neutrino free-streaming scale, for which the
neutrinos cannot cluster. {\bf This mass-dependent suppression of the matter power
spectrum on small scales can be measured with CMB lensing even for
masses close to the hierarchical targets}.
We illustrate the capabilities of COrE to distinguish the minimal-mass
inverted hierarchy from a model with three massless neutrinos via its
reconstruction of the lensing deflection field in
Fig.~\ref{fig:COrE_proposal_invnu_diffplot_fskyp7}.
If all other parameters were known, lensing with COrE could constrain the
summed neutrino mass to $0.012\,\mathrm{eV}$ (1$\sigma $).
However this ignores the issue of
uncertainties in the other parameters.
While future
distance-indicator measurements may help, to be conservative here we
consider only constraints from COrE alone in a joint analysis of the 
unlensed temperature and $E$-mode polarization and the reconstructed
deflection field, following~\cite{lens:perr}.
In our MCMC simulations, we vary the standard seven
parameters of flat $w$CDM models plus neutrino-mass parameters for three cases.
First, we consider a minimal-mass normal hierarchy and use an oscillation
prior (ignoring the errors in the squared-mass differences) to fix
$m_2$ and $m_3$ in terms of the lightest mass $m_1$ which we allow to vary.
Second, we assume a minimal-mass inverted hierarchy and vary $m_3$.
Finally, we assume degenerate neutrinos and vary the total mass about
a model with massless neutrinos.
We summarise our results as follows.

\begin{itemize}
\item If neutrinos have hierarchical masses, COrE  will bound
the lightest mass to $m_1 < 0.034\,\mathrm{eV}$ (normal) and
$m_3 < 0.045\,\mathrm{eV}$ (inverted) at 95\% confidence.
\item The minimal-mass inverted hierarchy could be distinguished
from a scenario with three massless neutrinos at the $3\sigma$ level.
\item If neutrino masses are degenerate, COrE  will measure the total mass
to $0.03\,\mathrm{eV}$ ($1\sigma$ error).
For comparison, the error expected from the Planck nominal mission
(including lensing) is $0.10\,\mathrm{eV}$~\cite{lens:perr}.
\end{itemize}

Current constraints on neutrino masses from cosmology are rather model
dependent and the tightest constraints come from combining multiple datasets
(e.g.\ CMB and large-scale structure data). The 95\%
limits on the total mass are in the range $0.3$--$2\,\mathrm{eV}$ (see~\cite{lens:hann_rev}
for a recent review). In models in which $w \neq -1$, CMB and large-scale
structure data indicate an upper limit at the $1\,\mathrm{eV}$ level. Our
forecasted constraints from COrE are for a single homogeneous dataset and
are much less model dependent than with current data since the
full shape information in the deflection power spectrum is accessible.
For example, Ref.~\cite{lens:cmbpol_stud} show that for a CMB lensing experiment comparable to
COrE, there are no significant degeneracies between $\sum_\nu m_\nu$,
$w$ and the spatial curvature. Our constraints from COrE are comparable
to forecasts for tomographic studies with future galaxy lensing surveys
(assuming a PLANCK prior), for example LSST~\cite{lens:LSST} or Euclid~\cite{lens:euclid},
but with quite different systematics.

\subsection{Primordial non-Gaussianity}

{\bf 

Cosmic microwave background non-Gaussianity is emerging as a powerful new probe of the 
origin of cosmic structures in the very early Universe and their subsequent evolution (see 
for example the reviews in \cite{r1,r2,r3,r4,r4a}). Non-Gaussianity is already the most stringent test of the 
standard model of inflation, the canonical slow-roll single field model, which predicts 
negligible primordial non-Gaussianity~\cite{mald,acquaviva}. The COrE satellite would raise
this confrontation to a new level, testing Gaussianity to one part in $10^5$. Furthermore, in a 
complementary way to probing primordial gravitational waves, COrE's probe of NG offers the 
prospect of dramatic new insights into fundamental physics and possible signals from the epoch 
of quantum gravity.

}

It has often been said that Gaussianity, along with spatial flatness and an approximately 
scale invariant spectrum of adiabatic cosmological perturbations, is one of the inexorable 
consequences, or tests, of inflation. While the WMAP data does suggest some hints of 
non-Gaussianity at low statistical significance, it is remarkable that the primordial signal 
observed in the CMB turns out to be very nearly Gaussian. Following a series of seminal papers 
in 2003 where the bispectral non-Gaussianity was calculated for the first time, it was 
realized that large classes of models predicting measurable non-Gaussianity exist and that CMB 
constraints of non-Gaussianity are far more robust to foreground systematics than was 
initially feared (e.g., \cite{yadav}). Much of the current research in theoretical cosmology 
has shifted toward exploring what patterns of non-Gaussianity are possible. These 
possibilities are closely linked to new physics near the Planck scale and modifications of 
gravity. Conservatively, we estimate that COrE will have a NG discovery potential (defined 
more quantitatively below) a factor of $\approx 20$ better than PLANCK, approaching the 
capabilities of an ideal CMB probe.

\begin{figure}[t]
\centering
\includegraphics[width=.3\textwidth]{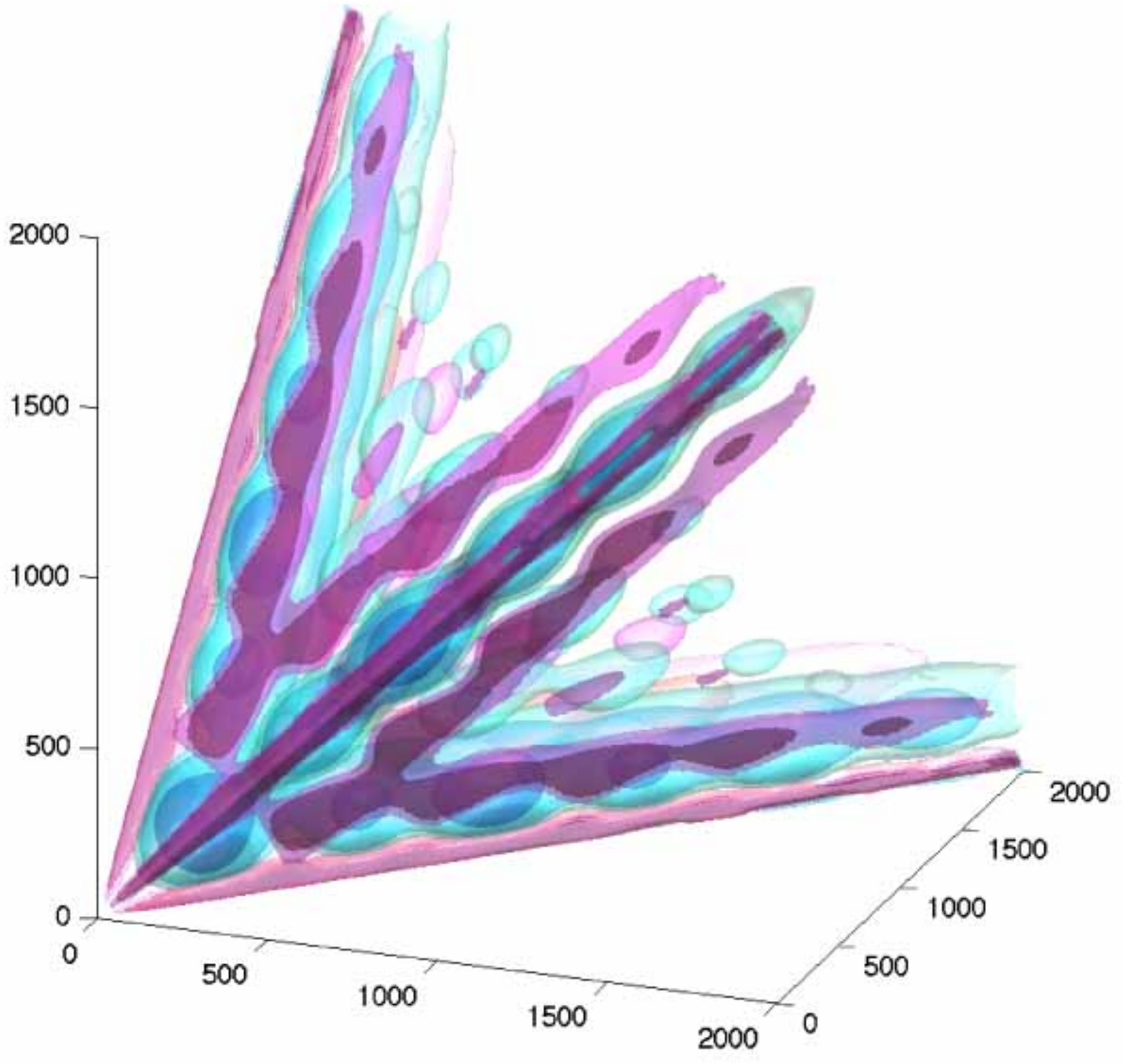}
\includegraphics[width=.3\textwidth]{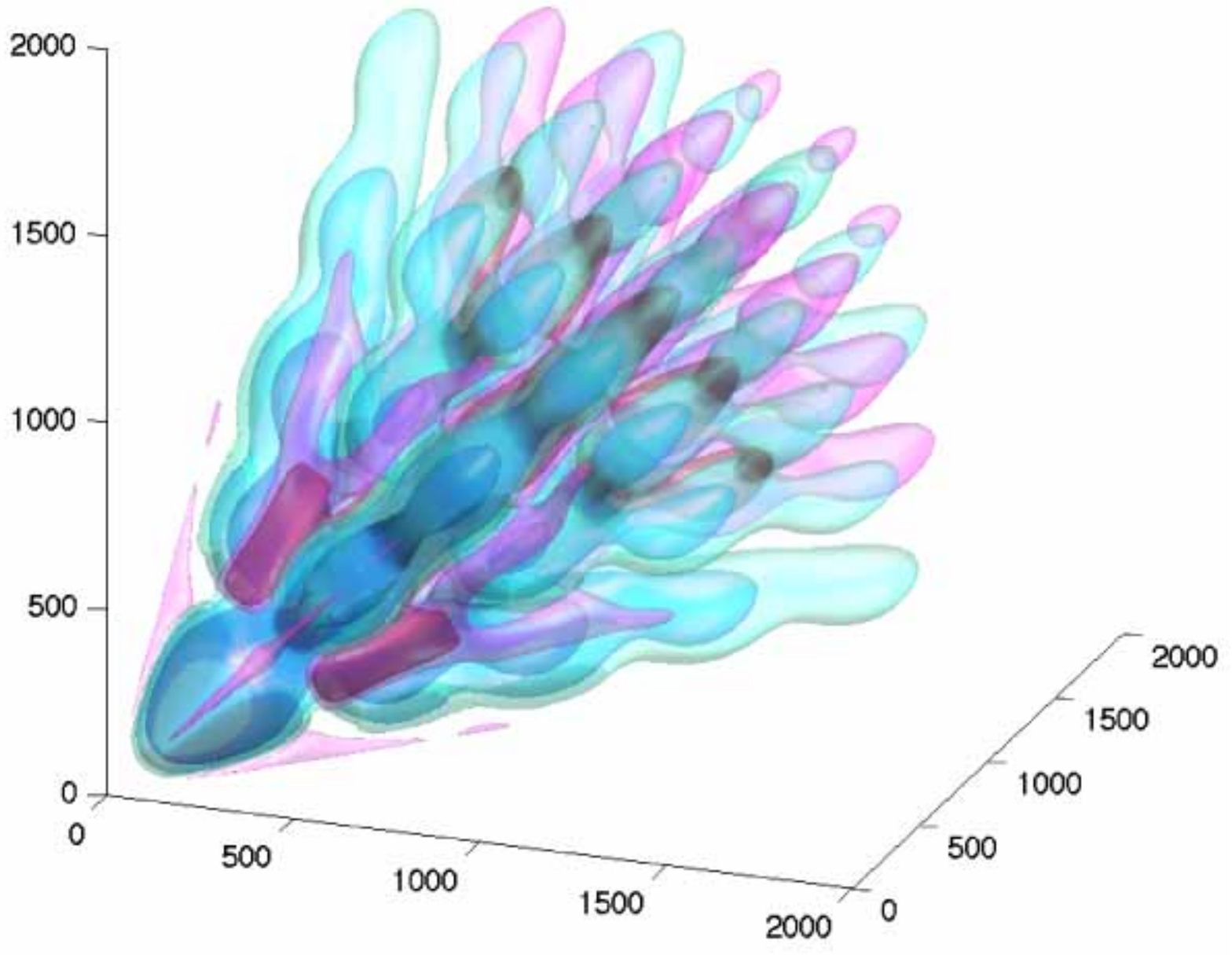}
\includegraphics[width=.3\textwidth]{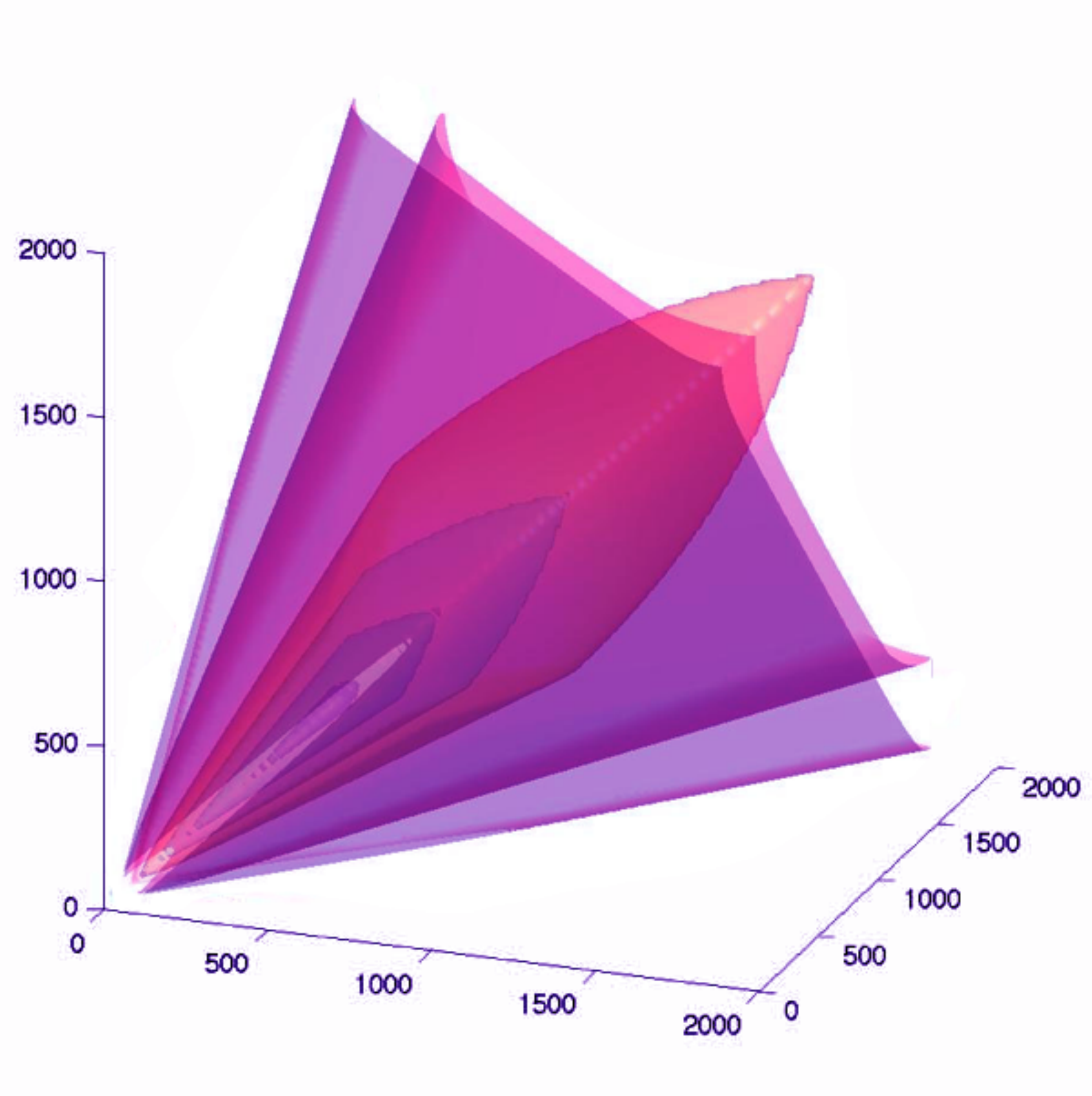} 
\small
\caption[Non-Gaussian shapes]{
{\bf Theoretical predictions for CMB bispectrum shape.}
The primordial CMB bispectrum depends on three multipole numbers $\ell _1, \ell _2, \ell _3$ subject to a triangle inequality constraint
as opposed to the standard 
`Gaussian' power spectrum $P(\ell )$ (which depends on only one multipole number). Hence the bispectrum contains rich shape 
information that can be exploited to confront observations with theoretical models and to probe the self-consistency of a possible
detection.
Here the CMB bispectra $B(\ell _1, \ell _2, \ell _3)$ from three theoretical models are plotted with positive (cyan) and negative (magenta)
isocontours \cite{shapes, strg2}.  The plots show (from left to right) the `local' model (e.g., multifield inflation), 
equilateral model (e.g., DBI inflation) and non-Gaussianity generated by cosmic strings, which are inherently highly nonlinear.}
\label{fig:NG}
\end{figure}

How does COrE compare to other CMB or non-CMB experiments? An important and unique advantage 
of COrE over non-CMB probes of NG results from its ability to recognize the distinct 
patterns which physical mechanisms leave in the \textit{shape} of higher order 
correlators, as illustrated in Fig.~\ref{fig:NG}. COrE's competitive strength will 
therefore be a vastly enhanced exploration of physically predicted NG shapes compared to any 
other projected probe of NG. While other probes claim competitive power for detecting `local'
bispectral non-Gaussianity (because the effects of nonlinearity can be cleverly factored out
for this special case), only the CMB, because of its cleanness and linearity, offers significant
sensitivity to other shapes of non-Gaussianity.

Only a high resolution polarization CMB mission such as COrE can provide the high
sensitivity for detecting and distinguishing differing NG bispectral shapes. In fact one can show that 
polarization maps contain more information than 
temperature maps (although the best constraints arise from combining the two).  Polarization is 
the more powerful probe because the ratio of the expected contaminant signal 
(mainly galactic dust) to the primordial signal is smaller for polarization 
than for temperature for the majority 
of resolved modes.  To 
construct a figure of merit comparing the predicted impact of COrE on physical forms of NG, we 
compare the predicted constraint volume in bispectrum space spanned by the local, equilateral 
and flattened bispectra. While PLANCK will reduce the constraint volume by a factor of 70 
compared to WMAP, COrE would reduce the volume by another factor of 20.  Considering the constraint 
volume based only on the polarization maps (which provide information independent of the 
temperature maps and hence provide an important consistency check), we find a volume reduction
factor from PLANCK to COrE of 
order $110$. COrE stands out on account of its full-sky coverage, polarization sensitivity, high resolution, 
strong rejection of systematics, and the benign environment in space at L2 making COrE the ideal CMB NG 
probe. On smaller angular scales $(\ell \gtorder 3000)$ the primordial CMB signal becomes subdominant
because other highly non-Gaussian compact contaminants (e.g., S-Z clusters and point sources, which are 
clustered). 

The improvement compared with PLANCK will be even more dramatic when considering combined bispectrum and 
trispectrum constraints on a larger range of NG shapes. The forecast precision with which 
local trispectrum parameters could be measured with COrE temperature data alone are $\Delta 
g_{NL}=3\times 10^4$ and $\Delta \tau_{NL}=1\times10^4$. COrE's polarization capability is 
estimated to sharpen these constraints by a factor of 2.  A trispectrum constraint satisfying 
$\tau_{\rm NL}\ll \frac12 (6f_{\rm NL}/5)^2$ would rule out large classes of multifield 
inflation models in addition to single field inflation, necessitating a fundamental 
reassessment of the standard field theory picture of inflation. Beyond searches for 
primordial NG, COrE will also see NG
imprinted on the CMB maps by the lensing/ISW correlation 
\cite{Hanson:2009kg,Mangilli:2009dr,Lewis:2011fk}. 
Such a signal can be exploited to yield the 
strongest dark energy constraints from the CMB alone. Furthermore, ancillary signatures will 
constrain modified gravity alternatives to the standard cosmological model.

In addition, there are many alternative inflationary scenarios for which an observable 
non-Gaussian signal is quite natural, including models with multiple fields, interactions, 
non-canonical kinetic energy, or remnants of a pre-inflationary phase. There are also more 
exotic paradigms which can create NG such as cosmic (super-)strings \cite{strg1,strg2} or a 
contracting phase with a subsequent bounce \cite{r5}.  Each of these scenarios leaves a 
distinct NG fingerprint, essentially the `shape' of the higher order correlators. This 
fingerprint can discriminate between many different inflationary models which are indistinguishable
based on the 
the CMB power spectra alone.  It can also be robustly distinguished from astrophysical NG 
fingerprints left by weak lensing, extra-galactic and galactic astrophysical signals that are 
interesting in their own right, and spurious NG fingerprints due to instrumental noise.


Non-Gaussianity is typically characterized by the parameter $f_{\rm NL}$, which is the 
amplitude of three-point correlations for the so-called `local' model (the best-motivated 
case which we will describe below).  The equivalent amplitude parameters for the local 
four-point correlations or trispectrum are $g_{\rm NL}$ and $\tau_{\rm NL}$ (for single-field 
inflation, we have $\tau_{\rm NL}= (6f_{\rm NL}/5)^2$).  Inflation predicts negligible 
primordial NG ($f_{\rm NL}\ll 1$) if the following minimal conditions are 
satisfied~\cite{kab}: i) a single field is responsible for driving inflation and generating 
the quantum fluctuations which are stretched on superhorizon scales to become the seeds for 
structure formation; ii) such a field has canonical kinetic energy so that its fluctuations 
propagate at the speed of light; iii) the inflaton obeys slow-roll, that is, it evolves slowly 
relative to the Hubble timescale; iv) all pre-inflationary state information has been erased 
during inflation. The violation of any of the above conditions can lead to the generation of a 
large detectable non-Gaussian signal. Each physical effect leaves its own recognizable 
fingerprint (see Fig.~\ref{fig:NG}).

Models with field interactions generate `local' primordial NG, which peaks in `squeezed' 
triangles (see, e.g., \cite{BU1, LR1, RSvT1, Byrnes1}); models with non-canonical kinetic 
term~\cite{Chengen} (as, e.g., DBI ~\cite{DBI} and ghost inflation~\cite{ghost}) generate 
`equilateral' NG, while trans-Planckian effects generate NG which peaks for 
`folded/flattened' triangles; inflationary models violating slow-roll can generate more 
complicated shapes (see, e.g., ~\cite{WangKamion,Chenfeat1,Chenfeat2}).  A linear combination 
of these shapes can also be realized~\cite{mix}.  All these models can naturally predict 
values of $f_{\rm NL} \gg 1$.

Whether the different shapes are observationally distinguishable can be determined by their 
cross-correlations, reducing the search categories~\cite{shapes}. Important tests of inflation 
include the following: 
\begin{itemize}

\item {\it 
A detection of any such primordial signal would rule out 
all standard slow-roll models of single-field inflation. 

\item 
Even more striking is the 
fact that a detection of local non-Gaussianity would rule out all classes of single-field 
(slow-roll) inflationary models irrespective of their Lagrangian.}~\cite{consistency}. 

\item {\it 
A significant trispectrum 
measurement satisfying $\tau_{NL} \ll \frac{1}{2} \left(6 f_{NL}/5\right)^2$ would rule out 
standard multifield inflation \cite{Suyama:2007bg, Suyama:2010uj,Sugiyama:2011jt}.} 

\end{itemize}
Beyond these special cases, general 
estimators have now been developed to search efficiently for arbitrary shapes in the CMB data, 
allowing {\it any} model or mechanism to be directly tested \cite{FLS2}. Another testable 
prediction of various non-standard models of inflation is that the non-linearity parameter 
$f_{\rm NL}$ can have a significant scale dependence, parameterized in terms of a running NG 
index $n_{\rm NG}$ as $f_{\rm NL}=f_{\rm NL}(k)(k/k_*)^{n_{\rm NG}}$ (see, 
e.g.~\cite{chenr,Chengen,loverde,Byrnes1,Byrnes2,Fasr}).

There are also models with a negligible bispectrum and a large trispectrum, including 
inflation with a parity symmetry \cite{SZ1,Byrnes:2006vq} and cosmic strings \cite{strg2}.  In 
the latter case, it is clear that the trispectrum will provide a significantly stronger 
constraint on the string tension than the power spectrum or bispectrum \cite{strg2}, though it 
will also be efficient to search for cosmic defects NG using other specially tailored methods 
or templates \cite{strg1}.

What if there is no detection of primordial NG?  In this case COrE would rule out all 
scenarios whose natural parameter space result in high values of the NG parameters such as 
$f_{\rm NL}$, while rendering many other early universe mechanisms cosmologically irrelevant. 
Examples include DBI inflation models typically predicting $|f_{\rm
  NL}| > 5$ and competitors to inflation like ekpyrotic/cyclic models
predicting measurable $| f_{\rm NL}| \sim10$ \cite{Lehners}.

{\it NG from anisotropic cosmologies and topology:} Deviations from isotropy as suggested by 
analyses of the WMAP temperature signal will be confirmed or otherwise by PLANCK  but COrE
will add the unambiguous consistency check of measuring the presence or otherwise of the 
corresponding signal in polarization, which could reflect the presence of anisotropy during 
inflation itself. Moreover, the polarization signature of simply-connected topologies will be 
stronger than for temperature, in contrast to the situation for PLANCK and WMAP where it is 
undetectable.

{\it `Agnostic' NG searches:} The much larger number of resolved modes in the COrE data 
allows detailed model-independent searches for anomalies in the temperature and polarization 
maps, which may signal new physics. The addition of high signal-to-noise polarization 
information with COrE allows for significant cross-checks since any model explanations of 
candidate anomalies seen in temperature data can be checked against the polarization data COrE 
will provide. This redundancy will offer significant discovery potential for relics of physics beyond 
the standard model of cosmology in the early universe.

\subsection{Other cosmology}

\subsubsection{Parameters}

{\bf 

Although in the previous sections we have highlighted three key areas---primordial
gravitational waves from inflation, gravitational lensing science, and non-Gaussianity searches---
as the mainstays of the cosmology part of the COrE science program, because of the leap in sensitivity afforded by COrE,
a host of other cosmology projects will be made possible and the potential for serendipitous discoveries
is very high. To quantify this potential, we stress that COrE will:
\begin{itemize}
\item Map the B mode polarization with $(S/N)\approx O(1)$ up to $\ell \ltorder 100$ no matter how small $r$ is,
owing to the gravitational lensing signal, which is essentially guaranteed.

\item Map the E mode polarization at the cosmic variance limit for $\ell \ltorder 2000$ at $(S/N)\gg 1.$
By contrast PLANCK measures the E mode with S/N oscillating near one up to $\ell \ltorder 800$. 

\item Map the T anisotropy at the cosmic variance limit for $\ell \ltorder 2000$ and measure the TT power
spectrum up to $\ell =3000.$ (By contrast the PLANCK T measurement is cosmic variance limited only up to $\ell \ltorder 1200$.)

\end{itemize}
This massive increase in precision will allow the cosmological parameters of a vanilla minimal cosmological
model to be determined approximately 2--3 better than possible with PLANCK and will probe with high precision
the consistency of the temperature and polarization anisotropies. Moreover, using the COrE data, many extensions
of the standard cosmological model including possible new physics will become testable within the interesting
range for the new parameters. One possible and not at all unlikely outcome is that thanks to this increased
precision we might have to abandon the vanilla model and we could see the ``concordance'' unravel. 

}

Since it is not possible to enumerate all the non-minimal models that could be tested, to illustrate the potential, 
we present results for the improvement in the determination of the minimal cosmological parameters as well as two example of
extensions showing how the limits would improve with the COrE relative to PLANCK.
Table~\ref{tabparams} shows the forecast for the constraints on 
the parameters of a ``minimal'' cosmological model described by the cosmological parameters:
\begin{equation}
 \label{parameter}
      \omega_b, 
   ~~   \omega_c,
   ~~   \Theta_s, 
   ~~   n_s, 
   ~~   \log[10^{10}A_{s}] 
\end{equation}
where $\omega_b\equiv\Omega_bh^{2},$ $\omega_c\equiv\Omega_ch^{2}$ are the
physical baryon and cold dark matter densities  relative to the critical
density, $\Theta_{s}$ is the ratio of the sound horizon to the angular diameter
distance at decoupling, $A_{s}$ is the amplitude of the primordial spectrum, and
$n_s$ is the scalar spectral index. The forecasts were computed using an MCMC approach as described in 
Galli et al. (2010) and the COrE experimental specifications given previously.
Together 
with the standard deviations on each parameter, we also report the improvement factor for each parameter 
defined as the ratio $\sigma_{\rm Planck}/\sigma_{\rm COrE}$.
The values in the Table show that the COrE satellite will improve by a 
factor $\sim 3$  the constraints on the baryon density, $H_0$ and $\theta_s$, while the 
constraints on parameters such as $n_s$ and $\tau$ are improved by a factor $\sim 2$. 

We have also considered a model with an extra background of 
relativistic non-interacting particles, $N^{\rm eff}_{\nu}$ and a model in which the Helium abundance $Y_p$
is allowed to vary. 
\footnote{When variations in the neutrino effective number and the primordial Helium 
abundance are considered, the constraints on the remaining parameters are also affected; 
however, the improvement respect to PLANCK is similar.} 
COrE will improve the constraints on these parameters by a factor of $\sim 3$ 
with respect to PLANCK.

\begin{table}[h!tbp]
\begin{tabular}{||r||c|c c||c|c c||c|c c||}
\hline\hline
Parameter& & & & & & &  & & \\
uncertainty& Planck & \multicolumn{2}{|c||}{COrE} & Planck & \multicolumn{2}{|c||}{COrE} & Planck & \multicolumn{2}{|c||}{COrE} \\
\hline
$\sigma(\Omega_b h^2)$ & 0.00011 & 0.000034 &(3.3) & 0.00017 & 0.000049 &(3.6) & 0.00016 & 0.000048 &(3.3) \\ 
$\sigma(\Omega_c h^2)$ & 0.00087 & 0.00037 &(2.4) & 0.0022 & 0.00073 &(3.1) & 0.0009 & 0.00036 &(2.5) \\ 
$\sigma(H_0)$ & 0.0039& 0.0014 &(2.8) & 0.011& 0.0034 &(3.3) & 0.0046& 0.0016 &(3.1) \\ 
$\sigma(\tau)$ & 0.0040& 0.0022 &(1.8) & 0.004 & 0.0022 &(1.8) & 0.0040& 0.0023 &(1.8) \\ 
$\sigma(n_s)$ & 0.0027 & 0.0014 &(1.9) & 0.0056 & 0.0025 &(2.3) & 0.0053& 0.0024 &(2.3) \\ 
$\sigma(10^{10}A_s)$ & 0.18 & 0.10 &(1.8) & 0.23& 0.11 &(2.1) &0.19& 0.10 &(1.9) \\ 
$\sigma(N_{\rm eff})$& $-$ & $-$ & $-$ & 0.14 & 0.044 &(3.3)& $-$ & $-$ & $-$  \\
$\sigma(Y_p)$& $-$ & $-$ & $-$ & $-$ & $-$ & $-$ & 0.0083 & 0.0027 &(3.1) \\
\hline\hline
\end{tabular}
\caption{{\bf Improvement of CORE relative to PLANCK on measuring cosmological parameters.}
$1\sigma $ errors on cosmological parameters. In parenthesis we give the improvement factor 
in the confidence level for the corresponding COrE configuration with respect to PLANCK. The second set 
of columns correspond to the case of an extra background of relativistic particles $N_{\rm eff}$. The third 
set of columns consider variations in the primordial $^4$He fraction abundance, $Y_p={}^4{\rm He}/(H + 
{}^4{\rm He})$.}
\label{tabparams}
\end{table}

COrE will provide valuable constraints on the physics of the neutrino 
decoupling from the photon-baryon primordial plasma. As it is well known,
the standard value of neutrino parameters $N_{eff}=3$ should be increased to
$N_{eff}=3.04$ due to an additional contribution from a partial heating of neutrinos during 
the electron-positron annihilations. This effect, expected from standard physics, can be tested by the COrE experiment, 
albeit at just one standard deviation.
However, the presence of nonstandard neutrino-electron interactions (NSI) may enhance the entropy transfer from electron-positron 
pairs into neutrinos instead of photons, up to a value of $N_{eff} \sim 3.15$. 
This value would be distinguished by COrE from $N_{eff}=3$ at $\sim 3$, shedding new light on NSI models.

COrE will also provide an independent determination of the
primordial Helium abundance. Current astrophysical  measurements of primordial Helium converge
towards a conservative estimate of $Y_p=0.250 \pm 0.003$.
Table~ \ref{tabparams} shows that the COrE experiment 
will reach a precision comparable to current astrophysical measurements, opening
a new window for testing systematics in current 
primordial helium determinations and further testing Big Bang Nucleosynthesis.

COrE will also be able to test the adiabaticity of the primordial scalar perturbation by probing 
for the presence of isocurvature perturbations of various types. (See for example \cite{bmt,mprl}
and references therein.) This is an important point because although the simplest possibility is that the 
perturbations were completely adiabatic, a myriad of models have been proposed that include 
some degree of isocurvature perturbations as well and the only way to rule out these models 
is through better observations \cite{Beltran:2005gr,Bean:2006qz,Komatsu:2011}.
COrE will improve constraints on isocurvature perturbations to the 
total CMB power spectrum. Considering a generic cosmological model with the addition of CDM, 
neutrino density and neutrino velocity isocurvature modes, a Fisher forecast of COrE shows an 
improvement of these constraints by approximately a factor of two over PLANCK. The most improved 
constraints will be those on the contribution of the neutrino density and velocity isocurvatures, 
which will be more than double that of PLANCK \cite{cg_pc}.



\subsubsection{Reionization history of the Universe}   


One of the most striking contributions of the WMAP space mission was
its measurement of the reionization optical depth $\tau$ of the microwave
photons through its characterization
of the E mode polarization on very large angular scales.
According to the seven-year WMAP analysis
\cite{2010arXiv1001.4635L}, the current uncertainty on $\tau$ is $\simeq
\pm 0.015$, almost independently on the specific
model considered.  Under various hypotheses
(simple $\Lambda$CDM model with six parameters, inclusion of curvature and
dark energy, of different kinds of isocurvature modes, of neutrino
properties,
of primordial helium mass fraction, or of a re-ionization width) the best
fit of $\tau$ lies in the range 0.086--0.089. On the other hand,
allowing for the presence of primordial tensor perturbations or (and) of a
running in the power spectrum of primordial perturbations
 the best fit of $\tau$ goes to 0.091--0.092 (0.096).
 PLANCK will certainly put new light on this topic thanks to its high
sensitivity to E mode polarization,
but this underlines the relevance of carrying out a combined analysis of
the re-ionization of the Universe and of primordial tensor perturbations,
firmly possible only through high accuracy measurements of both E and B
modes.
Re-ionization results from ionizing radiation emitted by very first
structures formed in the high-redshift
Universe -- either by very massive stars or by quasars -- and a better
characterization of the full re-ionization history as a function of
redshift would provide
important clues for understanding the scenarios of structure formation and
radiative feedback processes \cite{2008MNRAS.385..404B}
resulting in a different suppression of star formation in low-mass haloes.
The precise polarization data that COrE will provide at very low
multipoles will be invaluable to this endeavor.
The principal component analysis can be used to provide model-independent
constraints on the re-ionization history of the Universe
\cite{2008ApJ...672..737M} from CMB polarization data at low multipoles.
Furthermore, the COrE resolution up to a few arcminutes will 
constrain alternative
(double-peaked or very high redshift) re-ionization models, invoking
non-standard processes such as evaporation of mini-blackholes or particle
decays and annihilations,
manifesting themselves also at high multipoles \cite{2004MNRAS.347..795N}.
Moreover, COrE will measure and constrain models of patchy
reionization and probe parameters such as the duration of the patchy epoch and the mean bubble
radius
\cite{2009PhRvD..79d3003D}.

\subsubsection{Primordial magnetic fields}

Synchrotron emission and Faraday rotation measurements provide increasing evidence that the large scale 
structures in the Universe, such as galaxies and clusters, are pervaded by magnetic fields which are of order 
a few micro-Gauss. An open question is whether these fields were seeded in the Early Universe (by fields of a 
few nano-Gauss that arise in inflation or cosmological phase transitions) or whether they originated later on through 
astrophysical processes. In either case they would have sourced CMB anisotropies leading to unique observational 
signatures.  Very large scale, quasi-homogeneous magnetic fields lead to very distinct cross correlations 
between E and B modes.  Smaller scale, stochastic magnetic fields lead to a characteristic peak in the 
B-mode power spectrum at $\ell \sim 2000$. COrE will detect magnetic fields at the sub nano-Gauss level 
permitting an accurate characterization of their origin and evolution.

\subsubsection{Topological defects}

One view of the Early Universe is that the fundamental interactions are unified at some high energy scale, typically presumed to be 
Grand Unified Theory (GUT) energy scale $\eta\sim 10^{16}\,{\rm GeV}$, where there exists a symmetry connecting the electromagnetic, 
weak nuclear and strong nuclear forces. Evidence for this comes from the running of the couplings of the individual interactions with 
energy and the fact that the Electroweak Symmetry forms the basis of the standard model. This symmetry would be broken at phase 
transitions during the Early Universe and topological defects can form when the topology of the vacuum manifold is non-trivial. This is 
likely to be the case for a large fraction of symmetry breaking schemes from popular GUT models such as 
$SO(10)$~\cite{Jeannerot:2003qv}.

An alternative is that the fundamental interactions are unified in String Theory,
which is self-consistent only when the number spatial dimensions is 
much higher than the standard 3, typically 10 or 11 in the most popular models. 
Such models allow for a plethora of higher-dimensional topological defects known as D-branes, which 
can behave like cosmic strings from the three-dimensional point of view~\cite{Jones:2003da}. 
These cosmic superstrings~\cite{Copeland:2003bj} exhibit a range of complicated dynamics such as 
multiple tension networks that are the subject of on-going investigations.

A number of inflationary models derived from fundamental physics 
lead to the formation of topological defects at the end of inflation.
Such models are known as {\it hybrid} inflation models~\cite{Linde:1993cn,Copeland:1994vg}. 
Since the energy scale associated with these models is typically the GUT scale,
the topological defects lead to potentially detectable signatures uncorrelated with those 
from inflation. These include superymmetric (SUSY) hybrid inflation~\cite{Dvali:1994ms,Binetruy:1996xj} 
and brane inflation~\cite{Dvali:1998pa}. In both cases, and possibly more generally, the inflation 
generated gravitational wave background will be undetectable, $r\ll 10^{-4}$, in such models, 
but there will still be a B-mode signature produced by the 
defects~\cite{Seljak:1997ii,Battye:1998js,Pogosian:2007gi,Bevis:2007qz,garcia} because
defects produce significant levels of vector modes. Moreover, non-topological defects 
from global O(N) symmetry breaking at the end of inflation can generate a 
scale-invariant spectrum of GWs with an amplitude
significantly larger than that from inflation~\cite{fenu,jonessmith}.

The most commonly studied models are cosmic strings,
 which might be formed, for example, if the symmetry breaking transition results from 
the breaking of a $U(1)$ symmetry. 
The amplitude of the anisotropies and polarization 
for the CMB (and all other gravitational effects) created by strings is governed by 
the dimensionless parameter $G\mu/c^2.$ $\mu\sim\eta^2$ is the mass per unit 
length of the strings. For a symmetry breaking transition that takes place at $\eta$,
we find that $\mu \sim \eta ^2.$ 
$G\mu/c^2$ is typically $\sim 10^{-6}$ for strings formed at the GUT scale. 
The zoo of possible topological defects models is large and includes 
textures~\cite{Bevis:2004wk}, semi-local strings~\cite{Urrestilla:2007sf} 
and strings with multiple tensions motivated by cosmic 
superstrings~\cite{Pourtsidou:2010gu}. The spectra for these models are 
qualitatively similar to those for strings and therefore we will concentrate 
specifically on strings.

There is some uncertainty in precise details of the anisotropies and polarization 
predicted by cosmic strings~\cite{Wyman:2005tu,Battye:2006pk,Bevis:2007gh,Fraisse:2007nu,Battye:2010xz,Bevis:2010gj,Landriau:2010cb}. 
The qualitative features of the anisotropy spectrum $(\Delta T_{\ell})^2=\ell(\ell+1)C_{\ell}/2\pi$ are: (i) a gradual increase at 
low- $\ell$, due to the Integrated Sachs-Wolfe (ISW) effect from strings along the line of sight,  
to a peak somewhere between $\ell=500$ and 800---the precise scale being set by the correlation length of the network; 
(ii) a significant contribution from vector modes due to the large velocity component of the energy-momentum 
tensor and no Doppler peaks due to the lack of coherence of the perturbations; 
(iii) a power law (as opposed to exponential) decay which results from sharp features induced 
by the Kaiser-Stebbins effect \cite{kaiser:stebbins}. The E-mode polarization spectrum is similar to the anisotropy spectrum, 
although it lacks the ISW contribution on large and small scales.

\begin{figure}
\begin{center}
\includegraphics[width=12cm]{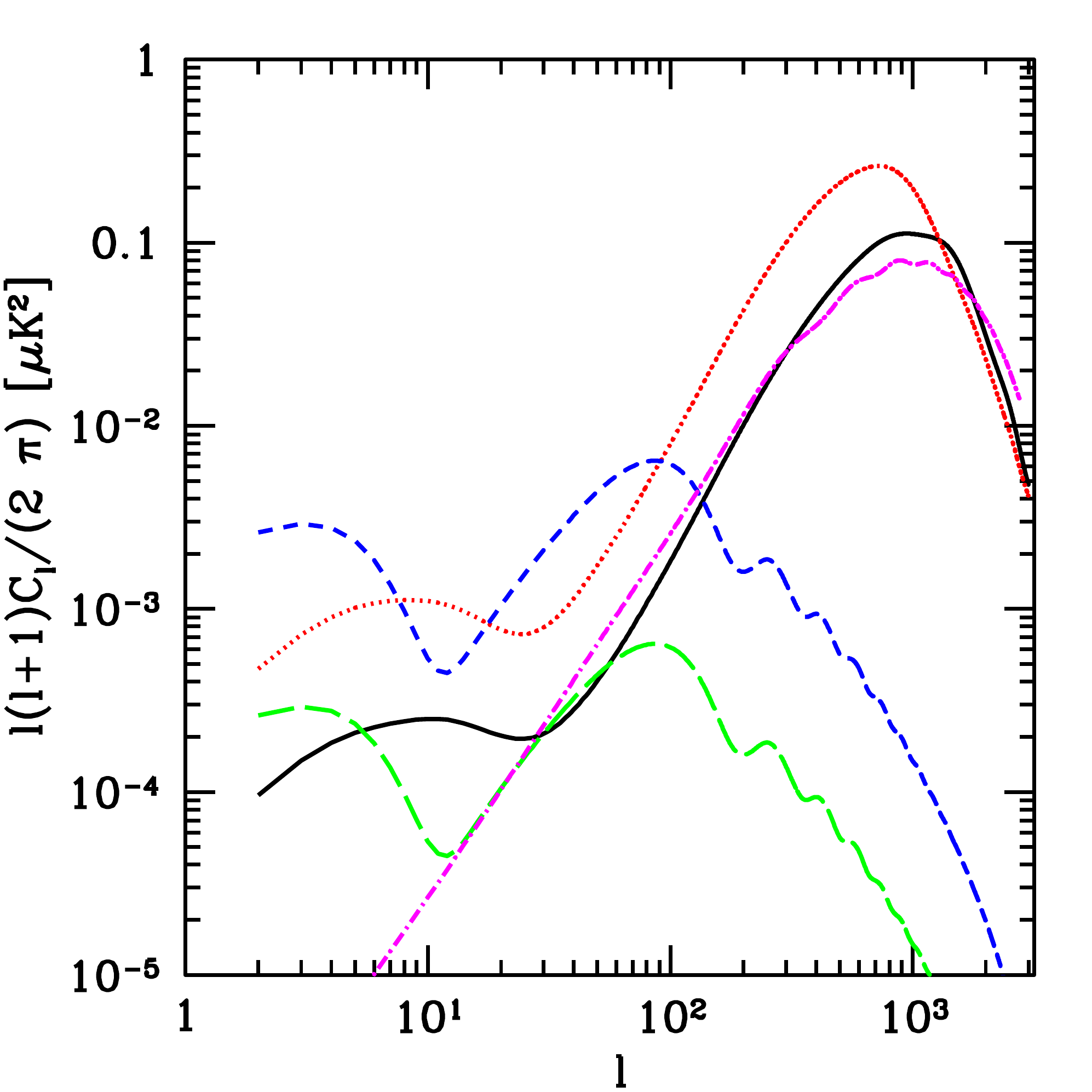}
\end{center}
\caption{\label{fig:bb}  $B-$mode polarization power spectra from strings, 
tensors and gravitational lensing presented in ref.~\cite{Battye:2010xz}. 
For the string spectra, we use values of $G\mu/c^2$ corresponding to the 
$2\sigma$ upper limit from CMB+SDSS data, that is. $G\mu / c^2=2.6\times 10^{-7}$ 
for the USM Nambu model (solid) and $6.4\times 10^{-7}$ for the USM AH case 
(dotted). We also show the inflationary primordial tensor spectrum with $r=0.1$ 
at $k=0.05 \, {\rm Mpc}^{-1}$ (short dashed) and $r=0.01$ (long dashed). 
Finally, we show the gravitational lensing spectrum generated from $E-$mode 
mixing (dot-dash) expected in the inflationary model.}
\label{td:fig}
\end{figure}

The B-mode power spectrum from strings is very different to that due to inflation induced 
gravitational waves. The spectra predicted by Abelian-Higgs (AH) simulations and Nambu 
simulations created using the `unconnected segment model' (USM)  are presented in Fig.~\ref{td:fig} with their 
amplitude set by the upper limits on $G\mu/c^2$ as discussed in the next paragraph. The qualitative features of 
the string and inflationary spectra are quite similar but the precise scales involved are very different. 
See ref.~\cite{Battye:2010xz} for a discussion of the differences between these two possible manifestations of 
the spectrum. On large scales there is a bump at low $\ell$ due to reionization and then there is an 
approximately white-noise ($C_{\ell}$ constant) portion of the spectrum rising to a peak. However, the peak 
is somewhere between $\ell=500$ and 1000 for strings, in contrast to $\ell\sim 100$ expected for inflationary 
spectra. 
The reason why the peak is at higher $\ell$ is because the defect models create perturbations inside the horizon as opposed to at horizon crossing,
which is the case in inflationary models. Unfortunately, there is some disagreement over the precise value of the correlation length and this leads 
uncertainty in the peak position.
There are no peaks in the high $\ell$ power spectrum. The spectrum due to 
strings is, in fact, very similar to that created by gravitational lensing of density perturbations and 
they could easily be confused, particular for the case of the Nambu spectrum.

The present $95\%$ upper limits on $G\mu/c^2$ come from a fit to the standard six
parameters plus an additional parameter to describe the amplitude of the 
string spectrum which is usually $G\mu/c^2$ or $f_{10}$, the contribution of strings 
to the $\ell=10$ multipole. Using properties of the string networks computed in 
Nambu simulations in the USM one finds that $G\mu/c^2<2.6\times 10^{-7}$~\cite{Battye:2010xz} 
using WMAP and SDSS and this can be improved to $1.6\times 10^{-7}$~\cite{Dunkley:2010ge} 
using results from ACT, whereas if one uses the spectra computed directly from Abelian Higgs 
simulations one finds that $G\mu/c^2<7\times 10^{-7}$~\cite{Bevis:2007gh} using just WMAP. 
In addition there is an even less certain limit of $G\mu/c^2<7\times 10^{-7}$ from the absence 
of a stochastic gravitational wave background created by the decay of cosmic strings 
loops~\cite{Battye:2010xz}.

We have performed an analysis of the prospects for detecting this background using COrE following the methods recently used for 
simulated CMBPOL observations~\cite{Mukherjee:2010ve}. We simulate CMB data for a flat $\Lambda$CDM model with the set of parameters 
used in ref.~\cite{Baumann:2008aq} and foreground subtraction residuals are also included~\cite{Verde:2005ff} with the aim of 
propagating their effect into parameter uncertainties. We use the cosmic variance of the weak lensing B-mode signal as additional 
uncertainty on the B-mode spectrum. We use the string B-mode spectrum computed directly from the AH simulations. We assume the 
inflationary consistency relation $n_{\rm t}=-r/8$ for the tensor spectral index and do not allow running of the scalar spectral 
index. The inflationary parameters are specified at a pivot scale of $k _*=0.05 \, {\rm Mpc}^{-1}$. We assume that 80\% of the sky 
can be used for cosmological analysis. The simulated data with noise expected for COrE are used as an input to MCMC analysis using 
COSMOMC~\cite{Lewis:2002ah}.

In an 9 parameter fit (the standard 6 parameters plus $r$ and the amplitudes of a string and texture components, for a model with no 
inflationary induced gravitational waves nor any defect component, we find limits $r<10^{-3}$, $f_{10}^{\rm str}<1\times 10^{-4}$ and 
$f_{10}^{\rm tex}<5\times 10^{-4}$ where $f_{10}^{\rm str}$ and $f_{10}^{\rm tex}$ are the contributions of the strings and textures 
to the anisotropy power spectrum at $\ell=10$ respectively. The correlation between tensors and strings is small, and there is little 
chance of signal misidentification (the quality of the fits, i.e., mean or best-fit likelihoods would point to the right component). A 
Bayesian model selection analysis that takes model priors into account will also be feasible to address model-level questions on the 
evidence for various defects, as demonstrated in ref.~\cite{Mukherjee:2010ve}. In absence of a detection COrE will deliver 
simultaneous 95\% confidence upper limits of $1\times 10^{-3}$ on tensors and $6 \times 10^{-8}$ on the $G\mu/c^2$ of strings compared 
to the present limit which is $7\times 10^{-7}$ for this model. Similar detection opportunities apply to other defect models such as 
textures and the fact that the different components are uncorrelated means that one can not just to detect defects, but to distinguish 
the type of symmetry breaking pattern through the type of defects generated. This would be an entirely unique measurement.

\subsection{Galactic science}


{\bf Undoubtedly, any claim of a detection of B-mode CMB polarization will face a critical assessment against alternative interpretations
involving foreground contributions. The necessity of reliable foreground removal 
links the search for a 
polarization signal from cosmic inflation to research in Galactic astrophysics.  Converserly, COrE hold promises of breakthroughs 
in our understanding of the dusty magnetized interstellar medium in the Milky Way. 

Dust and synchrotron radiation from the Galaxy  are the dominant contributions to the microwave sky polarization. 
Synchrotron radiation traces the magnetic field over the whole volume of the Galaxy, while 
dust polarization traces the field within interstellar clouds. 
COrE will image these two complementary tracers of the Galactic magnetic field  
with the sensitivity and angular resolution needed to gain a complete view of Galactic magnetism.
The improvement in sensitivity of COrE over Planck is factor of 10 
for the dust polarization and a factor of 30 for the synchrotron emission.  
At the highest frequency COrE will map the field with an
angular resolution of 1.3', a factor 3 better than Planck, which complements the small angular scales 
probed with ground based instruments. This quantitative jump forward in the observations
will provide the missing data needed to map continuously the Galactic magnetic field. The data analysis will 
unravel the interplay between the magnetic field 
and interstellar turbulence, and the role that the field plays in the formation of interstellar structure and star formation.
COrE polarization observations will 
provide complementary information to advance our understanding of the nature of interstellar dust, 
its evolution across interstellar space, and grain alignment with the field direction. }

\subsubsection{Galactic magnetic field}

The Galactic ecological system cannot be understood
without knowledge about the strength and structure of its magnetic field.
COrE will make possible to understand this elusive but dynamically important
component of galaxies.  

The magnetic field is dynamically tied to  cosmic rays and  the turbulent interstellar gas.  
These three components have  
comparable energy densities indicating that they are in
continuous interaction and providing feedback to each other. 
The importance of the magnetic field for the energetics of the interstellar medium (ISM),  
star formation, and galaxy evolution is widely  
recognized, but its actual role remains quantitatively debated due to the paucity of data on its 
its correlation with interstellar turbulence and the density structure of the ISM. 
 We detail some of the prime questions that the COrE data will resolve in a unique way.  \\

\noindent
{\bf (1) What is the impact of the magnetic field on the interstellar medium energetics?}\\

Synchrotron emission is a prime tracer of the galactic magnetic field. 
The synchrotron intensity allows  estimation of the total magnetic field
strength, while its polarization fraction gives information about the field structure on both 
global and local scales.  The COrE sensitivity will be far superior to other missions.
It will provide a major step forward in our ability to study  magnetic
fields in the Milky Way disk  and  halo.  The
very accurate, highly resolved and detailed view at the polarization of synchrotron emission will be free of Faraday rotation.
Faraday rotation causes the synchrotron polarization vector
to turn as it propagates through the 
magneto-ionized medium. This greatly complicates
analyzing synchrotron emission and can lead to depolarization. At COrE frequencies, the
observed polarization angle of the synchrotron emission directly
relates to the direction of the magnetic field at the location of
emission. This gives a diagnostic of the Galactic magnetic field
complementary to ground-based observations at lower frequencies 
subject to significant Faraday rotation. 

Fluctuations in the synchrotron flux in intensity and polarization provide insight into magnetic turbulence. 
Several characteristic statistical properties are encoded in these data, for example the magnetic 
energy and helicity spectra as well as the spectrum of the magnetic tension force.
The accuracy of the COrE data will thus provide a unique opportunity to  
study magneto-hydrodynamical turbulence and dynamo action, which govern magnetic field amplification and
the energetics of the ISM. It will drastically increase the spectral range of accurately probed magneto-hydrodynamical modes. Therefore 
the detection potential for relevant plasma physics processes and their characteristic scales such as turbulent energy 
injection and dissipation will be considerably increased.  \\

\noindent
{\bf (2)  What role does the magnetic field play in the evolution and structuring of interstellar 
matter and the onset of star formation?} \\

The interplay between turbulence and magnetic fields is also key to star formation.  To investigate this question one needs
to study the polarization of the dust emission, which traces the magnetic fields within interstellar clouds.
Dust polarimetry in emission traces the interstellar magnetic field projected on the plane of the sky. It is 
possible to estimate the magnetic field intensity from polarimetric maps by relating the local dispersion of polarization angles 
and the local amplitude of the turbulent gas motions, a method first proposed by \cite{Chandrasekhar53}. 

The COrE survey will provide the combination of sensitivity and angular resolution required 
to map continuously the Galactic magnetic field associated with interstellar matter, 
from the diffuse interstellar medium to molecular clouds where pre-stellar cores are formed.
No other experiment offers such a capability. Planck data will provide much information on the Galactic magnetic field on large
scales, but does not have the required sensitivity to map the field across the diffuse interstellar medium.  
Nor do upcoming balloon-borne experiments such as Polar-BLAST and PILOT have the required sensitivity.
Stellar polarization observations will 
continue to improve, but they are intrinsically limited to discrete sets of sight lines. 
Ground based telescopes at sub-mm and millimeter wavelengths including ALMA will
image at sub-arcminute angular scales a variety of compact sources including pre-stellar
condensations, but they cannot map polarization from diffuse emission.    

Star formation occurs as a result of the action of gravity, which is 
counteracted by thermal, magnetic and turbulent pressures (\cite{Larson03} and \cite{McKee07}).  
In the diffuse interstellar medium, outside star-forming molecular clouds,  
the kinetic energy from interstellar turbulence and the magnetic energy 
are comparable. Both are much larger than the cloud gravitational binding energy 
and the gas internal energy.  For stars to form, gravity must become, locally at least, the dominant force. 
This occurs where the turbulent energy has dissipated and matter has condensed 
without increasing the magnetic field flux in comparable proportions. When and how frequently does this occur? This question
is key to our physical understanding of what regulates the efficiency of star formation.  
There is a broad consensus that the answer will follow once we have a physical understanding of the 
interplay between interstellar turbulence and the magnetic field structure in interstellar clouds.

The consumption timescale of the molecular gas in the  Milky Way, $\rm \sim 10^9 \, yr$,  is two orders of magnitude 
larger than the dynamical timescale of its giant  molecular clouds (i.e., their cloud crossing time).
The explanation of this observational fact, established in the early 80s with the advent of CO observations, is still a subject of debate.
The first scenario proposed to regulate star formation combines 
long cloud lifetimes and a low star formation efficiency.  According to this scenario, 
molecular clouds are prevented from collapsing on large scales by turbulent and magnetic 
pressures, while star formation is locally controlled by the rate at which material can cross field lines through
ambipolar diffusion.  Observations and numerical simulations are challenging this view (\cite{Ballesteros07}).  
It is increasingly believed, but still debated,
that the bottle-neck to star formation is the formation of magnetically supercritical condensations where the
magnetic pressure is too weak to
counteract gravity. Once formed, these condensations rapidly collapse to form stars.
Star formation would generally be inefficient, because only a small fraction of the matter
reaches this stage within a cloud lifetime.
Under this second paradigm, molecular clouds are 
transient, and dynamically evolving gas concentrations are produced by compressive
motions out of diffuse 
interstellar matter. This view stresses the importance of the unique ability of 
the COrE survey to map the field structure over the extended regions 
of molecular clouds and of the diffuse interstellar medium. 
{\bf COrE will provide the missing observations that we need to understand 
how the interplay between turbulence and magnetic fields determines the initial conditions for star formation 
within the diffuse  interstellar medium.}  \\


\noindent
{\bf (3) What is the impact of the magnetic field on the large-scale 
structure and evolution of the Galaxy?}\\

COrE is expected to provide important new clues concerning 
the effects of the Galactic magnetic field on larger scales.
We know that the magnetic field helps support the interstellar gas 
against its own weight in the Galactic gravitational potential 
and that it confines cosmic rays to the Galactic disk.
In this manner, both the magnetic field and cosmic rays partake 
in the overall hydrostatic balance of the interstellar medium,
causing the gaseous disk to become much thicker.
In turn, the thickening of the disk tends to make it unstable to 
the Parker instability (a type of magnetic Rayleigh-Taylor instability).
As this instability develops, magnetic field lines undulate,
and the interstellar gas slides down along the field lines into the magnetic troughs 
accumulating there.
The whole process, it has been suggested, could give birth to new
molecular cloud complexes and ultimately trigger star formation.

 
The magnetic field is also believed to play an important role
in disk-halo interactions, although the exact mechanisms involved
are still poorly understood and debated.
{\it A priori} one expects the magnetic field to impede mass
exchange between the disk and the halo.
The magnetic field could tend to contain super-bubbles
and prevent them from venting their gas into the halo. This
view is challenged by observations of edge-on galaxies which
show evidence for a multiphase halo in galaxies like the Milky Way (e.g., NGC~4631 \cite{Wang01}).
In contrast to previous studies that used a static medium,
numerical simulations that follow the evolution
of superbubbles within a supernova-driven turbulent magnetized medium
suggest that blowout is more likely than previously thought, mainly because
bubbles evolve in an inhomogeneous background medium
\cite{Avillez05}.  Theoretical studies also suggest that the magnetic field
might stabilize infalling clouds and enable them
to reach the disk without being disrupted by their dynamical interaction
with the hot halo gas \cite{Heitsch09}.
To test these ideas and numerical simulations, it is crucial to gain a better observational
knowledge of the Galactic magnetic field, particularly in the halo.
Similarly, a better knowledge of the Galactic magnetic field,
both in the disk and in the halo, will make it possible to place
stronger constraints on the existing dynamo scenarios.\\

\subsubsection{Polarization properties of interstellar dust}

The unique combination of spectral and spatial information provided by the 
COrE survey will open a new dimension to our understanding of dust, its 
nature, and its evolution within interstellar space. 

The Galactic interstellar medium is dusty. Large dust grains (size $> 10$~nm) dominate the dust mass. 
Within the diffuse interstellar medium, the grains are cold ($\sim10-20$~K) and emit at far-IR to millimeter wavelengths. 
Dipolar emission from small carbon dust particles is a main emission component, often refered to as the anomalous microwave emission.
Both dust components are relevant to CORE observations.  

Dust properties (size, temperature, emissivity) are found to vary from one line of sight to another 
within the diffuse interstellar medium and molecular clouds. These observations indicate that dust grains evolve through 
the interstellar medium. They can grow through the formation of refractory or ice mantles,  or 
by coagulation into aggregates in dense and quiescent regions. They can also be destroyed by 
fragmentation and erosion of their mantles under more violent conditions.  The composition of interstellar 
dust reflects the action of interstellar processes, which contribute to break and re-build grains over timescales 
much shorter than the timescale of injection by  stellar ejecta. While there is wide consensus on 
this view at interstellar dust, the processes that drive its evolution in space are still poorly understood (\cite{Draine09}).   
Understanding interstellar dust evolution is a major challenge  in astrophysics
underlying key physical and chemical processes in interstellar space. 

Polarization observations are a new, essentially unexplored, means
to study interstellar dust.  The polarization of dust emission  is related to grain 
properties (size, shape, and magnetic susceptibility) and the efficiency of grain alignment, which depends 
on local physical conditions (gas density, radiation field). Dust evolution modifies 
the optical properties of dust grains as well as the efficiency of the mechanism that tends to 
align their rotational axis along the magnetic field lines.  
The physical processes that couple dust grains and their alignment with local physical conditions 
make the separation of the Galactic and CMB polarization intrinsically difficult. 
These physical couplings break the simplest assumption by which the spectral frequency 
dependence of the Galactic polarization and its angular structure on the sky are separable.  

There are several questions that we need to answer in order
to characterize the spectral dependence of dust polarization and its variations across the sky. 
Which dust components are polarized? 
How does one account for grain alignment, and what is its dependence on local 
physical conditions? 
The wide spectral coverage and the large number of spectral bands are unique assets of the CORE project 
to answer these questions.  Success in this challenging endeavor  will be key 
to understand fully how dust polarization acts as a tracer of 
the magnetic field structure within interstellar matter and how
to achieve an optimal accuracy in component separation for the detection of B-modes. 

\begin{figure}
\centering
\includegraphics[scale=0.4]{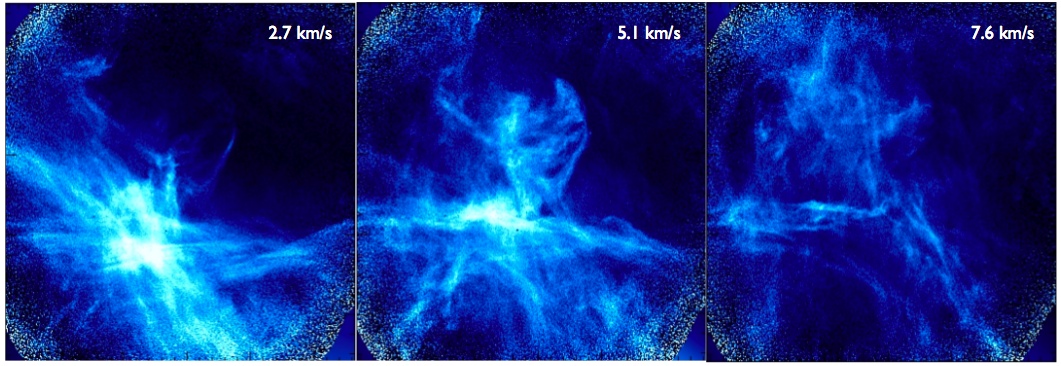}
\caption{The COrE data will allow us to map the turbulent component of the Galactic magnetic field, and thereby 
characterize its role in shaping the diffuse interstellar medium and regulating star formation. 
The interstellar medium filamentary structure is illustrated here using HI spectroscopic observations obtained 
with the Dominion Radio Astrophysical Observatory. The three images correspond to three different velocity 
channels of one $4^\circ \times 4^\circ$ field (data from the Planck Deep Field HI survey, PI Peter Martin).}
\label{fig:HI}
\end{figure}

\subsubsection{Galactic center}

\noindent
{\bf {Nonthermal emission in the galactic center}}\\

The center of the Milky Way is the locus of high energy processes
and a source of high energy cosmic rays which
interact with the Galactic magnetic field as well as with interstellar matter.
The polarization measurements of the plasma produced and
diffusing from the Galactic Center (GC) region can set crucial constraints on
the nature of the non-thermal emission from the GC region, on the
large-scale structure of the magnetic field stretched along the
outflowing regions (thought to have an intensity $\sim 50 - 500$
$\mu$G), on cosmic ray sources (likely SNRs), on the
properties of the diffusive transport (and its effective
time scale in comparison to advective time scales), and on the
relative importance of hadronic and leptonic processes for the
production and acceleration of nonthermal particles in the GC
region.

\vspace{2mm}
\noindent
{\bf {Evidence for Dark Matter annihilation}}\\

%
Annihilation of Dark Matter (DM) particles ( $\chi$)  followed by diffusion of secondary particles in the
magnetic field of the GC region could be producing detectable
$\gamma$-ray emission (due to $\chi \chi \to \pi^0 \to
\gamma + \gamma$) together with a diffuse halo (haze) of 
synchrotron and inverse Compton 
scattering of CMB (and other background) photons by energetic
electrons (due to $\chi \chi \to X + \pi^{\pm} \to X + e^{\pm}$).
The intensity, spatial extension, and
spectrum of such emission depend on the details of the DM particle
composition and annihilation cross-sections as well as on the
Galactic magnetic field establishing the diffusion properties of
the secondary particles produced in DM annihilation (\cite{Colafrancesco06}).

The existence of an anomalous microwave ``haze" (in the WMAP data) coinciding 
with gamma-ray bubbles (in the Fermi-LAT data)  centered around the GC
has been largely debated and remains a controversial
issue (\cite{dobler10,Gold11}. Emission from DM secondary particles 
was originally invoked to explain the anomalous WMAP microwave ``haze". 
%
%
This interpretation is at variance with an astrophysical scenario, where
the putative haze would be associated with star formation in the GC region 
producing a fast
wind with substantial quantities of cosmic ray ions
that diffuse out to large angular scales on the sky \cite{Crocker11}.

The broad frequency and high sensitivity polarization survey of the
GC region obtainable with COrE will provide an unprecedented view of
the GC region and may lead to the discovery
of a microwave ``haze" associated with DM annihilation.
Polarization is likely to play a key role in discriminating among competing
possible interpretations.  
While strong polarization features (both in synchrotron
and in inverse Compton emission) are expected in a scenario of SNe-driven 
outflows and winds, a DM-produced microwave ``haze" is expected to have
low polarization, because the secondary electrons
are produced {\it in situ} along the DM density profile of the Galaxy,
and are subject to a smooth and rather isotropic spatial diffusion
in the GC magnetic field.

\subsubsection{COrE sensitivity to polarized synchrotron}

Accurate measurements of the polarized synchrotron radiation provide a unique probe to 
understand the structure of the Galactic magnetic field and to study the energy 
distribution of cosmic rays. The WMAP mission has provided first full-sky sensitive measurements of 
synchrotron-dominated sky emission polarization in the 20-90 GHz frequency range
\cite{2007ApJS..170..335P}, providing also an estimate of the level of contamination by polarized galactic emission
outside of the galactic plane, in regions useful for CMB studies.

The observing power of the low frequency channels of COrE 
will be ideal to extract the rich information encoded in the Galactic diffuse 
synchrotron component. Synchrotron radiation is intrinsically highly polarized, up to 
70-75\% in a completely regular field. The observed synchrotron polarization depends on 
the uniformity of the field orientation within the resolution element. The typical 
synchrotron sky temperature at 45 GHz---the lowest-frequency COrE channel---is $\sim 35 
\mu$K at intermediate Galactic latitudes. Assuming a polarization fraction of $\sim 
60\%$, we predict the typical polarized signals to be of order $\sim 20\mu$K, with 
values depending on the features observed. Significantly higher values may be found in 
the loops and spurs, remnants of old supernova events close to the Galactic plane. At 
high latitudes, away from such local features, the expected degree of polarization 
should be only a few percent, resulting in polarized signals of a few $\mu$K. At 
intermediate latitudes, the polarization will lie somewhere between these two extremes.

The design sensitivity of the COrE 45, 75, and 105 GHz channels is approximately 
0.4$\mu$K per resolution element (with beam-widths of 23$'$, 14$'$ and 10$'$), 
corresponding to signal-to-noise ratios (SNRs) for synchrotron polarized emission of 
$\sim 50,$ $\sim 15$, and $\sim 4$, respectively. This will represent an enormous 
step forward compared to the 30 months of Planck data, which is expected to achieve 
SNRs for synchrotron polarimetry of $\sim 2$, $\sim 0.3,$ and $\sim 0.2$ at 44, 70, and 
100 GHz, respectively, at similar angular resolution. The polarization maps from the 
Planck LFI 30 GHz channel and from the WMAP 23 GHz channel, both with high SNR ($\sim 10$) 
for polarized synchrotron, will constitute useful low-frequency ancillary data for the COrE 
analysis, in particular for disentangling synchrotron radiation from the contribution of 
the polarized component of interstellar dust.{\bf 
The exquisite polarization sensitivity of COrE will therefore deliver a high 
definition extraction of the synchrotron component in a Faraday-rotation free frequency 
domain.}

\subsubsection{Statistical analysis of Galactic magnetic fields with COrE}

The accuracy of COrE polarization information on Galactic magnetic fields will provide a 
unique opportunity to study magneto-hydrodynamical turbulence and dynamo action in great 
detail within our Galaxy. It would greatly improve the reliability of any polarization 
data on small angular scales of the Milky Way in synchrotron light and thereby increase 
the spectral range of accurately probed magneto-hydrodynamical modes. Therefore the 
detection potential for relevant plasma physical processes and the characteristic 
scales of turbulent energy injection and dissipation would be increased considerably. 
Furthermore, accurate Galactic polarization data will be of eminent value for upcoming 
Faraday tomography measurements with telescopes like LOFAR, eVLA, ASKAP, and especially 
the SKA, which has a comparable timeline as COrE.

\subsubsection{Faraday rotation free polarization}

The COrE 45 or 60 GHz channels will provide a very accurate, highly resolved and 
detailed view on the synchrotron emission of our own galaxy. The polarization data 
permits to study the morphology of Galactic magnetic fields on global and local scales. 
CMB experiments measure at high frequencies and reveal there the original polarization 
structure, which do not suffer from Faraday rotation. Planck is currently increasing the 
angular resolution of such maps to a level comparable to ground based radio telescopes 
and COrE will boost the sensitivity and spectral information.

\begin{figure}[t]
 \centering
\includegraphics[bb=0 0 502 320,width=\columnwidth]{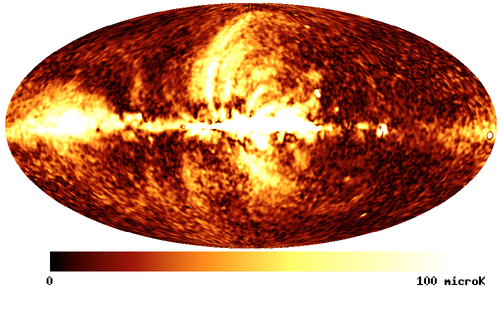}
 \includegraphics[bb=0 0 589 294,width=\columnwidth]{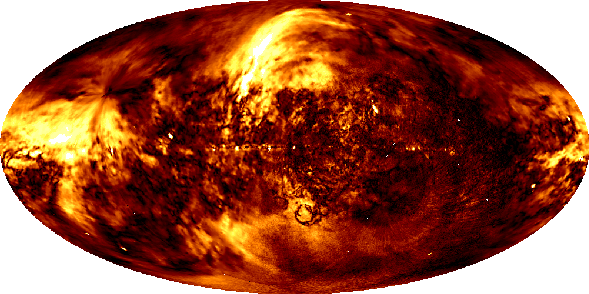}
 \caption{The WMAP 22.8 GHz all-sky polarized intensity map (upper
panel) and the 1.4 GHz all-sky polarized intensity map (lower panel).
The polarized intensities are shown greyscale coded from 0 to 100 $\mu$K
for 22.8 GHz and from 0 to 570 mK for 1.4 GHz. Galactic Faraday-depolarization structures 
are visible in the lower frequency map. Data from 
\cite{2006A&A...448..411W,2007ApJS..170..335P,2008A&A...484..733T} 
and figures from \cite{2008A&A...477..573S}.
}
 \label{fig:galPol1400MHz}
\end{figure}

All ground-based radio-band synchrotron measurements of Galactic magnetic fields to date
were carried out at much lower frequencies. Therefore they exhibit a high level of Faraday rotation and 
depolarization within the Galactic plane, hampering a direct analysis. COrE, by contrast, will 
unveil the polarization structure with essential no Faraday rotation (see Fig.~\ref{fig:galPol1400MHz}). {\bf This data will 
make possible disentangling depolarization phenomena arising from Faraday 
rotation by the Galactic magnetic field and the superposition of contributions with 
different polarization orientations along the line of sight within the beam.} Having this 
unrotated polarization information with high precision will be of importance for at 
least two scientific research directions:

\begin{itemize}

\item Accurately measured fluctuations in the synchrotron flux in intensity and
polarization provide insight into magnetic turbulence. Several characteristic
statistical properties are encoded in this data, for example the energy
and helicity spectra as well as the spectrum of the magnetic tension force.

\item Such data will allow Faraday rotation tomography measurements of
our galaxy, which will be pursued with upcoming instruments like
LOFAR, eVLA, ASKAP, and especially the SKA. Faraday tomography data
can be expected to reveal further statistical information on Galactic fields,
with very promising potential for detecting signatures of magneto-hydrodynamical
processes.
\end{itemize}

\subsubsection{Magnetic spectra}

The statistical properties of Galactic magnetic fields imprint themselves on observables
such as the synchrotron intensity, polarization, and Faraday rotation measure.
Methods to extract this information from observational data already exist and are being 
improved.  Quantities highly relevant for an understanding
of Galactic turbulence and dynamo processes are encoded in polarimetric
data. Examples include:
\begin{itemize}
 \item The {\bf magnetic energy spectrum}, which is imprinted on intensity, polarization spectra and 
cross spectra \cite{1982ApJ...261..310S,1983ApJ...271L..49S,1989AJ.....98..244E,1989AJ.....98..256E}.

 \item The {\bf magnetic helicity spectrum}, which can be measured from polarimetric data in 
combination with extragalactic Faraday data \cite{JunklewitzInThesis,JunklewitzInPrep}. 
Magnetic helicity is a key quantity to understand the inner workings of large-scale 
Galactic dynamos \cite{2009PPCF...51l4043B,2006A&A...448L..33S,2007PPCF...49..447S}.

 \item The {\bf magnetic tension force spectrum}, which is encoded in polarimetry data alone, and is 
powerful in discriminating between different magneto-hydrodynamical scenarios 
\cite{2009MNRAS.398.1970W}.

\end{itemize} 
More physical relevant information might be encoded in the data. The above 
examples are only the quantities that are known today, for which a reconstruction from 
COrE data should be possible.

Due to our location within the galaxy, the angular fluctuations in observables 
correspond to physical magnetic field structures of different sizes. Disentangling these fluctuations 
in order to separate different physical scales such as the turbulent injection scale (see e.g. 
\cite{2008ApJ...680..362H}) or dissipative scales will be challenging. Highly accurate 
polarimetric data, with full-sky coverage and {\bf precise} calibration
will be invaluable for this endeavor. The probing of a large range of 
physical scales simultaneously with high angular resolution
makes it possible to monitor diagnostics
of Galactic turbulence and dynamo theory.

{\bf Statistical analysis of the polarized diffuse 
synchrotron emission from WMAP and from radio surveys disagree.
WMAP finds similar E-mode and B-mode angular power spectra
while those obtained from radio surveys differ significantly 
\cite{2006cmb..confE..16B}. Only more accurate measurements free from 
potential systematic effects can resolve
this discrepancy. COrE will provide the necessary data and test the proposed 
theoretical models \cite{2011A&A...526A.145F} at different angular scales to understand if the 
origin of this observational disagreement is explained by residual systematics, data analysis errors, 
or results from subtle astrophysical mechanisms.}



%

 
\begin{figure}
 \centering
 \includegraphics[width=\columnwidth,bb=0 0 1204 888]{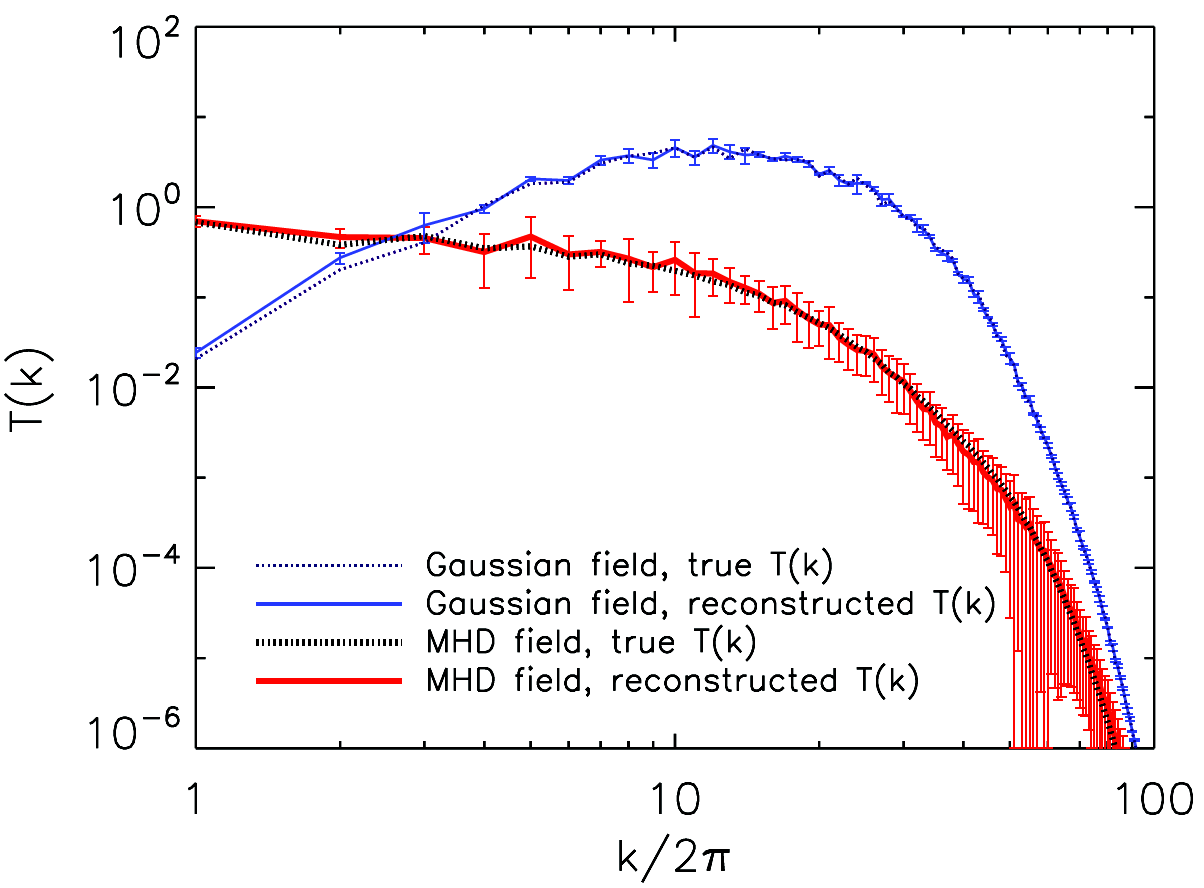}
 \caption{Tension force spectrum reconstructed from mock polarimetry data using the method of Stokes correlators 
(from \cite{2009MNRAS.398.1970W}). The Gaussian random field was constructed to exhibit the same magnetic energy 
spectrum as the magneto-hydrodynamical simulation, but it has a different fourth-order statistic as measured by the Stokes correlators.}
 \label{fig:tension}
\end{figure}

\subsubsection{Faraday tomography}

Faraday tomography provides three-dimensional information on magnetic 
fields and thereby restores some of the information lost in the line-of-sight projection 
of most astronomical observations. The depth information of a synchrotron emitter is 
encoded in the rate of rotation of its polarization angle as a function of wavelength 
$\lambda$. For different sources at different physical depth, and therefore different 
Faraday depth, this rate differs. Mathematically, the observed polarization as a 
function of wavelength squared is the Fourier transformed polarized emission per Faraday 
depth \cite{1966MNRAS.133...67B}.

An inverse Fourier transformation can therefore reveal the polarization per unit Faraday 
depth \cite{2005A&A...441.1217B}. This technique has already been successfully applied to 
radio data \cite{2006AN....327..545B} and is extremely promising for upcoming radio 
telescopes. However the problem for many of these measurements, especially for long 
wavelength telescopes like LOFAR, is that the full $\lambda^2$-space can not be 
examined. In particular, the negative part of this space cannot be probed by 
any instrument. However information in the full $\lambda^2$-space would be required for 
a direct inversion of the Fourier relation. Therefore inverse methods have to be 
applied which benefit from any available information, in particular close to the 
unaccessible negative $\lambda^2$ range. Thus the information close to $\lambda=0$ as 
will be provided by the COre low frequency channels will be of greatest importance 
for Faraday tomography.

Since the Faraday tomographic data will provide much deeper insight into the details of 
Galactic magnetism than the two dimensional information discussed above, it is obvious 
that the scientific return will be even larger. Accurate all-sky COrE data will be 
crucial for the succuess of this technique for exploring the richness of magnetic phenomena in the Galaxy.

 

\clearpage

\subsection{Extragalactic science}

\subsubsection{The Sunyaev Zel'dovich effect in galaxy clusters}

The Sunyaev-Zel'dovich effect (SZE) \cite{SZ1980} in which CMB
photons suffer spectral distortion, or Comptonization, as a result
of scattering off (thermal and nonthermal) electrons of the
cluster atmosphere provides a powerful probe of the energetics,
spectra, and stratification of the electron distribution in
clusters \cite{Colafrancesco2007,Colafrancesco2011} due to its
sensitivity to the details of the electron distribution of the
cluster atmosphere. When scattered off hot electrons by Thomson
scattering, CMB photons are boosted blueward in energy while the
total number density remains constant. This effect leads to a
distortion relative to the CMB blackbody spectrum, which is seen
in the direction of a galaxy cluster according to the relation
\begin{equation}
\Delta I(x)=2\frac{(k_{\rm B} T_{CMB})^3}{(hc)^2}y g(x)
 \label{eq.deltai}
\end{equation}
where $\Delta I(x)= I(x) - I_0(x)$, $I(x)$ is the observed
spectrum in the direction of the cluster, and $I_0(x)$ is the
unscattered CMB spectrum in the direction of a sky area contiguous
to the cluster.
Here $x \equiv h \nu / k_{\rm B} T_{CMB}$ is the re-scaled frequency where $h$ is
the Planck constant and $T_{CMB}=2.726$ K the CMB temperature.
The Comptonization parameter $y$ is given by the line-of-sight integral
\begin{equation}
y=\frac{\sigma_T}{m_{\rm e} c^2}\int P_{\rm e} d\ell ~
 \label{eq.y}
\end{equation}
in terms of the pressure $P_{\rm e}$ contributed by the electron
population. $\sigma_T$ is the Thomson cross section, $m_e$ the electron
mass, and $c$ the speed of light.
The spectral shape of the SZE distortion $g(x)$ is given by
\begin{equation}
 \label{gnontermesatta}
 g(x)=\frac{m_{\rm e} c^2}{\langle \varepsilon_{\rm e} \rangle} \left\{ \frac{1}{\tau_e} \left[\int_{-\infty}^{+\infty} i_0(xe^{-s}) P(s) ds-
i_0(x)\right] \right\}
\end{equation}
in terms of the photon redistribution function $P(s)$ and
\begin{equation}
i_0(x) = I_0(x)/[2 (k_{\rm B} T_{CMB})^3 / (h c)^2] = x^3/(e^x -1)
\; .
\end{equation}
The quantity
\begin{equation}
 \langle \varepsilon_{\rm e} \rangle  \equiv  \frac{\sigma_{\rm T}}{\tau_e}\int P_e d\ell
= \int_0^\infty dp f_{\rm e}(p) \frac{1}{3} p v(p) m_{\rm e} c \;
,
 \label{temp.media}
\end{equation}
where $f_e(p)$ is the normalized electron momentum distribution
function, is the average energy of the electron plasma
\cite{Colafrancescoetal2003}.
The optical depth of the electron population along the line of
sight $\ell$ is
\begin{equation}
\tau_{\rm e} = \sigma_T \int d \ell ~n_{\rm e} \; .
 \label{tau}
\end{equation}
The photon redistribution function $P(s)$, where $s =
\ln(\nu'/\nu)$ is the logarithm of the CMB photon frequency change
$\nu' / \nu$, can be calculated by repeated convolution of
the single-scattering redistribution function, $P_1(s)= \int dp~
f_{\rm e}(p)~P_{\rm s}(s;p)$, where $P_s(s;p)$ expresses the
physics of inverse Compton scattering.
The previous description is general enough to be applied to
both thermal and nonthermal plasmas as well as to a combination of
the two (see
\cite{Colafrancescoetal2003,ColafrancescoMarchegiani2010} for
details).

A velocity (or kinematic) SZE also arises if the scattering medium
causing the thermal (or non-thermal) SZE is moving relative to the
Hubble flow. In the reference frame of the scattering gas the CMB
radiation appears anisotropic, and the effect of the
inverse-Compton scattering is to re-isotropize the radiation
slightly. Back in the rest frame of the observer the radiation
field is no longer isotropic, but shows a structure towards the
scattering atmosphere with amplitude proportional to $\tau_e V_z /
c$, where $V_z$ is the component of peculiar velocity of the
scattering atmosphere along the line of sight
\cite{SZ1972,RephaeliLahav1991,Phillips1995}.
The intensity change due to the kSZE is given by
\begin{equation}
{\Delta I \over I} = - \tau_e \beta_z h(x)
\end{equation}
with $\beta_z \equiv {V_z \over c}$ and $h(x)={x e^x /e^x
-1}$ in the nonrelativistic regime \cite{Birkinshaw1999}. A
relativistic generalized derivation of the kSZE has been given in
the framework of the general Boltzmann equation and in the
covariant formalism \cite{Nozawaetal2010}.

Direct numerical integration of the Boltzmann collision term have been
done in this context 
\cite{Noz2006} that confirm the validity
of analytic formula for the description of kSZE
(see, e.g., \cite{itoj1998,Noz1998,Lasenby1998,sazonov})
the Rayleigh-Jeans limit. 

Due to its redshift-independent nature, the SZE is also a powerful
cosmological probe
\cite{Birkinshaw1999,Barbosaetal1996,Aghanimetal1997,Diegoetal2002}.
and can check for potential traces of cluster evolution better
than other cosmological probes \cite{Delsartetal2010,Adeetal2011},
in particular for galaxy clusters in the medium mass regime.
The SZE can be used to determine accurately the main parameters of
a $\lambda$CDM cosmology and the dark energy equation of state
\cite{Mohr2004}, and also to set constraints on modified gravity
scenarios \cite{Tsutomu2009} and on the properties of primordial
magnetic fields \cite{Tashiroetal2010}.
To realize this program the SZE in galaxy clusters must be
determined with very good accuracy in order to derive accurate and
unbiased cosmological probes.

\begin{figure}[ht]
\begin{center}
\hbox{
 \includegraphics[width=9cm]{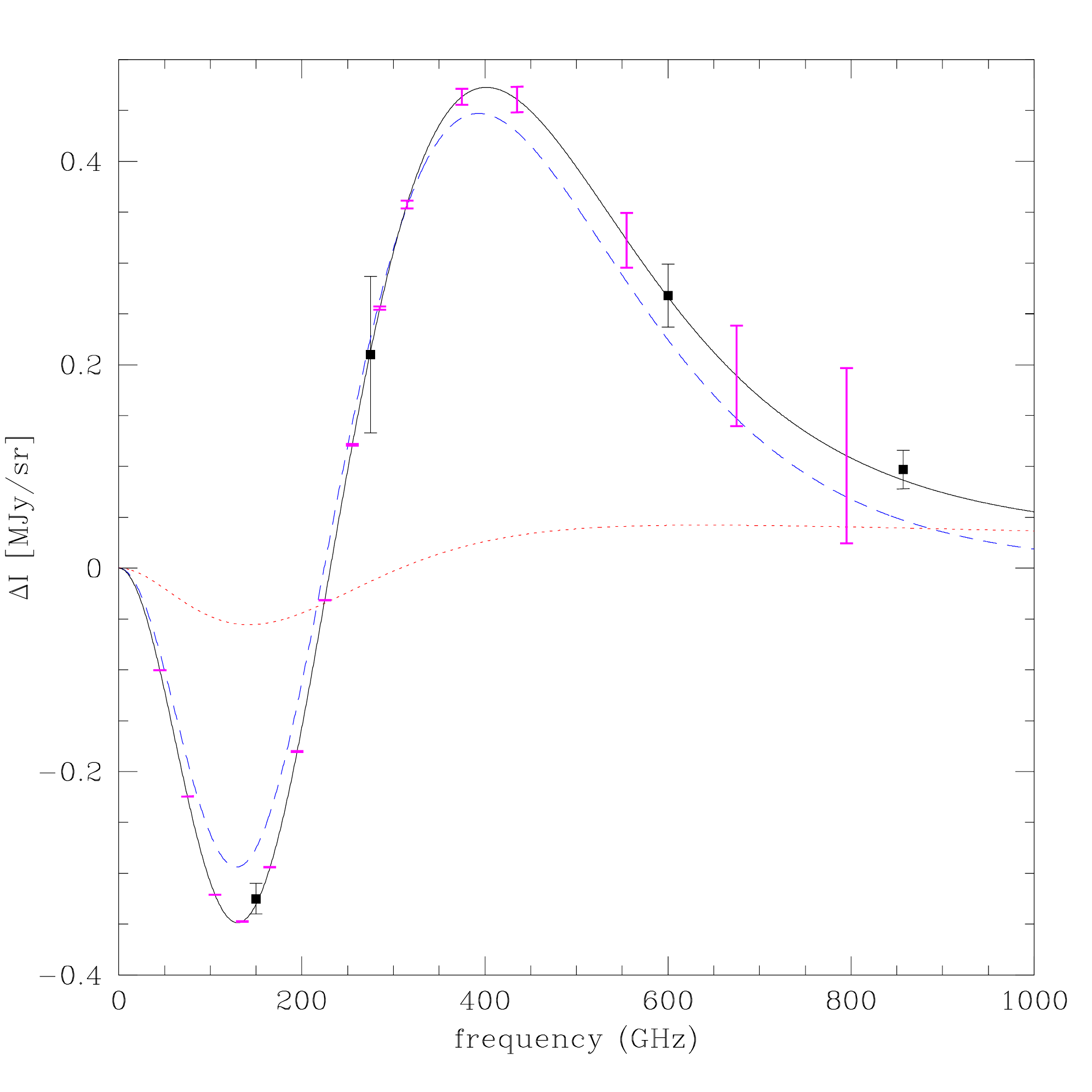}
 \includegraphics[width=9cm]{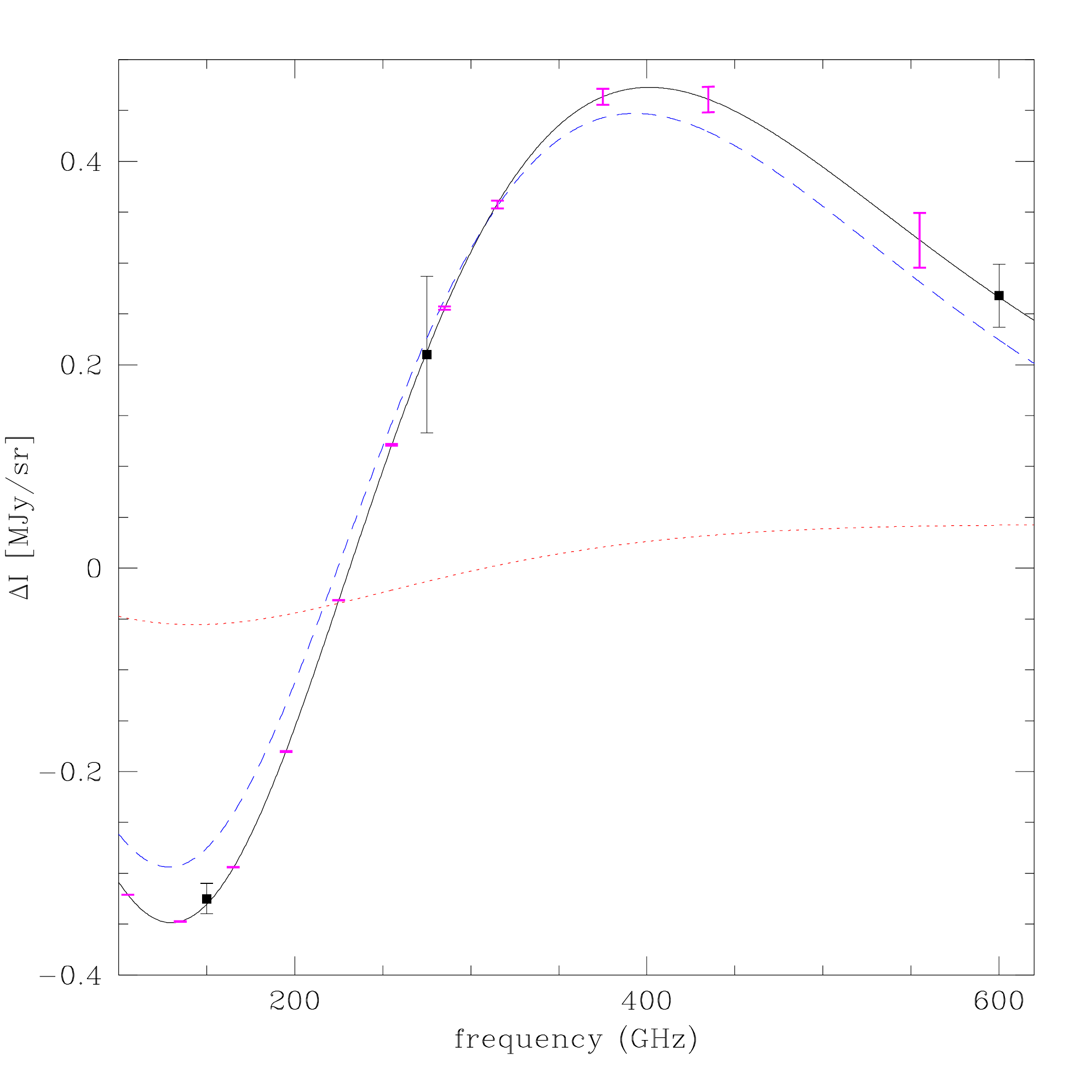}
}
\end{center}
 \caption{\footnotesize{Left: The SZE spectrum at the Bullet
 cluster center modelled with a thermal plus nonthermal plasma. Shown are:
 thermal plasma with $kT=13.9$ keV,
 $\tau=1.1\times10^{-2}$ (blue dashed); nonthermal plasma with $p_1=1$, $s=2.7$ and
 $\tau=2.3\times10^{-4}$ (red dotted); total SZE produced by the sum of the two
 plasmas (black solid). The observed data are shown as black
 squares plus uncertainties and the COrE sensitivity at the
 various frequencies are shown by magenta error bars.
 Right: Same as in the left panel but zoomed in the region where
 the SZ effect spectrum of the Bullet cluster is sensitive to the
 relativistic effects and the COrE sensitivity is highest.
 }}
 \label{fig.bullet_core}
\end{figure}

In this context, precise observations of the SZE at microwave and
millimeter wavelengths are crucial for unveiling the detailed structure of
cluster atmospheres, their temperature distribution, and the
possible presence of suprathermal or nonthermal plasma because
the high-frequency part (i.e., $\nu \simgt 350$ GHz) of the SZE
spectrum is more sensitive to the relativistic effects of the
momentum distribution of the electrons plasma
\cite{Colafrancescoetal2003,Colafrancesco2007,ColafrancescoMarchegiani2010}.
This is even more so for galaxy clusters with a complex plasma
distribution as found in powerful merging clusters such as
the exemplary case of the Bullet cluster (1ES0657-56)
\cite{Colafrancescoetal2011} (see Fig.~\ref{fig.bullet_core}).

\begin{figure}[ht]
\begin{center}
 \includegraphics[width=7.0cm]{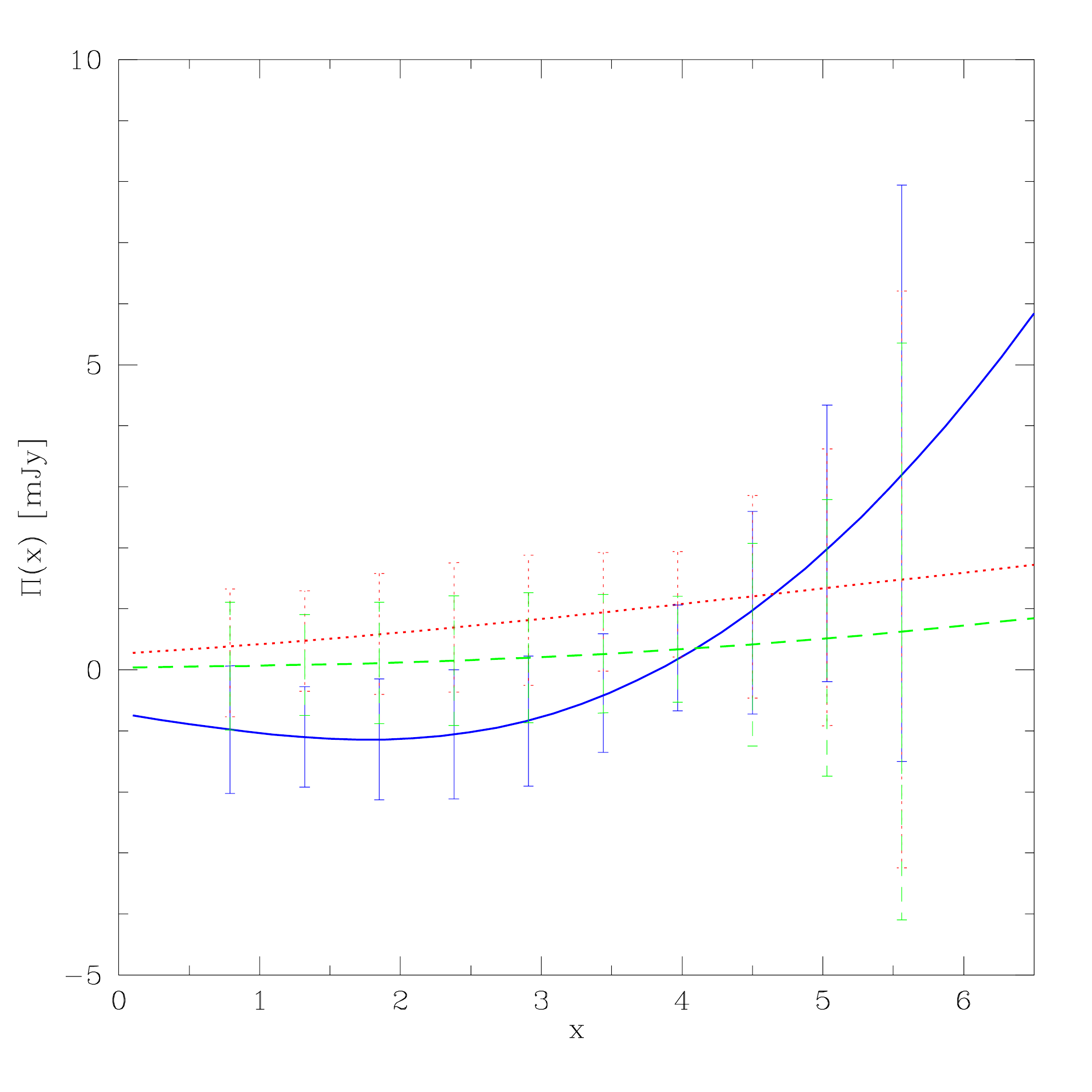}
 \includegraphics[width=7.0cm]{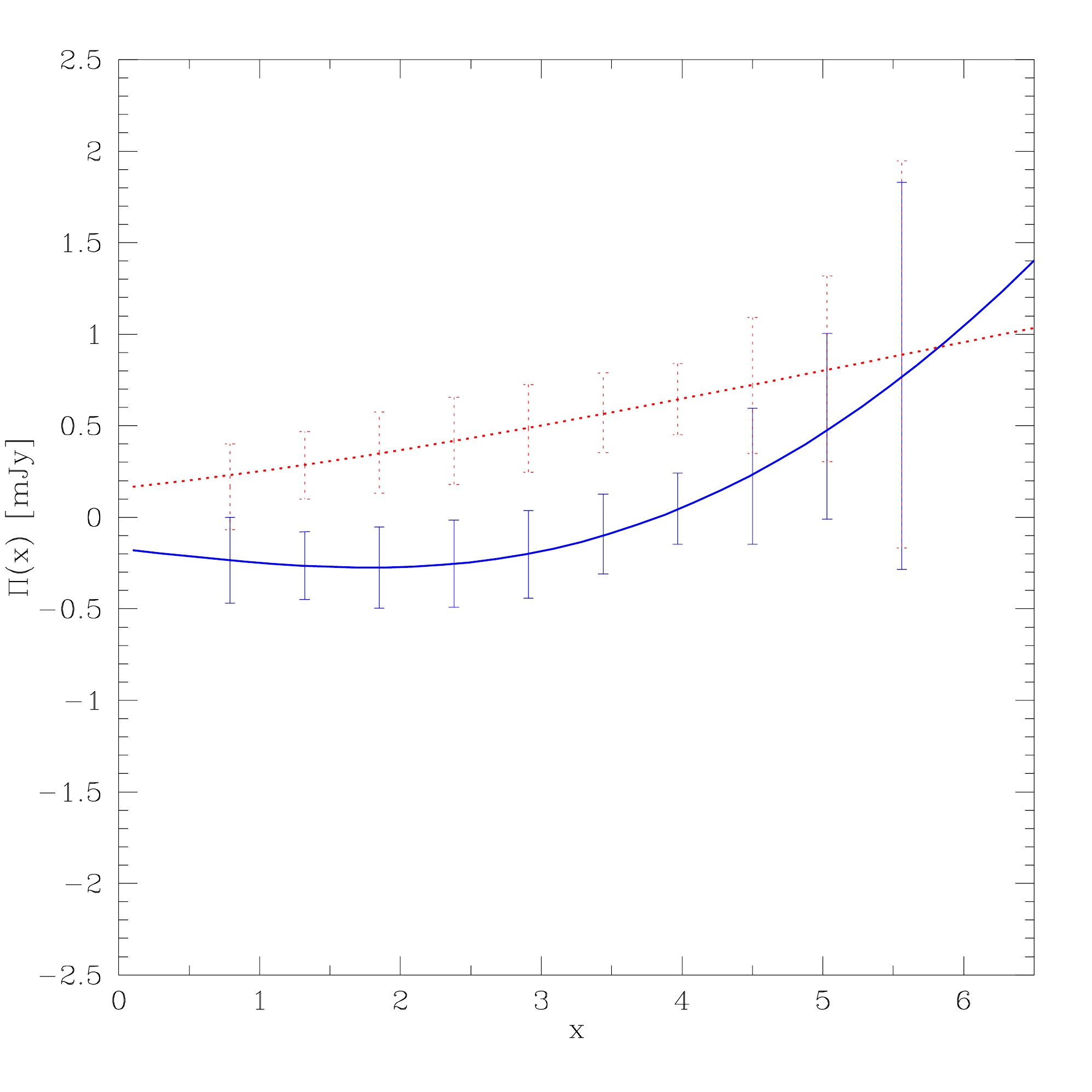}
\end{center}
 \caption{\footnotesize{Left. The spectral shape of the maximum SZE polarization
 signal for a cluster with $kT = 15$ keV and $\tau= 5\cdot 10^{-2}$ are compared
 to the COrE polarization sensitivity (shown as error bars) at the various
 frequencies: thermal SZE polarization due to double scattering $\Pi-{th} \propto
 (kT_e / m_e c^2) \tau^2$ (blue curve and COrE error bars), polarization due to
 tranverse motion of gas inside the cluster $\Pi_v \propto \beta_t
 \tau^2$ (green curve and COrE error bars), polarization due to
 transverse bulk motion of the cluster $\Pi_k \propto \beta_t^2
 \tau$ (red curve and COrE error bars). A cluster size of 3 arcmin
 radius has been considered. Right. The thermal SZE polarization
 spectrum (blue solid) from a stacking analysis of 20 clusters
 with $kT > 10$ keV and $\tau > 0.03$ is compared with the same
 analysis for the kinematic SZE polarization spectrum (red
 dotted). The statistical uncertainties refer to a stacking
 analysis produced with COrE.
 }}
 \label{fig.sze_pol_core}
\end{figure}

Polarization $\Pi$ of the SZE arises from various dynamical and
plasma effects related to plasma transverse velocities and
multiple scattering processes. These effects include galaxy
cluster transverse motion ($\Pi_k \propto \beta^2_t \tau$ in the
Rayleigh-Jeans (RJ) regime with $\beta_t = V_t/c$), transverse
motions of plasma within the cluster ($\Pi_v \propto \beta_t
\tau^2$ in the RJ regime), multiple scattering between electrons
and CMB photons within the cluster ($\Pi_{th} \propto \Theta
\tau^2$ in the RJ regime, with $\Theta = k T_e/m_ec^2$)
\cite{Sazonov1999,Challinor,ColafrancescoTullio2010}.
The general covariant relativistic derivation of the SZE
polarization for both thermal and nonthermal plasma
\cite{ColafrancescoTullio2010} generalizes the nonrelativistic
derivation \cite{Sazonov1999} taking into account the direct
dependence of the SZE polarization spectra on the properties of
the electron distribution in the atmospheres of galaxy clusters
and other cosmic structures. We stress that the SZE polarization
signals in clusters have spectra quite different from the
intensity SZE spectra (see Fig.~\ref{fig.sze_pol_core}).

Combining intensity and polarization observations of the SZE with
the frequency dependence and spatial distribution of the Stokes
parameters of the SZE can uncover unique details of the 3d
(projected and along the line of sight) velocity structure of the
intracluster plasma, of its 3d pressure structure and of the
influence of a structured magnetic field in the stratification of
the intracluster plasma, and therefore provides a full 3d
tomography of the structure of cluster atmospheres. Analogously,
the combination of the intensity and polarization observations of
the kinematic SZE (and its frequency dependence) can yield crucial
information on the 3d distribution of the cosmological velocity
field traced by galaxy clusters.
Specifically, the ratio $\delta I/\Pi_{th}$ yields direct
information on the plasma optical depth $\tau$, and the ratio
$\delta I/\Pi-{v}$ on the combination $\tau \cdot \beta_t$, thus
allowing to use intensity and polarization SZE measurements to
fully disentangle the pressure and velocity structure of the
cluster atmospheres.

Recent observational results and theoretical analyses have clearly
shown that an effective study of the SZE (both in intensity and
polarization) should be carried out over a wide frequency range
including especially high frequencies where the effects of the
electron distribution function are larger and allow millimeter
observations to be used to derive the fundamental parameters of
the cluster atmosphere and its plasma stratification and dynamics.

SZE observations in the millimeter range are becoming available
with PLANCK-HFI (in the $\sim 100-850$ GHz range) and
HERSCHEL-Spire (in the $\sim 600 - 1200$ GHz range). HERSCHEL,
however, lacks spectral coverage in the low-$\nu$ part of the SZE
spectrum that is important for an unbiased determination of the
main cluster parameters (i.e., the gas density, temperature, and
cluster peculiar velocity), while PLANCK-HFI suffers from a
limited sensitivity and spatial resolution needed to determine and
resolve the cluster parameters for the majority of medium- and
small-mass systems. At best very limited information on the SZE
polarization can be obtained with these instruments. Other
experiments from the ground (e.g. SPT and ACT) are producing large
area surveys of cluster with SZE measurements in the low-$\nu$
part of the spectrum (i.e. at $\nu \approx 150$ Ghz) but cannot
provide an all-sky SZE, do not have access to the high-$\nu$ part
of the SZE spectrum (which is crucial to obtain physical
information on the structure of cluster atmospheres) and strongly
suffer (as in any ground experiment) from atmospheric opacity and
frequency band inter-calibration effects.

The wide frequency coverage (from 45 to $\sim$ 800 GHz) and the
unprecedented sensitivity in both intensity and polarization of COrE
will provide a substantial improvement in the study of the SZE in
galaxy clusters and in other cosmic environments.

\begin{itemize}
\item COrE will detect more than $\sim 5 \cdot 10^4$
clusters at 135--165 GHz and $\sim 3 \cdot 10^3$ clusters at 375
GHz with thermal SZE above the sensitivity threshold, assuming
that PLANCK will detect $\approx 10^3$ clusters in the HFI 143 GHz
channel. These clusters will also be resolved better owing to the
improved spatial resolution of COrE (i.e., $\sim 6.4$ arcmin at
165 GHz and $\sim 2.8$ arcmin at 375 GHz) compared to the $\sim 7$
and $\sim 5$ arcmin resolution of Planck at similar frequencies.

\item CorE will provide full coverage of the SZE spectrum with a
sensitivity that allows a systematic study of the thermal,
multi-temperature, and nonthermal properties of many galaxy
clusters (see Fig.~\ref{fig.bullet_core}). Such studies will be
possible for several hundreds of galaxy clusters with $kT \simgt 2-3$
keV.

\item COrE will provide a spatially resolved spectral study of many
nearby clusters, especially at high $\nu$ where relativistic
effects of single- and multi-temperature, or supra-thermal, plasmas
are relevant. In about 100 clusters with angular size larger than
$\sim 5$ arcmin radius, COrE will also be able to perform
spatially resolved spectral studies of the SZE in the frequency
range $\sim 200 - 600$ Ghz.

\item COrE will provide relevant constraints on the polarization of
the thermal SZE in many nearby, hot clusters.
Fig.~\ref{fig.sze_pol_core} shows that even clusters with high
temperatures $kT \approx 15$ keV and high optical depth $\tau
\approx 5 \cdot 10^{-2}$ produce maximum SZE polarization signals
that are marginally detectable by COrE in single cluster
observations. However a stacking analysis of even small samples
($\sim 20$) of hot and dense galaxy clusters observed
with COrE at several frequencies would allow one to determine statistically
up to $\sim 200$ GHz the polarization signals of the thermal SZE
for clusters with $kT > 10$ keV and $\tau > 0.03$ (see
Fig.~\ref{fig.sze_pol_core}).
A sample of powerful merging clusters would set limits on the
polarization features due to transverse gas velocities $V_t \simgt
500$ km/s.
Finally a relatively large sample ($\sim 10^2$) of clusters could be
used to set limits on the three-dimensional velocity field
by observing the kinetic SZE polarization.
\item COrE will provide a more detailed description of the millimeter band
point-like source contamination to determine the
spectral and spatial properties of the cluster SZE (and of its
polarization) especially in the frequency range probed by COrE.

\item COrE will provide both detailed observations and
unprecedented constraints on
the SZE expected in other astrophysical plasmas such as the
lobes of radio galaxies and the plasma halos of galaxies
\cite{Colafrancesco2011}.
In fact, as a consequence of the pervading presence of
inverse Compton scattering of CMB
photons in extended radio galaxy lobes, a SZE from the lobes of
radio galaxies is inevitably expected (as first proposed and
discussed by \cite{Colafrancesco2008}). This nonthermal,
relativistic SZE has a particular spectral shape that depends on the
shape and on the energy extent of the spectrum of the electrons
residing in the RG lobes (see
\cite{Colafrancesco2008,Colafrancescoetal2011}).
The SZE in RG lobes has not yet been detected. Only loose upper
limits have so far been derived on the SZE from these sources (see
\cite{Birkinshaw1999}; see also \cite{Yamada2010} for a recent
attempt to detect this effect at radio wavelengths). The wide
spectral range of COrE and its enhanced sensitivity (also in
polarization) will endow this experiment with the unique ability to
detect the nonthermal SZE in a sample of more than 10 giant
radio galaxies.
\end{itemize}


%
%
%
%
%
%
%
%
%

\subsubsection{Non-polarized point sources}


In total intensity COrE will be so sensitive that it will reach the source 
confusion limit. We can use galaxy counts from large area Herschel surveys 
(H-ATLAS and HerMES, Clements et al. 2010; Oliver et al. 2010)
\cite{2010A&A...518L...8C,2010A&A...518L..21O}
at COrE wavelengths to determine the confusion limits for the COrE beamsize and 
thus the number of sources that will be detected over the sky.
For a beam FWHM of $1.3'$ at 795 GHz and of $1.9'$ at 555 GHz, we find $5\sigma$ detection limits of 
$\simeq 96\,$mJy and of $89\,$mJy respectively, if we take into account only the Poisson 
fluctuations. If we allow for the effect of clustering (Negrello et al. 2007) \cite{2007MNRAS.377.1557N},
the $5\sigma$ confusion limits increase to 119 mJy and to 127 mJy at 795 and 555 GHz, respectively.

\begin{SCfigure}
\includegraphics[angle=0,width=7cm,height=7cm]{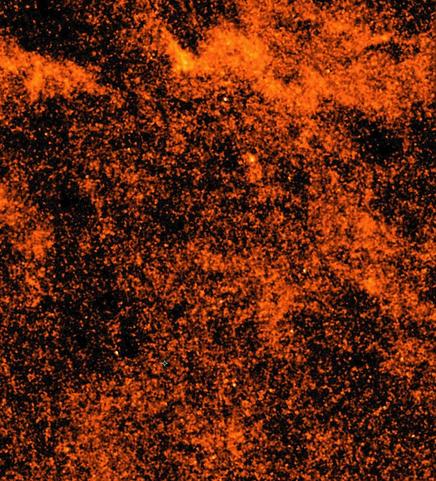}
\caption{
\baselineskip=0pt
\small
{\bf H-ATLAS 350$\mu$m SDP field ($14.4\,\hbox{deg}^2$) smoothed to COrE resolution.}
The brightest sources in this image include an unusual clump of Galactic dust emission, possibly a high Galactic latitude Bok globule, a WMAP blazar (Gonzalez-Nuevo et al. 2010) 
and five strongly lensed high redshift ($z\sim 2$--3) dusty galaxies (Negrello et al., 2010), 
all potentially detectable by \core. \core\  will select such extremely interesting sources over most of the whole sky.
}
\label{fig:smoothed}
\end{SCfigure}

Surveys of small areas of the sky are already available at 350$\mu$m (857 GHz) with somewhat better 
angular resolution and depth from \Herschel. In Fig.~\ref{fig:smoothed} we show the $4^\circ\times 4^\circ$ 
region surveyed by H-ATLAS (Eales et al. 2010; Rigby et al. 2010) 
\cite{2010PASP..122..499E,2010arXiv1010.5787R}
smoothed to the 1.3 arcmin FHWM beam of \core\  at 795 GHz.
Among the brightest objects in this H-ATLAS image are five strongly lensed background galaxies 
lying at $z\sim 2$--3, which can easily be identified by their red far-IR color and absence from radio surveys (Negrello et al. 2010).
\cite{2010Sci...330..800N}.
COrE  will be able to use its highest frequency channels to provide a similar selection of high redshift lensed 
far-IR sources, but this will be possible beyond the limited areas covered with Herschel. 

Using an updated version of the Negrello et al.~(2004) model \cite{2007MNRAS.377.1557N},
we estimate that COrE will be able to detect at 795 GHz approximately $1.3\times 10^4$  strongly lensed 
high-$z$ dusty galaxies above $|b| \sim 20^\circ$, where Galactic confusion should be of 
relatively minor importance. In addition, it will detect, at the same frequency, 
$\simeq 4\times 10^4$ normal and starburst galaxies and $\simeq 540$ unlensed high-$z$ 
proto-spheroidal galaxies. Interestingly, the strongly negative K-correction that makes 
(sub-)mm surveys extraordinarily well suited for detecting high-redshift star-forming galaxies, 
shifts their expected distribution to higher redshifts at longer wavelengths. We thus expect that 
millimeter surveys with COrE will lead to the detection of ultra-bright strongly lensed galaxies at very high redshifts.

Furthermore, COrE will detect, at $|b|>20^\circ$, some 1500 blazars in
the poorly explored sub-millimeter range, close to the synchrotron peak of
the most luminous objects of this class.

\subsubsection{Characterizing extragalactic sources}

\begin{figure}[h]
\begin{center}
\includegraphics[width=8.5cm]{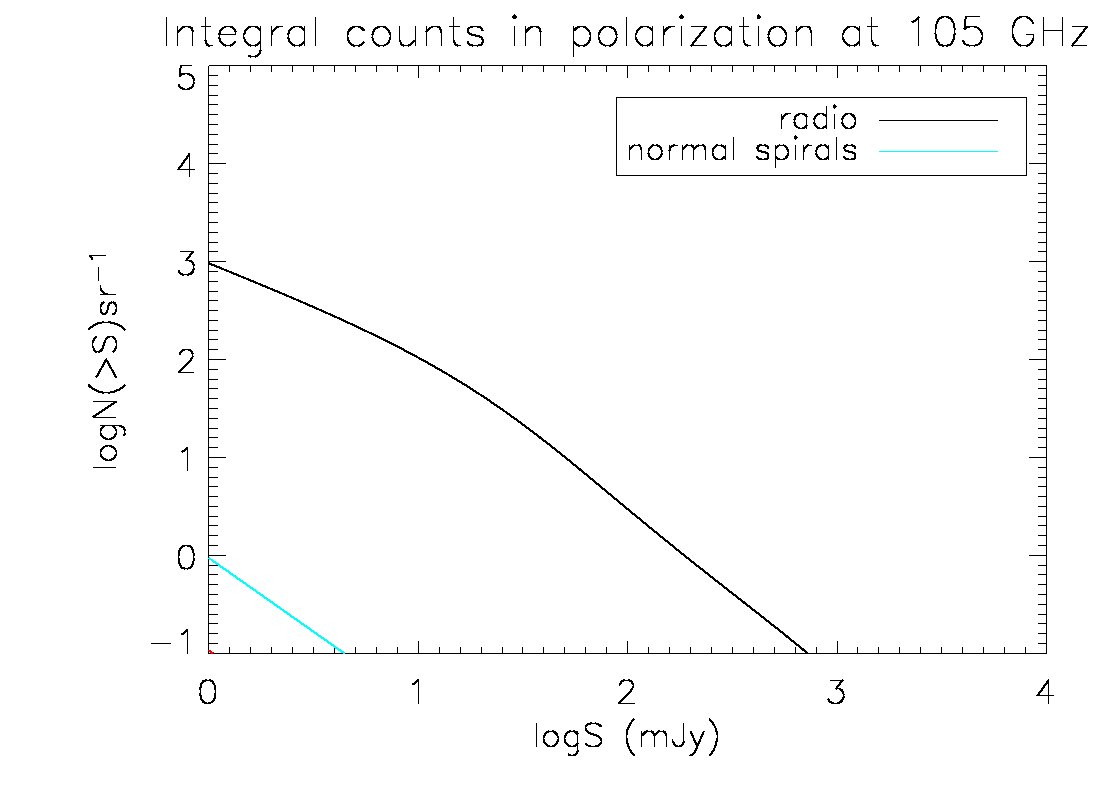}
\includegraphics[width=8.5cm]{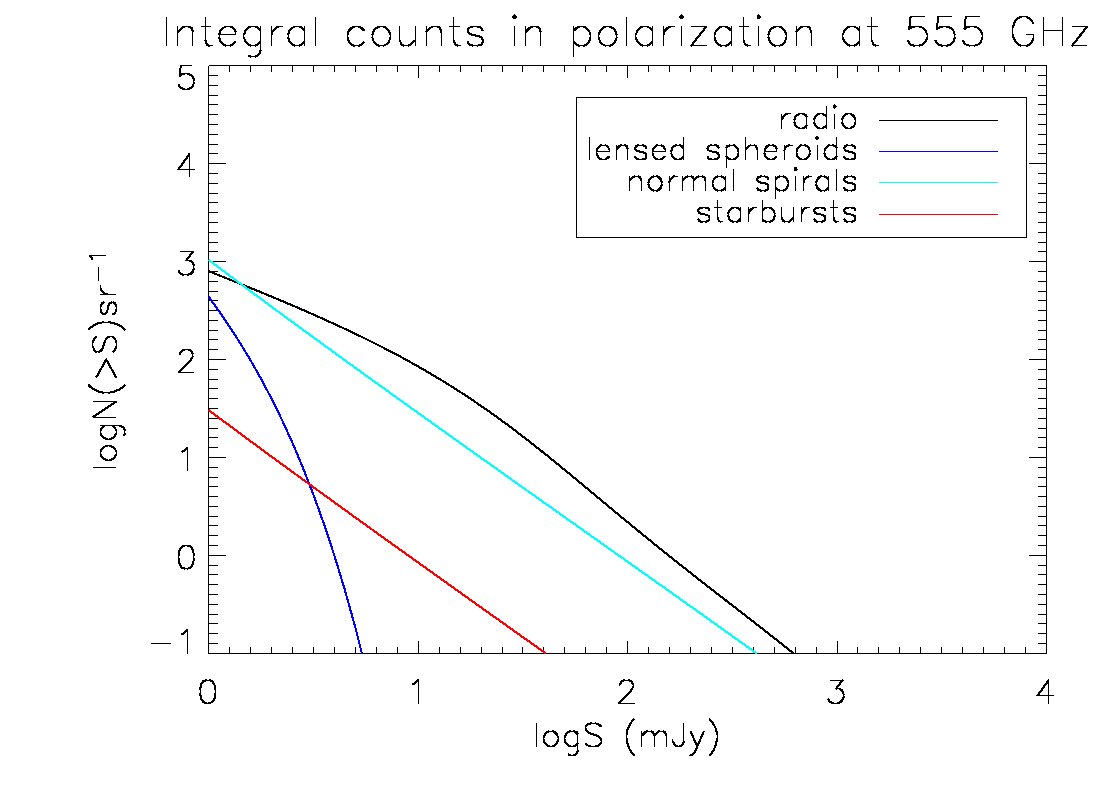}
\end{center}
\vskip -1cm
\caption{\baselineskip=0pt
\small
Expected integral counts as a function of the polarized flux of radio sources and of 
different populations of dusty star-forming galaxies at 105 and 555 GHz. }
\label{exgal1}
\end{figure}

Measurements of the polarization properties of radio sources at millimeter and sub-millimeter 
wavelengths provide important insights into the physical properties (and in particular into the 
structure of magnetic fields) of the innermost compact regions of relativistic jets in AGNs. 
These measurements will be also essential to assess the point sources contamination of CMB 
polarization maps. In fact, at high Galactic latitudes,
radio sources are expected to be the main contaminant on scales $\lsim 0.5^\circ$ up to 
frequencies of a few hundred GHz (De Zotti et al. 1999; Tucci et al. 2004, 2005). 
\cite{1999NewA....4..481D,2004MNRAS.349.1267T,2005MNRAS.360..935T}.

The only blind polarization surveys presently available are those produced by WMAP (Wright et al. 2009). 
\cite{2009ApJS..180..283W}.
L\'opez-Caniego et al. (2009) 
\cite{2009ApJ...705..868L}
reported the detection in polarization, in at least one WMAP channel;
however, of 22 objects five are  doubtful.
PLANCK can double the number of detections, but the sample will still be too limited for 
a meaningful study of polarization properties of the several radio source populations.

Ground based studies of high-frequency polarization properties of radio sources have been carried out 
by following-up samples drawn from surveys in total intensity. In spite of recent efforts 
(Jackson et al. 2010; Agudo et al. 2010) \cite{2010MNRAS.401.1388J,2010ApJS..189....1A},
polarimetric data at millimeter wavelengths are limited, and essentially non-existent at sub-millimeter 
wavelengths.

The most extensive polarization information on complete samples selected at high radio frequencies 
was obtained at $\simeq 20\,$GHz (Ricci et al. 2004; Massardi et al. 2008; Jackson et al. 2010) 
\cite{2004A&A...415..549R,2008MNRAS.384..775M,2010MNRAS.401.1388J}.
Radio sources are found to be significantly polarized (up to $\gsim 30\%$, Shi et al. 2010, 
\cite{shi_et_al}
although the median polarization degree is approximately 3\%).

Polarized emission in giant radio galaxies is an
exceptionally powerful tool to understand the origin of relativistic
particles, ultra-high-E cosmic rays, and magnetic fields (see, e.g., Kronberg 1994)
\cite{1994RPPh...57..325K}).
In regions of the RG lobes where shocks are strongest, the inflation of the lobe is
likely accelerating particles to very high energies, producing
also quite flatter synchrotron spectra (see e.g. \cite{2009MNRAS.395.1999C}),
and relatively high degrees of polarization (20-40\%) 
that could be observed with COrE in many nearby objects. In addition, Inverse Compton scattering of CMB
photons in giant radiogalaxy lobes produces an SZ effect (Colafrancesco 2008) \cite{2008MNRAS.385.2041C},
whose spectrum (both in  intensity and polarization at a level of 10-20\%) is best
observed at high frequencies where it shows a maximum and a flat
spectrum in the range 500-1000 GHz that could be observed with COrE
in a dozen of nearby extended objects. The combination of polarized synchrotron 
and SZ effect in giant radio lobes offers the unique opportunity to disentangle efficiently 
the distribution of the relativistic electrons from that of the magnetic field and
to provide, for the first time, a complete 3d tomography of the atmospheres of these cosmic structures.

Very little is known about the polarization of dusty galaxies that dominate the counts in total intensity above 
$\simeq 100$--200 GHz, but there are indications that it is probably $\le 1\%\,$ (Seiffert et al.  2007; Greaves \& Holland 2002)
\cite{2002AIPC..609..267G,2007MNRAS.374..409S},
as expected because of the complex structure of the magnetic field resulting in 
cancellations of polarization directions.

The spectacular sensitivity of COrE provide the first 
large sample of sources blindly selected 
in polarization. Because of the polarized flux is on average a small fraction of the total flux, 
source confusion is far 
less of a problem for polarization than for total intensity. 
Adopting an average polarization degree of 3\% for radio sources and of 1\% for dusty star-forming galaxies, 
and using the model by De Zotti et al. (2005) \cite{2005A&A...431..893D}
for the former population and the model by Negrello et al. (2007) \cite{2007MNRAS.377.1557N}
for the latter, we find that the confusion noise is always sub-dominant and becomes negligible above 100 GHz, 
at least in high Galactic latitude regions where polarized diffuse emissions are low
provided that the CMB can be efficiently removed.

As illustrated by Fig.~\ref{exgal1}, at 105 GHz we expect to detect the polarized flux of 
$\simeq 320$ radio sources per steradian above the $5\sigma$ detection limit 
$S_{\rm lim} = 5(\sigma_{\rm noise}^2+\sigma_{\rm conf}^2)^{1/2}\simeq 3\,$mJy. At higher frequencies 
the polarized dust emission from dusty star-forming galaxies should also be detectable. Under the 
above assumptions for the mean polarization degree, we have $\simeq 12$ radio sources and 
$\simeq 5$ normal spiral galaxies per sr with polarized flux larger than the $5\sigma$ 
detection limit of $\simeq 32.5\,$mJy at 555 GHz. At 795 GHz the $5\sigma$ detection limit 
is $\simeq 48.5\,$mJy and the expected numbers of detections per sr are of  $\simeq 6$ 
radio sources and of $\simeq 16$ dusty galaxies.

\section{Separation of the polarized microwave sky into components}
\label{sec:compsep}

The COrE mission is designed to achieve unprecedented sensitivity in fifteen frequency
channels ranging between 40 and 800 GHz. 
COrE is designed to serve as a comprehensive explorer of all astrophysical emissions in this wavelength range.
The sensitivity of the mission in each of these 
channels is designed to reach the many science goals discussed in Sect.~\ref{sec:science}.
Achieving these science goals requires the capability to avoid astrophysical confusion---that is, 
the capability to separate the emissions from different astrophysical sources. For a recent overview of 
diffuse component separation methods see \cite{2009LNP...665..159D}. 
A comparison of 
existing methods in preparation for the PLANCK mission can be found in \cite{2008A&A...491..597L}. 

In the tour of the microwave sky in the bands covered by
COrE, we described how the observed emission in a 
particular frequency channel is the linear superposition
of several components arising from distinct physical processes.
The various diffuse components identified---mainly emissions from 
the galactic ISM and the primordial CMB---have differing frequency
dependences. This feature allows their separation, 
which is a prerequisite for all the science described
in the preceding sections. 

Since the COBE detection, the development and application of
component separation methods have been a very active area of
CMB research, and many methods have been successfully applied
to temperature component separation, both on real data and 
on simulations for validation. In fact, until now foregrounds
have been much feared but have turned out to be less of a 
problem than anticipated, at least for the measurement of 
the temperature power spectrum. 
In practice, many methods have been developed and optimized.
At their core, these methods derive from a number of simple ideas, including:
\begin{itemize}
\addtolength{\itemsep}{-6pt}
\item Masking the brightest point sources and modelling the residuals;
\item Cutting out the galaxy, less so for small wavenumbers and more
aggressively for larger wavenumbers;
\item Extrapolating the frequency dependence of foreground components
using physically motivated emission laws, or empirical frequency scaling
obtained from correlation studies;
\item Using `blind' methods to let the data express its natural
decomposition into components through the hypothesis of
statistical independence.
\end{itemize}

For polarization studies, a much higher degree of cleaning
through component separation is required because
for the B modes the polarized foregrounds outshine the 
targeted signal in all frequency bands by about two orders 
of magnitude or more (see Fig.~\ref{Fig:Jacques1}). 

Below we review in more detail existing work, discussing 
its relevance to the component separation for COrE, and justify the 
rationale for the COrE frequency bands. We also demonstrate that the expected 
final performance for measuring the CMB B modes in the presence of 
polarized foregrounds matches the requirements of the mission.

\subsection{Sky emission modelling and component separation methods}

Over almost all the sky except very near the Galactic plane, astrophysical emissions in the millimeter 
wavelength range originate from sources which are optically thin. The total 
observed emission is consequently simply the sum of all emissions along the 
line of sight, without obscuration of one source by another. 
At each frequency $\nu$, the total observed emission in direction $p$ is then, 
for $N$ distinct components:
\begin{equation}
x(\nu,p) = \sum_i^N x_i(\nu,p)
\end{equation}
where $x_i(\nu,p)$ denotes the emission from the component $i$.

\subsubsection{Linear mixtures}

The coherence of the astrophysical emissions across frequencies is what
allows for the identification and separation of the contributions
of the individual components. In an idealized case, each component is modelled 
with a single template $s_i(p)$ scaling with frequency according to a single 
emission law $a_i(\nu)$. We then have
\begin{equation}
x(\nu,p) = \sum_i^N a_i(\nu) s_i(p).
\label{eq:linear-mixture}
\end{equation}
If the frequency dependence $a_i(\nu)$ is known (from physical principles) for all components, 
then separating the different contributions amounts simply to inverting the linear system to 
measure the intensity $s_i(p)$ of each emission in each pixel. Methods for doing so in the 
presence of instrumental noise and under various hypotheses about the number and statistical 
properties of the component emission maps have been discussed by numerous authors in the context 
of CMB temperature measurements \cite{1996MNRAS.281.1297T,1998MNRAS.300....1H,1999NewA....4..443B,2002MNRAS.330..807D}. 

In practice, the frequency scaling for the astrophysical components is known only to limited precision, either because 
of imperfect calibration of the measurements (e.g., for the CMB component), or because of intrinsic uncertainty as to 
the parameters of the physical emission processes (e.g.i, for emission from the galactic ISM). 
In this case component separation for CMB applications can be achieved using 
Independent Component Analysis (ICA) methods 
\cite{2000MNRAS.318..769B,2002AIPC..617..125S,2002MNRAS.334...53M,2003MNRAS.346.1089D,2007ITIP...16.2662B}. 
Estimates $\widehat{s}_i(p)$ of the component emission templates $s_i(p)$ are obtained, for instance, 
as linear mixtures $\widehat{s}_i(p) = \sum_j w_{ij} x(\nu_j,p)$, where the weights $w_{ij}$ maximize 
a given measure of independence between the reconstructed components $\widehat{s}_i(p)$. 
ICA methods can also be tuned to estimate directly the elements $A_{ij} = a_i(\nu_j)$ 
of the ``mixing matrix'' $A$ and then to invert the system. In addition to $A$, some methods 
also estimate second-order statistics of the components \cite{2003MNRAS.346.1089D,2003MNRAS.345.1101M}, 
which is of direct cosmological interest for the measurement of the CMB power spectrum and also allows using
these estimates to reconstruct maps of the individual components by Wiener filtering.

However astrophysical foregrounds are not always independent. Galactic emissions are all concentrated in 
the galactic plane, which gives rise to 
correlations among the components. 
Moreover, the components are physically interdependent.
Spinning dust emission arises from dust particles that also emit greybody radiation. 
For polarized emissions, 
the same magnetic field confines trajectories of electrons responsible for synchrotron emission and 
controls the alignment of dust grains emitting polarized radiation at submillimeter wavelengths. 
Hence the assumption of independence, which is the statistical foundation of the ICA methods, does 
not strictly hold. Methods exists which deal also with partially correlated foreground emissions 
\cite{2005EJASP2005.2400B,2006MNRAS.373..271B,2008ISTSP...2..735C}.

Most of the methods developed for CMB temperature can be applied with minor modifications to separate polarization maps
\cite{2006MNRAS.372..615S,2007MNRAS.376..739A,2007PhRvD..75h3508A,2009A&A...503..691B}.

\subsubsection{More complex sky emission}

In reality, the linear mixture model of eqn.~\ref{eq:linear-mixture} holds only approximately \cite{1998ApJ...502....1T}. 
As a next level of complexity, the emission law $a_i(\nu)$ of many of the components is not strictly 
the same at different pixels.
As observed by several authors 
\cite{1988A&AS...74....7R,2006MNRAS.370.1125D,2008A&A...479..641L,2009ApJS..180..265G}, 
synchrotron spectral indices between 408 MHz and few GHz vary from pixel to pixel. In addition, the steepening 
of the synchrotron emission law at high frequencies (above $\sim$30 GHz) due to the aging of cosmic rays is 
expected to be inhomogeneous on the sky. Similarly, the emission law of galactic dust depends on the local distribution of 
the dust grain properties (size, chemical composition, temperature, etc.).
Finally, the sky emission comprises also integrated emission from a large population of extragalactic objects, each of which has 
its own emission law, red-shifted (for the farthest ones) by the cosmological expansion.
In summary, astrophysical foreground emission is complex, and the linear mixture model is at best a crude approximation, 
which is totally inadequate for sensitive observations. For COrE, which targets a sensitivity 20 to 30 times better than 
the previous CMB space mission, the difference between real foregrounds and their approximation on the basis of a linear model 
is expected to be orders of magnitude larger than the sensitivity. Component separation must deal with this complexity. This 
impacts the design of the mission, which must provide the information needed to extract the very tiny B-mode signal from a sky 
emission dominated by poorly characterised foregrounds. 

Two main approaches have been explored for the separation of such complex foregrounds. The first is based on multi-frequency linear filtering, and specifically targets CMB science. The second is based on a pixel dependent fit 
of all physical or phenomenological parameters describing the emission of all components, using as input data the observation of 
the sky in all the available frequency bands. 
Both approaches and variants are discussed below.

\subsubsection{Linear filters}

One of the most challenging component separation problems is the extraction of 
CMB polarization B modes down to a level $r \simeq 10^{-3}$ as targeted by COrE. Even at high galactic latitude, 
the power spectrum of CMB B modes is expected to be several orders of magnitude below galactic contamination 
\cite{2005MNRAS.360..935T,2009AIPC.1141..222D}.

Even if ICA methods are not well suited to separate components for which the linear mixture model of 
eqn.~\ref{eq:linear-mixture} does not hold, component separation methods exist which exploit the independence 
of the CMB from other emissions, and address specifically the problem of separating this single component from 
a complex mixture of unidentified, and possibly correlated, foregrounds. 

The SMICA method \cite{2003MNRAS.346.1089D,2008ISTSP...2..735C}, in particular, is a very powerful method to directly estimate either the tensor-to-scalar ratio $r$, or the power spectrum of CMB pseudoscalar modes $C_\ell^{BB}$, as well as the errors of the estimated CMB parameters in the presence of foregrounds. So far this approach, for polarzation, has been tested on simulations made with synchrotron and dust 
polarized foregrounds as simulated with version 1.6.4 of the PSM. The simulated synchrotron is modeled with pixel-varying 
amplitude and spectral index, and the simulated dust with two greybodies with varying amplitudes and temperatures. With
the CMB included and point sources neglected, this makes a total of 7 different maps which model the total polarized emission.

Using these simulations,
the work described in \cite{2009A&A...503..691B} shows that for various configurations of a space mission 
with high sensitivity, high angular resolution, and at least 8 frequency channels, the tensor to scalar ratio can be measured 
down to $r \simeq 10^{-3}$ with a signal-to-noise ratio of order 5. Moreover, these results are shown to be robust against minor changes 
of the simulated foregrounds (adding synchrotron curvature, some point source contamination, and increasing the level of 
polarized dust emission). 

It should be noted that the simulated polarized sky uses 7 maps, and the successful instruments have at least 8 frequency channels.
As discussed in \cite{1999NewA....4..443B}, the impact of a varying emission law can be linearized to first order, 
so that the total sky emission (CMB and foregrounds) is, if modeled as a linear mixture of fixed templates, about 
7-dimensional. In practice, however, the number of templates used to marginalize over foreground emission in 
\cite{2009A&A...503..691B} is only 4, which means that some of the complexity induced by varying spectral parameters have a 
negligible impact on the ability to measure the CMB. Loosely speaking, it also means that about 4 remaining channels 
(eight minus four) contribute to the sensitivity to the CMB.

As an alternate approach to SMICA, several authors have used the simple ``internal linear combination'' ILC method to investigate component separation 
for measuring CMB B-modes. Instead of fitting directly for the B-mode power spectrum, the ILC first builds a CMB map by 
forming a linear combination of the input observations that minimizes the total variance of the reconstructed map, up 
to the constraint that the response to the CMB should be unity. The power spectrum of the reconstructed CMB map 
is then computed, taking into account biases and additional variance induced by foreground and noise 
residuals in the reconstructed map.

Although less powerful than SMICA, ILC has the virtue of simplicity. It is however prone to subtle errors, for example biases 
resulting from correlation between CMB and foregrounds \cite{2007ApJS..170..288H,2009A&A...493..835D}. Calibration errors can become amplified
in a high signal-to-noise regime \cite{2010MNRAS.401.1602D}. These biases, however, will be small (as compared to errors due to noise, 
cosmic variance and residual foregrounds) for measurement of B modes with COrE.

\subsubsection{Parametric fitting}

The parametric method directly fits a parametric model of the emission of all components in each pixel 
\cite{2006ApJ...641..665E,2009ApJ...705.1607D,2009MNRAS.392..216S}.
In current implementations, the method offers only moderate capability to disentangle low frequency foregrounds due to a lack of 
sensitive low frequency observations. Accuracy in the measurement of the parameters is limited by the instrument noise. 
The capability to constrain many parameters is limited by the insufficient number of measurements at different frequencies.
For suborbital experiments less sensitive than COrE, observing clean regions of the sky, a frequency coverage
limited as compared to that of COrE may be sufficient to reach $r \simeq 4\times 10^{-2}$ after polarised component separation with a parameter-fitting technique \cite{2010MNRAS.408.2319S}. 
COrE is, however, targetting a sensitivity fourty times better than that on the tensor to scalar ratio, which requires significantly better foreground cleaning in these regions of low foreground contamination, but also a significantly larger sky area.

For accurate component separation with COrE, the emission law of the synchrotron should then be represented with 2 or 3 parameters 
(amplitude, spectral index, and running due to the aging of the relativistic electrons), CMB with 1 parameter (temperature), 
and thermal dust is usually fit with 3 to 6 parameters (amplitude, spectral index, and temperature for one or two greybodies). 
Free-free emission requires one or two parameters per pixel (amplitude and average electron temperature), and thermal SZ effect 
also one or two (amplitude and electron temperature). Spinning dust emission is predicted to have a smooth spectrum, limited to 
low frequencies below the frequency coverage of COrE. Molecular line emissions, in full generality, require one parameter per 
line, but impact only frequency bands containing strong molecular lines.
The main polarized emissions being the CMB, synchrotron, and dust, an accurate fit for polarization requires measuring sky 
emission in at least 6-10 frequency bands. For temperature, about 5 additional bands are necessary.

\subsection{The choice of frequency bands}

Proper selection of frequency bands for the space mission is essential to successful component separation. 
For each pixel of the sky, it is important to have enough independent
measurements to constrain all parameters of all the emission model. 
Given the level of rejection targeted, the bands must provide the capability of checking consistency 
internally. For polarized component separation, the frequency channnels more numerous 
than the number of linear components or model parameters.

Since 6-10 parameters are necessary to describe the polarized components (synchrotron, 
CMB and dust), at least as many frequency channels are required. Redundancy is 
achieved by providing several additional channels beyond this minimum.

For the intensity maps, the spectral coverage must also allow for fitting 
unpolarized emissions: free-free, spinning dust, 
thermal Sunyaev Zel'dovich effect, and molecular lines, with one or more parameter for each.
Hence we require a conservative four to five additional frequency channels for adequate
frequency coverage in temperature, for a
total of ten to fifteen parameters to fit unpolarized emission. 
In order to measure as many parameters as 
required to fit and interpret the data, we design COrE  to observe the sky in 
15 different frequency bands. This is also useful 
to avoid using the CMB sensitive channels as polarized foreground tracers.

Broad spectral coverage is important not only for addressing the COrE science objectives,
but also
to test and guarantee {\it a posteriori}
the effectiveness of component separation.
This broad spectral coverage is a strength of the COrE design 
as compared to previously proposed or currently operating CMB polarization experiments.

\subsection{Measuring $r$ in the presence of foreground emission}

To study the COrE capabilities more specifically, several independent foreground removal simulations were carried out.
We now describe the preparation of simulated data sets for this study and the 
methods developed to extract the CMB B mode signal.
We first describe simulations performed for this analysis, and then results obtained with two methods: The first approach is an
ILC in needlet space (NILC), very similar to what has been done in \cite{2009A&A...493..835D} for WMAP temperature maps.
The second is based on a pixel-based linear component separation method 
(pix-LCS), for which errors due to noise and foreground residuals have been forecast for COrE with the method 
described in \cite{2011arXiv1101.4876B}.

\subsubsection{Simulations with the PLANCK Sky Model}
\label{sec:simus-psm}

The PLANCK Sky Model (PSM) is a data and software package developed in preparation for the
PLANCK mission. It serves as a tool that predicts or simulates temperature and polarization emissions of the 
different astrophysical components relevant to sky observation in the 3~GHz--3~THz frequency range. 
The PSM implements models of the following emissions: primordial CMB, the galactic ISM (synchrotron, free-free, 
thermal dust, spinning dust, molecular CO lines), the thermal and kinetic Sunyaev-Zel'dovich effects, 
known radio and infrared sources, unresolved 
infrared galaxies (the CIB). Of these, CMB, synchrotron, thermal dust and emission 
by known point sources are polarized in the PSM version used for the present study (v1.7).

CMB emission is modeled as a multivariate ($T$, $E$, $B$) Gaussian random field, the statistics of which are fully 
described by power spectra $C_\ell^{TT}$, $C_\ell^{TE}$, $C_\ell^{EE}$, and $C_\ell^{BB}$, computed using 
the CAMB software for WMAP 7-year best-fit cosmological parameters. Tensor modes at the level of 
$r=10^{-3}$ or $r=5 \times 10^{-3}$ have been added (see below). B modes from lensing shear are modelled as Gaussian for this application.

The polarized galactic emission used includes polarized synchrotron and polarized dust. A 23~GHz synchrotron template 
is obtained from the analysis of WMAP data described in \cite{2008A&A...490.1093M}. The polarisation fraction ranges from 0 to 70 \%, with an average over the sky of about 18\%. Two 100~$\mu$m polarized dust 
templates are scaled assuming greybody emission laws with pixel-dependent amplitude and temperature
to predict polarized dust emission at all frequencies. A dust polarization fraction ranging from 0 to 12\% 
(slightly less than 3\% on average) is assumed, in agreement with measurements from  
Archeops \cite{2004A&A...424..571B}. The direction of dust polarization matches that of the synchrotron polarization, 
because it is the same galactic magnetic field that sets both orientations (at least to first order, 
neglecting differences in the distribution of the emission along the line of sight). 

\subsubsection{Masking}

Component separation for CMB reconstruction involves a delicate balance between masking contaminated regions
(essentially the galactic plane) and keeping as much sky as possible for minimising the cosmic variance. This balance depends on the level of $r$.
The masks used in the final analysis of COrE data still have to be defined. Here, we use simple galactic cuts. Two of the masks used have been apodized, 
blanking out galactic latitudes below some minimum galactic latitude $|b| \leq b_{\rm inf}$, 
leaving untouched regions at galactic latitudes $|b| \geq b_{\rm sup}$ and weighting the maps with a smooth transition in between. 
A conservative mask (mask 1) uses $b_{\rm inf} = 20^\circ$ and $b_{\rm sup} = 40^\circ$, for about 50\% of sky masked. A less conservative mask (mask 2) uses $b_{\rm inf} = 15^\circ$ and $b_{\rm sup} = 30^\circ$ (equivalent to 30\% of sky masked).
Finally, a sharp 20 degree galactic cut ($b_{\rm inf} = b_{\rm sup} = 20^\circ$) is also used as an alternative (mask 3, which blanks out 35\% of sky). 

\subsubsection{B-mode needlet ILC on data subsets}

The first approach to component separation with COrE is based on the ILC method. For each frequency band, 
it is assumed that two independent B mode maps are made, using each half of the data (either half the 
observing time, or half the detectors). Each map has twice the noise power of the total mission in the corresponding band. 

We determine ILC weights in needlet space, in a similar way as in \cite{2009A&A...493..835D}, from the total band 
averaged data obtained by coadding the two maps at each frequency. This is done after masking the maps with mask 1 for 
the case where $r=0.001$, and mask 2 for the case where $r=0.005$ (a choice which could yet be optimised, as discussed later-on).

\begin{figure}[bht]
\begin{center}
\includegraphics[width=12cm]{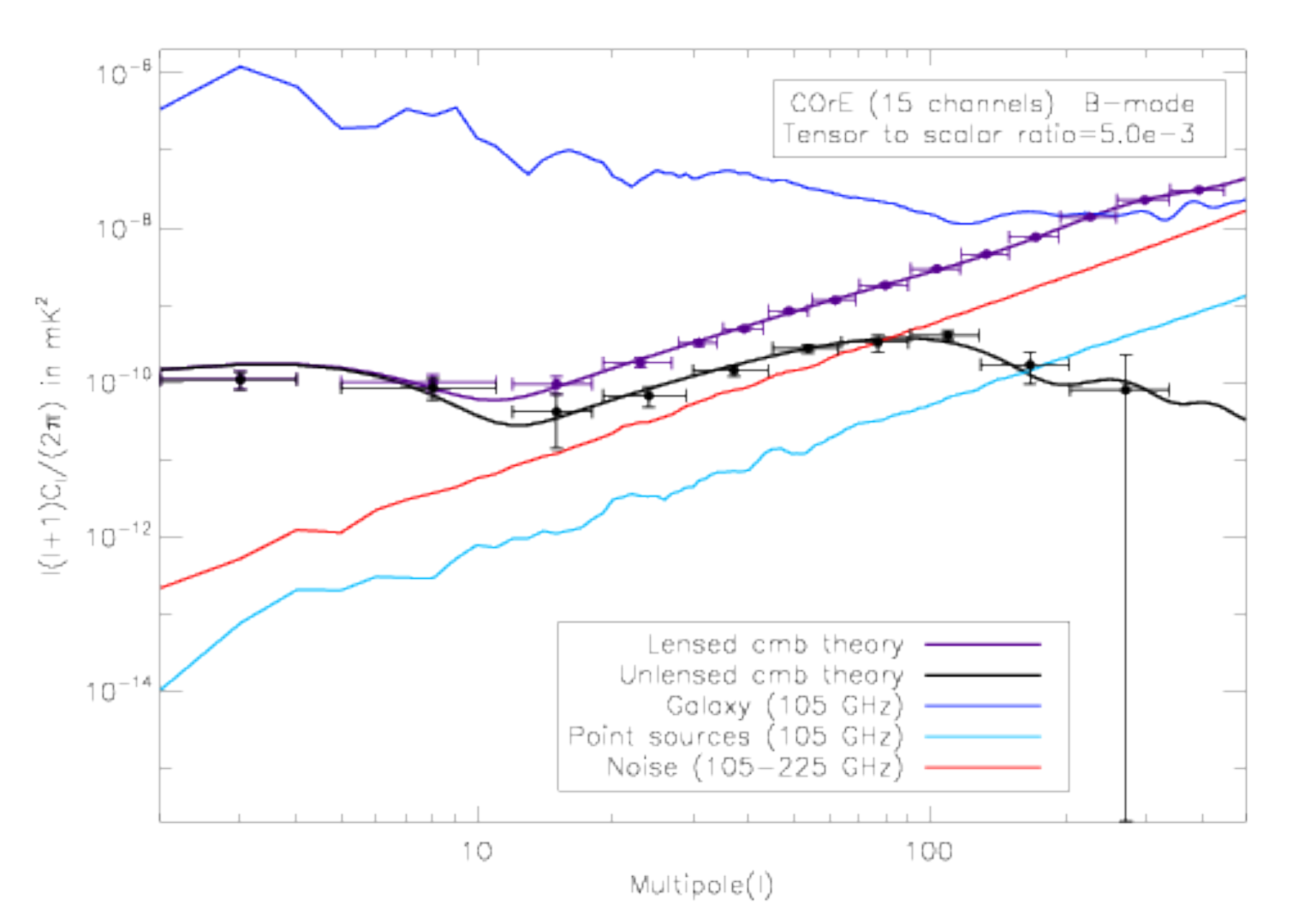}
\end{center}
\caption{\baselineskip=0pt
\small{
Component separation exercise for
B mode detection assuming $(T/S)=5\times 10^{-3}$. The solid black curve 
shows the predicted blackbody B mode power spectrum,
which is a combination of the tensor B modes (black
curve) and a gravitational lensing background (not shown)
making primordial E modes appear partially as B modes. The 
upper solid blue curve shows the contribution of diffuse galactic emission
in one of the `cleaner' channels (here 105~GHz) after masking. The red
curve indicates the instrument noise that would be 
obtained by combining five CMB channels, and the light blue curve indicates contamination 
by point sources after the brightest ones ($S>100\,$mJy at 20~GHz and $S>500\,$mJy at 
100 microns) have been cut out.
The purple data points indicate the recovered raw primordial spectrum
measurements, as compared to the theoretical spectrum (purple line).
The black points result after the gravitational lensing
contribution has been removed, leaving only the recovered tensor
contribution.
Here mask 2 (an apodised, galactic cut with $f_{sky} \simeq 0.70$) has been used. 
}}
\label{Fig:Jacques1}
\end{figure}

\begin{figure}[bht]
\begin{center}
\includegraphics[width=12cm]{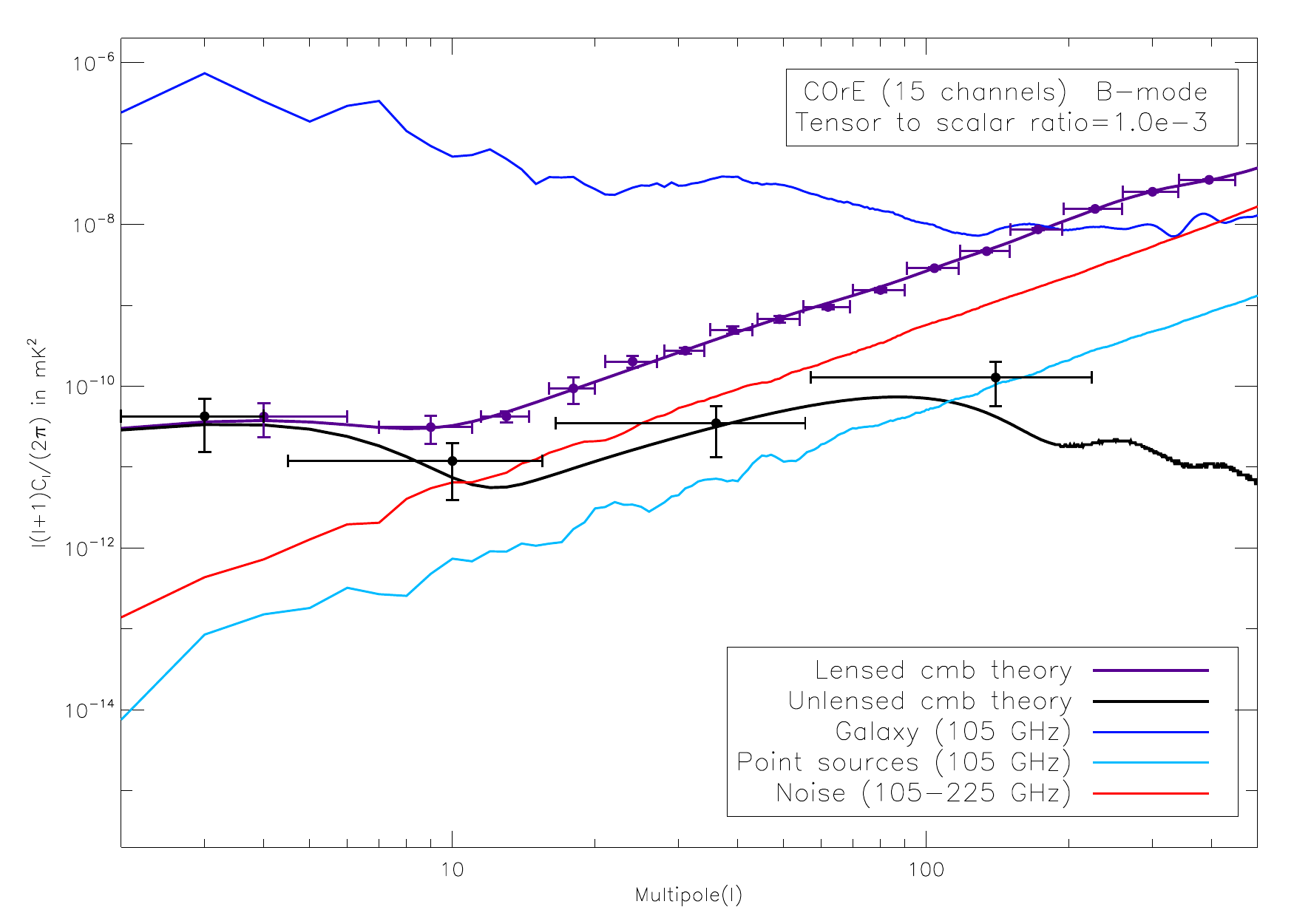}
\end{center}
\caption{\baselineskip=0pt
\small{
Same as figure \ref{Fig:Jacques1} but now with $(T/S)=10^{-3}$ and using mask 1 instead, a larger, very conservative, apodised, galactic cut with $f_{sky} \simeq 0.50$. 
}}
\label{Fig:Jacques2}
\end{figure}

The ILC weights obtained in this way in the needlet domain are then applied independently to each of the 
two subsets (each of them masked), to obtain two independent foreground-reduced maps. The cross-power 
spectrum of these two maps is used to estimate the CMB B mode power spectrum. The outcome of the 
processing, for the cases $r=0.005$ and $r=0.001$ respectively,  are shown in Fig.~\ref{Fig:Jacques1} and Fig.~\ref{Fig:Jacques2}.

The total B mode spectrum (purple line, and purple data points) contains a contribution from 
lensing of the E mode. Assuming that the power spectrum of this conversion of E into B is measured 
on small scales and extrapolated based on a cosmological model, 
the measured B-mode spectrum is debiased from the lensing part to yield a measurement of the 
primordial B modes (black line and measurements).
As can be seen from the figures, this very simple blind component separation method
is adequate to measure the primordial B mode spectrum for $r=10^{-3}$ over 50\% of the sky. 
Further optimisation is likely to improve these results.

\subsubsection{Forecast of errors due to residual foregrounds for linear component separation}

As an independent confirmation of the previous results, errors due to noise and foreground residuals have been forecast for 
COrE with the method described in \cite{2011arXiv1101.4876B}. The component separation approach considered in 
the forecast relies on two steps: a ``model learning'' phase, in which a linear model is matched to the data 
exploiting a blind method, and a ``source reconstruction'' phase, in which the linear system is inverted.

We start from the data model described in Sect.~\ref{sec:simus-psm}, with a tensor to scalar ratio of either $r=1 \times 
10^{-3}$ or $r=5 \times 
10^{-3}$. We assume conservative errors in the estimation of the spatially-varying spectral emission law of 
diffuse synchrotron and dust emissions. These errors are calibrated on the predicted performances of the harmonic 
correlated component analysis (CCA) component 
separation method as applied to simulated PLANCK polarization data in \cite{2010MNRAS.406.1644R}. This level of residual 
error is to be considered as conservative: with a much higher sensitivity and larger frequency number of 
frequency bands, COrE is expected to allow a significantly more accurate foreground characterization.

Given the true spectral emission laws and their estimation errors, we obtain a set of estimated spectral 
dependencies by random generation. Each set is considered as a possible output for the model 
learning phase of our component separation pipeline and exploited for the source reconstruction phase, which is 
done with ``generalized least squares'' solution in pixel space. Noise and foreground residuals errors at the 
power spectrum level are obtained for each set and finally averaged over the different sets to get our final 
error estimates. 

\begin{figure}
\begin{center}
 \includegraphics[angle=90,width=11 cm]{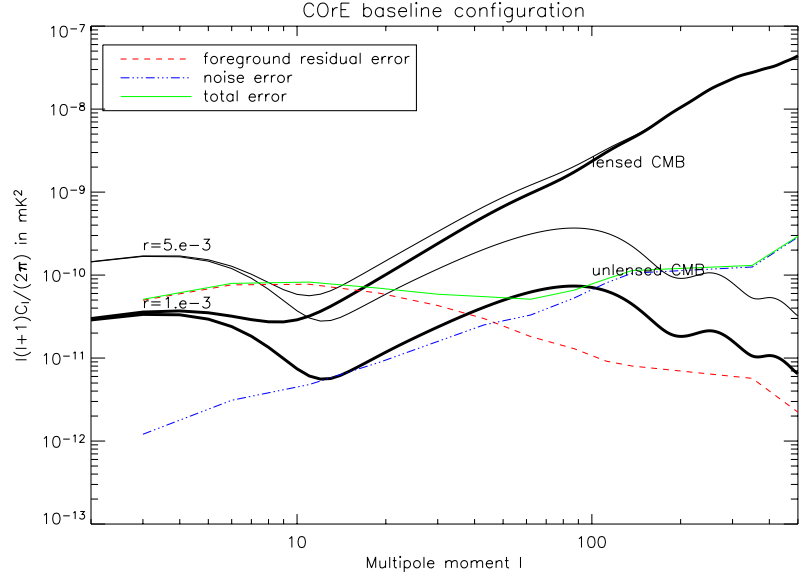}
\end{center}
\caption{
\small{
Forecasted errors on the CMB map due to foreground residuals (red dashed line), noise (blue dot-dashed line) and total 
(green solid line) for a linear component separation technique in pixel space (pix-LCS) assuming conservative errors on 
the recovery of the spectral model. Errors are compared to input CMB model with $r=(T/S)= 5 \times 10^{-3}$. 
Here mask 2 (a galactic cut with a smooth transition, corresponding to $f_{sky} \simeq 0.70$) has been used. The foreground rejection is good, but foreground residuals still dominate the error on large scales.}}
\label{fig:foregrounds-forecast}
\end{figure}

\begin{figure}
\begin{center}
 \includegraphics[angle=90,width=11 cm]{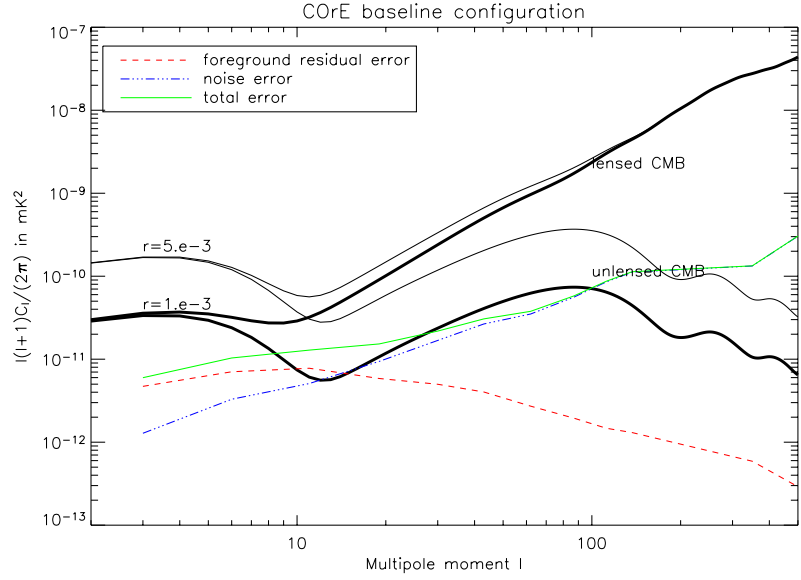}
\end{center}
\caption{
\small{
Same as figure \ref{fig:foregrounds-forecast}, but errors are compared to CMB models with both $r=(T/S)= 10^{-3}$ and $5 \times 10^{-3}$. 
Here a more conservative galactic cut with $f_{sky} \simeq 0.65$ has been used (mask 3). The difference between the level of galactic residuals here and in figure \ref{fig:foregrounds-forecast} shows the importance of masking regions close to the galactic plane. The more conservative mask allows for very little residual foreground emission, adequate for measuring $r=0.001$.}}
\label{fig:foregrounds-forecast2}
\end{figure}

This estimation of residual error due to inaccurate estimation of spectral emission laws is performed for mask 2 and mask 3. 
Figures \ref{fig:foregrounds-forecast} and  \ref{fig:foregrounds-forecast2}  display the resulting error on CMB B modes, compared to the theoretical CMB B-mode power spectrum. As shown in the figure, we are able to keep both noise (blue dot-dashed 
line) and foreground contamination (red-dashed line) under control, thus making a significant detection of the 
B-modes even under very conservative hypotheses. 

The amount of residual foregrounds is seen to depend significantly on the mask used.
From the relative level of residual foregrounds and noise 
at low $\ell$, we see that mask 2 is not as good as mask 3 for the detection of B modes in our simulations.
The level of foreground contamination is reduced by increasing the galactic mask. Optimization will be necessary, in particular
to measure low values of $r$.

\subsection{Foregrounds and component separation: summary}

Astrophysical confusion has become the major source of error for sensitive CMB observations. The development of component separation methods has become a very active topic in the context of the analysis of WMAP and Planck data. The design of COrE builds on the expertise developed for the previous space missions.

The complete spectral coverage provided by 15 different frequency bands is an important property of COrE. In any region where all components contribute (e.g. a galaxy cluster at moderate galactic latitude, with contamination by intra-cluster sources), or in regions with complex galactic foregrounds,  all of these channels are needed for accurate characterization of all the emissions, which is an important part of the scientific programme of our proposed space mission. 

This spectral coverage also important for securing adequate component separation for CMB science. 
Significant effort indeed is put into making COrE a very sensitive surveyor of the sky emission.
For benefitting fully from the mission sensitivity, it is necessary that component separation be performed at an accuracy such that residual confusion is about as low as the noise level. This sets very demanding requirements for COrE: by reason of the very high S/N, component separation must be done to exquisite precision.

We have performed component separation and residual foreground contamination estimates specifically for measuring B modes with COrE, using for this PSM simulations which comprise polarised synchrotron and dust (both with emission laws varying over the sky), polarised point sources, and CMB. We show that a low level of primordial B-modes ($r=0.001$) is detectable with COrE, although this requires masking a substantial region of the galaxy. For higher values of $r$, we would gradually increase the sky coverage, for optimisation of the trade-off between cosmic variance and residual foreground contamination.

\label{sec:fgs}

\clearpage
\section{Science requirements}
\label{sec:requirements}

\subsection{Sensitivity}
\label{sect:Sens}

The primary science goal of COrE is to investigate the physics at play in the early universe and to 
test extensively the standard cosmological model through high precision mapping of the CMB 
temperature and polarization anisotropies. Just as with previous CMB missions, this requires the 
sensitive measurement of sky emission in the millimeter wave domain where CMB emission is the 
strongest. It also requires additional channels to constrain and understand all astrophysical 
emissions contributing to the total sky emission. The level of polarization of CMB B modes from 
gravitational lensing, which is both an important scientific objective and an ultimate limitation 
for measuring primordial tensor B modes, is the main driver for the sensitivity of COrE. It sets 
the requirement that the COrE mission should measure sky polarization with a sensitivity better 
than 5$\mu$K$\cdot $arcmin in a set of frequency channels at the maximum level of CMB emission 
relative to foregrounds. The baseline COrE design includes among its fifteen frequency bands, six 
frequency channels between 75 and 225 GHz each having a sensitivity better than this requirement. 
This allows not only for the target sensitivity, but also provides the redundancy needed to check 
that the measured fluctuations indeed have the expected color of CMB fluctuations after these 
channels have been cleaned of foreground contamination.

The deployment of the detectors among the 15 frequency channels has not yet been optimized and 
should be studied further before a final configuration is fixed. For such optimization, which 
depends somewhat on the science goals and their relative priorities, there are essentially three 
constraints to be considered: focal plane area, cooling power, and telemetry rates. What 
performance can be expected for detecting B modes in the presence of galactic and extra-galactic 
foregrounds has been studied in Sec.~\ref{sec:compsep}. There we show what values of $(T/S)$ can be 
achieved for the present configuration. For the galactic science, the most important channel is the 
795 GHz because it has the best angular resolution.

%

For the COrE Galactic science program, the goal 
is to map the Galactic polarization down to approximately one arcminute resolution to 
characterize the interplay between interstellar turbulence and the magnetic field in structuring the cold 
ISM. We want to probe the formation of $H_2$ clouds
within the diffuse interstellar medium as well as pre-stellar cores in 
star forming molecular clouds. 
COrE will complement ground-based observations in 
a unique and essential way. Single-dish telescopes (IRAM, CCAT) and interferometers (NOEMA, ALMA) will 
map the magnetic field within compact sources on arcminute scales but will not be sensitive to extended 
emission.

Table~\ref{table:s_over_n_galactic} shows the polarization sensitivity of COrE for the emission from the 
extended envelopes of molecular clouds and bright clouds at high Galactic latitude (corresponding
to a visible extinction of 1 magnitude). 
The sensitivity to the polarized fraction $\sim 1\%$ compares well with an expected 
polarization fraction that it is expected to vary across clouds within the range 3-15\%. Data averaging 
over coherent structures in the intensity maps (e.g., interstellar filaments) will allow extending the 
CORE mapping of the Galactic polarization to fainter clouds at high Galactic latitude.  The on-going 
analysis of PLANCK polarization data will allow us to sharpen the sensitivity 
requirements and decide on an optimal distribution of detectors among the frequency bands. 
We see that $S/N\gtorder 7$ is achieved in a pixel of the highest-frequency channel, which has 
the highest angular resolution $(1.3')$. 
In the last entry of the table we indicated the effect of increasing
the number of detectors in the highest frequency channel by a factor of five relative to the previous
baseline. It should be noted that for areas of very diffuse Galactic emission (H21cm emission
of approximately $150~K~km~s^{-1}$), a signal-to-noise of approximately unity 
is achieved in a fwhm square pixel of the highest frequency channel (assuming a polarization fraction of 3\% ).

\begin{table}[htd]
\begin{center}
{\small
\begin{tabular}{
|c|c|c|c|c|c|c|c|c|c|c|
}
\hline
\hline
 & & & &
\multicolumn{3}{|c|}{$(\Delta P)$/arcmin}& 
\multicolumn{2}{|c|}{Pixel sensitivity}&
$(\Delta P)_{A(V)=1}^{forecast}$  &  \\
\cline{5-10}
$\nu $ & $\Delta \nu $ & $n_{det}$ & $\theta _{fwhm}^{arcmin}$ &  
$(\muK)_{thermo}$& $(\muK)_{RJ}$& $MJy/st$&
$(\muK)_{RJ}$& $MJy/st$&
$MJy/st$ & $(S/N)_{pol}^{pix}$\\
\hline
255  &  15 &  575  &   4.10 &   $1.05\times 10^{1}$  & 2.43  &  $4.85\times 10^{-3}$   &  0.59  &  $1.18\times 10^{-3}$   &  $6.30\times 10^{-3}$ &  5.33\\
285  &  15 &  375  &   3.70 &   $1.74\times 10^{1}$  & 2.94  &  $7.33\times 10^{-3}$   &  0.79  &  $1.98\times 10^{-3}$   &  $8.20\times 10^{-3}$ &  4.13\\
315  &  15 &  100  &   3.30 &   $4.66\times 10^{1}$  & 5.62  &  $1.71\times 10^{-2}$   &  1.70  &  $5.19\times 10^{-3}$   &  $1.13\times 10^{-2}$ &  2.20\\
375  &  15 &   64  &   2.80 &   $1.19\times 10^{2}$  & 7.01  &  $3.03\times 10^{-2}$   &  2.50  &  $1.08\times 10^{-2}$   &  $2.12\times 10^{-2}$ &  2.00\\
435  &  15 &   64  &   2.40 &   $2.58\times 10^{2}$  & 7.12  &  $4.14\times 10^{-2}$   &  2.97  &  $1.72\times 10^{-2}$   &  $3.82\times 10^{-2}$ &  2.20\\
555  & 185 &   64  &   1.90 &   $6.26\times 10^{2}$  & 3.39  &  $3.21\times 10^{-2}$   &  1.78  &  $1.69\times 10^{-2}$   &  $7.53\times 10^{-2}$ &  4.47\\
675  & 185 &   64  &   1.60 &   $3.64\times 10^{3}$  & 3.52  &  $4.92\times 10^{-2}$   &  2.20  &  $3.08\times 10^{-2}$   &  $1.28\times 10^{-1}$ &  4.13\\
795  & 185 &   64  &   1.30 &   $2.22\times 10^{4}$  & 3.60  &  $6.99\times 10^{-2}$   &  2.77  &  $5.38\times 10^{-2}$   &  $1.65\times 10^{-1}$ &  3.07\\
~~795**& 185 &   64  &   1.30 & $1.00\times 10^{4}$  & 1.61  &  $3.13\times 10^{-2}$   &  1.24  &  $2.41\times 10^{-2}$   &  $1.65\times 10^{-1}$ &  6.86\\
\hline
\hline
\end{tabular}
}
\end{center}
${}^{**}$ represents the new modified  baseline with the number of detectors in the 795 GHz channel increased by a factor of five as discussed in the main text.
\caption{{\bf \core\ performance for mapping polarized dust in the highest frequency channels.}
For the eight highest frequency channels for the baseline defined in 
Table 1, we indicate in three different ways the sensitivities scaled to an arcmin square pixel
for the polarization $(Q,U)$ anisotropies, first
as a thermodynamic temperature fluctuation 
relative to $T_{CMB}=2.73 K$---that is, as a fluctuation
in $\Delta T_{CMB},$ then as a Rayleigh-Jeans temperature fluctuation, and finally
in terms of radiance units---that is, megaJansky per steradian. 
The polarization sensitivity is then given for a square pixel of dimension 
$\theta _{fwhm}$ on a side, as well as the prediction of the rms signal
in Q and U in a pixel expected in a region with 
$A_V=1.$ Finally the resulting 
signal-to-noise within a pixel with unit magnitude visual extinction 
is indicated. 
}
\label{table:s_over_n_galactic}
\end{table}%

We propose to amend the baseline configuration defined in the proposal document submitted in 
December 2010 by deploying five times more detectors in the 795 GHz channel.
What is required to effect this increase would take away a small part of the focal plane from the 
central `primordial cosmology' channels and thus will leave intact the primordial cosmology capability 
forecasts. In the former baseline configuration 
the detectors devoted to the 795 Ghz channel occupy less than 
1\% of the focal plane area and contribute less than 4\% to the data rate. Therefore their number could 
easily be increased by a factor of five with only modest reductions in the other channels. The 
keep the data rate transmitted to Earth down, one could also study the option of placing five detectors 
one behind the other in the scan direction so that on-board compression could be used to keep the data 
rate unchanged.

\subsection{Angular resolution}

The observation of the primordial tensor CMB B-mode spectrum nominally requires only a modest resolution, 
of order 1$^\circ$ on the sky, significantly lower than already achieved by WMAP and PLANCK. CMB lensing, 
however, generates deflections of order 2.7 arcminutes. Since the lensing signal is concentrated on the 
smallest scales of the survey, one wants to sample the primordial CMB with S/N of approximately unit on 
as small scales as possible where foreground removal is feasible. Beyond the COrE resolution one rapidly 
enters into the Silk damping tail and contamination from compact sources 
rapidly becomes dominant. With its 
resolution COrE is sensitive to the interesting range of absolute neutrino masses suggested by current 
neutrino oscillation data.

For Galactic science, the high resolution mapping of the galactic magnetic field in star-forming regions 
is the key for understanding the interplay between turbulence and magnetic field, which leads to the 
formation of pre-stellar condensations. A resolution of order 1 arcminute is adequate to complement what 
can be done from the ground with ALMA and single dish sub-mm telescopes, which will provide polarization 
imaging at sub-arcminute angular scales of a variety of compact sources including pre-stellar 
condensations.

As a final constraint, we note that galactic and extragalactic point sources can be a significant 
contaminant for measuring CMB B modes, should the tensor-to-scalar ratio be low. As each point 
source has its own emission spectrum, 
this contamination cannot be effectively removed using multi-frequency 
observations unless the sources can be isolated in the maps. This calls for 
the best possible angular resolution, which sets the lower limit on the flux limit of detectable point 
sources.

A limiting constraint for the COrE resolution, however, is set by the diffraction limit due to the size 
of the telescope (set in turn by the size of the fairing for a space mission). We thus require the 
highest possible resolution available from space, and demonstrate, as discussed in the science section, 
that it is adequate for achieving the main science objectives of COrE.

\subsection{Control of systematics}

Rigorous control of systematic effects is an essential requirement for any experiment aiming to measure
CMB polarization. Systematic effects are essentially of two kinds. 
Additive errors (such as glitches due to 
cosmic ray impacts, stray signals due to temperature drifts of the payload, low frequency noise), and 
multiplicative errors (errors on the instrumental response such as inaccurate beam shapes, unknown
cross polarization leakage, sidelobes in the optical response, gain drifts).
Minimizing such systematics requires both a very careful instrument design and  
the capability to measure instrumental imperfections in flight.

Extensive simulation of the impact of systematic effects on polarization measurements has been 
carried out for PLANCK and for the US EPIC project. This work has determined the limits needed on gain 
mismatch and beam asymmetry and have been taken into account in the COrE instrument design. In addition, 
COrE mission provides four layers of redundancy: 
at the detector level through polarization modulation, which provides coverage of many polarization 
angles by the same pixel viewed with the same detector while keeping the telescope attitude fixed 
with respect to the sky and thereby rejecting systematics fixed with respect to the spacecraft; at 
the scanning strategy level through a tightly connected coverage of the sky resulting from a well 
designed scanning strategy allowing the same detector to revisit the same sky pixel at different time 
intervals; at the band level, through many different detectors observing the same sky pixel over different 
time intervals; at the focal plane level, through six CMB sensitive frequency channels. This many levels 
of redundancy offer many tests for unexpected systematics, and many ways to correct for them. 

\subsubsection{Additive errors}

Previous experience from PLANCK HFI demonstrates that cosmic ray hits can be detected, and their effect 
can be subtracted from bolometer timelines \cite{2011arXiv1101.2048P}. The effectiveness of such data 
correction depends on the rate of cosmic ray hits to be taken into account in each bolometer time 
stream (too many of them results in blending of the glitches). As most of the improvement of sensitivity 
between PLANCK and COrE comes from the increased number of channels (rather than from the sensitivity of 
the individual detectors themselves, which is nearly photon shot noise limited), 
the PLANCK experience for glitch 
correction can be adapted to COrE without major change.
 
Similarly, low frequency noise in bolometer timelines will be kept low enough, so that the so-called 
``knee frequency'' (at which low frequency noise starts being dominant) will be lower than the typical 
modulation frequency provided by the RHWP and the scan strategy. As discussed by numerous authors in the 
context of the PLANCK mission 
\cite{1998A&AS..127..555D,2000A&AS..142..499R,2002A&A...387..356M,2005MNRAS.360..390K,2007A&A...467..761A,2009A&A...493..753A,2009A&A...506.1511K,2010A&A...510A..57K}, 
this measure 
will allow almost complete suppression of residual low frequency noise after data processing to 
produce maps of temperature and polarization. Most other additive errors induce signals which are either 
very sparse (e.g., microphonic lines) or very low frequency (e.g., thermal drifts) and can be dealt with 
similarly in a data analysis stage.

\subsubsection{Multiplicative errors}

Multiplicative errors have been identified as a potential source of error for measuring CMB polarization. 
This category includes all errors in characterizing the detector response: calibration, 
beam shape, band shape, transfer function, detector orientation, pointing, depolarization (or 
cross-polarization leakage), etc.
In essence, the problem is that each detector measures a linear combination of three Stokes parameters 
$I$, $Q$ and $U$. Coefficients of the linear combination depend on the detector response. The recovery of 
all Stokes parameters requires combining measurements from either different detectors, or from the same 
detector, but with a different orientation (or different polarization angle as set by the RHWP). This 
linear combination of different measurements is an inversion of a linear system which connects measured 
quantities to sky temperature and polarization. If the coefficients in the linear system are not 
well known, the inversion is imperfect, and the measured Stokes parameters still contain 
admixtures of unwanted components. Because of the strong hierarchy in power among $I$, $E$ and $B$, these
errors
induces leakage of $I$ into $E$ and $B$ (from errors in calibration, beam, frequency bands, etc.), and of 
$E$ into $B$ (e.g., from polarization orientation errors) \cite{2002AIPC..609..209K,2003PhRvD..67d3004H}.

This potentially severe problem has been identified during the preparation of the PLANCK mission and has 
motivated method papers to handle the problem during 
the data analysis stage \cite{2007A&A...464..405R} as well as 
several studies for defining the EPIC mission proposal in the US 
\cite{2008PhRvD..77h3003S,2009PhRvD..79f3008M}. While detailed studies of the problem for COrE 
specifically will be part of the ongoing activities for the preparation of the space mission, the COrE 
design takes into account the experience already gathered on these issues. The RHWP provides a selection of the 
polarization as the first optical element, which contributes to the cleanliness of the measurement. As a 
second step, data processing has been shown to allow correction of effects to first order. Finally, the 
many detectors provide an effective averaging of imperfections (an effective averaged beam from 1000 
detectors is more symmetric than the beam from a single detector) as well as redundancy to check that 
all measurements agree.

To take a simple example, current measured beams have an ellipticity of order 5\%, well above the 
requirements for an accurate measurement of the CMB $B$ modes \cite{2009PhRvD..79f3008M}. Both 
instrumental optimization and the RHWP, however, can reduce this ellipticity (or its impact on the 
measurement) by a factor 2--5 at least. Corrections during the
data analysis stage can gain an order of magnitude, as shown 
for PLANCK in \cite{2007A&A...464..405R}. The averaging of residual errors from O(400-1000) detectors 
gains another factor 20-30, for a final beam ellipticity effect of order 400 to 1500 times lower (of 
order $2-6 \times 10^{-5}$), well within requirements for the measurement of primordial and lensing 
B-modes.

\subsection{Why space?}
\label{sec:why:space}

Balloon experiments such as EBEX and SPIDER along with ground-based telescopes such as PolarBear and SPT
\footnote{For an up to date list of suborbital experiments, see 
{http://lambda.gsfc.nasa.gov/product/suborbit/su\_experiments.cfm}}), 
with thousands of detectors in one 
or two frequency bands and sufficiently long integration time, are just now reaching levels of 
instantaneous sensitivity to achieve white-noise levels of $3\mu K$ per arcminute at a single frequency, 
arguably sufficient to reach $r\simeq0.05$ in the absence of contaminating systematics, with angular 
resolutions varying from ten arcmin to one degree. These experiments have already demonstrated 
the technical feasibility of measuring the CMB with thousands of detectors.

However the full science goals outlined above require a yet higher sensitivity to 
cosmological and astrophysical polarization over a wider range of angular scales and frequencies. To 
realize this program in practice requires control of systematics at a level that cannot be achieved in 
suborbital experiments. A satellite experiment like COrE is therefore a necessary next step. COrE will 
observe the full sky at multiple bands spanning over a decade in frequency and take data for very 
long integration times in the very quiet environment of the L2 Lagrange point of the Sun-Earth system.

\subsubsection{Full sky coverage}

COrE's science goals are best achieved with high sensitivity over the full sky. Full-sky 
coverage will enable 
measurement of the reionization signal in polarization, which is
crucial for confirmation of the last-scattering 
B-mode signal at medium scales. Moreover, full-sky coverage will accumulate sufficient 
signal-to-noise for the non-Gaussian 
signal, constrained by the number of independent triples (or quadruples,$\ldots $) of modes
available. Full-sky coverage will be 
crucial if we want to go beyond detection of $f_\textrm{NL}$ to detailed non-Gaussian models of the 
higher-order moments as discussed above.


Situated between the strongly radiating ground and Sun, suborbital experiments are highly constrained in 
the amount of sky that they can observe as well as how they can observe it. They must scan at 
constant elevation to avoid rapidly-changing ground pickup while simultaneously avoiding the Sun. Hence 
these experiments can typically observe a maximum of a few thousand square degrees with only mild 
cross-linking of different scans, with details strongly dependent upon the location of the experiment.

\subsubsection{Avoiding the atmosphere}

The atmosphere allows observation in just a few atmospheric windows, incompatible with the wide frequency 
coverage required to fulfill the scientific objectives of COrE. If only for this reason, COrE must be a 
space mission. Even for less ambitious science goals, the atmosphere turns out to be a severe problem for 
polarized CMB observations from the ground. The atmosphere indeed is a strong source of microwave 
emission. Although largely unpolarized, the atmosphere adds to the photon shot noise, reducing detector 
sensitivity and hence making necessary longer integration times. Moreover, Zeeman splitting due to the 
Earth's magnetic field leads to both linear and circular polarization. The linear polarization may be 
small $(\approx10\textrm{nK}$), but the circular polarization is much larger $(\approx 10^2 
\mu\textrm{K})$, and great care must be taken to avoid conversion into linear polarization within the 
instrument. Another effect leading to significant polarization is the back-scattering of thermal 
radiation from the hot ground by the atmosphere. This results primarily from large ice crystals (i.e., 
100--1000 $\mu\textrm{m}$) in high-altitude clouds, often optically thin (along with a smaller effect 
from molecular scattering). Simulations demonstrate that at both the South Pole and in the Atacama desert 
such polarized emission will likely swamp the B mode signal from primordial tensor modes. \cite{Pietranera2007}.

\subsubsection{Benign environment (especially L2)}

Even with highly constrained, constant elevation scans, suborbital experiments would need to reject far 
sidelobes from the $250\;\textrm{K}\times2\pi\;\textrm{sr}$ ground at a level of roughly $10^{-10}$ to 
remain below the desired noise on large angular scales. From a distant point in space such as L2, the 
Earth will instead subtend an angle of only $0.5^\circ$, reducing its emission by a factor $10^5$. 
Moreover, at L2 the Earth, like the Sun, can be more easily shielded, limiting the emission from both to 
diffraction around (and temperature variations within) the shielding.

These same considerations apply to temporal variations in the ambient temperature, which can be up to 
several degrees per day on Earth at mid-latitudes, compared to the nearly completely stable environment 
in space. For COrE, any change in thermal loading is due to the behavior of the cooler itself and can be 
strictly controlled and monitored, as well as kept to specific times during the mission.

One potential worry in deep space is the increased flux of cosmic rays. The rates at L2 for 
high-sensitivity bolometers have been measured by PLANCK HFI and while not negligible, do not prevent 
PLANCK from reaching (and indeed surpassing) its sensitivity goal. Hence the achieved PLANCK detector 
sensitivity gives the proper estimate for COrE, multiplied by the vast increase in detector numbers. We 
are thus very confident that any extra noise induced by cosmic rays will not challenge the COrE 
sensitivity goal.

The L2 environment also allows much longer integration times than can be achieved from balloons (with an 
ultra-long duration limit of roughly 40 days at present) and a higher duty cycle than ground-based 
experiments.

Only from space can an experiment like COrE achieve its simultaneous goals of full-sky measurements of 
astrophysical polarization over a decade in frequency and an angular resolution of a few arcmins, and 
$\muKarcmin$ sensitivity coupled with thermal stability and $10^{-5}$ sidelobe rejection.

\section{Proposed COrE instrument}
\label{sec:instrument}


{\bf 

The technology deployed for the PLANCK mission is performing exceptionally well as reported in the papers released in 
January 2011 \cite{lfi_perf, hfi_perf},
and breakthrough science is in the process of being delivered. Although the cosmological results are still being analyzed,
the quality of the data collected has been characterized and documented in these papers \cite{planck2011_papers}. Much of the technology of the COrE
instrument has already been tested on PLANCK and the expertise accumulated from PLANCK will be of great value for COrE.
COrE builds on the PLANCK and Herschel heritage and on the experience gained from a generation of suborbital experiments. 
In order to reduce the risks and develop an 
instrument in a timeframe compatible with the next ESA M3 mission, COrE builds on the 
subsystems from PLANCK. The optical system, the passive as well as some of the active 
cooling system and most of the service module use the same technology, leading to a payload with 
similar characteristics (mass, volume power budget and cost) as shown in Table~\ref{tab:summary}. Moreover, 
whereas PLANCK was based on two different detector technologies leading to instruments with two separate focal 
planes---the LFI instrument based on HEMT technology and the HFI instrument based on bolometers---, COrE uses the same 
technology for all its spectral bands, reducing the 
risks associated with potential incompatibilities and integration problems.

The great advance in science output from PLANCK to COrE is made possible by
a massive increase of number of detectors. COrE will have altogether about 6400 detectors whereas PLANCK only
has 71 detectors. The use of large arrays of dual-polarized detectors divided in 15~spectral bands ranging from 45 to 
795~GHz allows for this massive increase.
The resulting performances have already been summarized in Table~\ref{tab:CorePerf} in the Introduction.
The spatial resolution ranges from 23 
arcmin to about 1.3 arcmin for the highest frequency band with a Q and U sensitivity of about 4.5~$\muK\cdot $arcmin 
in each of the main science channels.

Another key difference with respect to PLANCK is the presence of a polarization modulator as the first element in
the optical chain. For the massive increase in raw sensitivity to be useful, it must be matched by a corresponding 
improvement in the control and characterization of systematic errors. Unlike PLANCK, in COrE
the incoming polarization is modulated using a common single rotating reflective half-wave plate.
In this way, most of the systematic errors arising within the telescope as well as those involving issues of control of
the beam profile are modulated away. While the rotating half-wave plate does add some complexity to the system,
its presence allows requirement on other aspects of the system to be relaxed. The SAMPAN design, for example, lacked modulation
but instead endeavored to control systematics using a complicated and difficult to realize scanning strategy.

The main challenges of the COrE design arise from the development of the focal plane components and the implementation of the 
polarization modulator. Several European institutes also supported by industry are able to produce the elements 
constituting each pixel (the feedhorn, orthomode transducer (OMT), and the bolometric detectors) as detailed
in the following sections. The production of the 
large number of pixels will be ensured by parallel production of the various components.

}

\subsection{COrE instrument overview} 
\label{sec:overview}

\begin{figure}[htbp]
\begin{center}
 \includegraphics[scale=0.7]{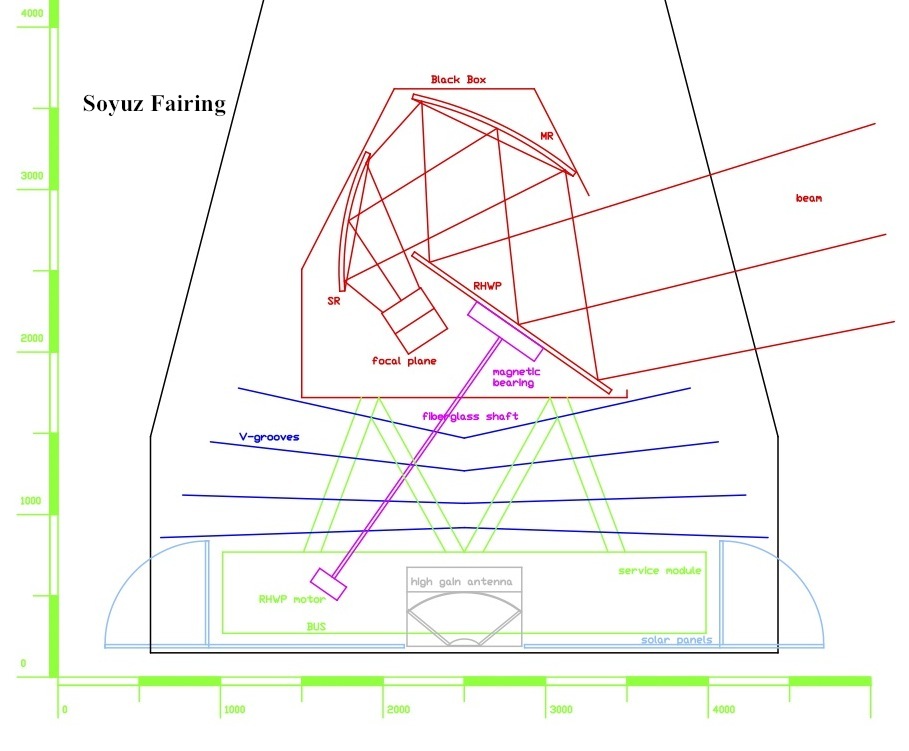}
 \caption{\baselineskip=0pt
\small
Sketch of the Soyuz fairing including the COrE payload}
\label{fig:infairing}
\end{center}
\end{figure}

\begin{figure}[tbp]
\begin{center}
 \includegraphics[scale=0.28]{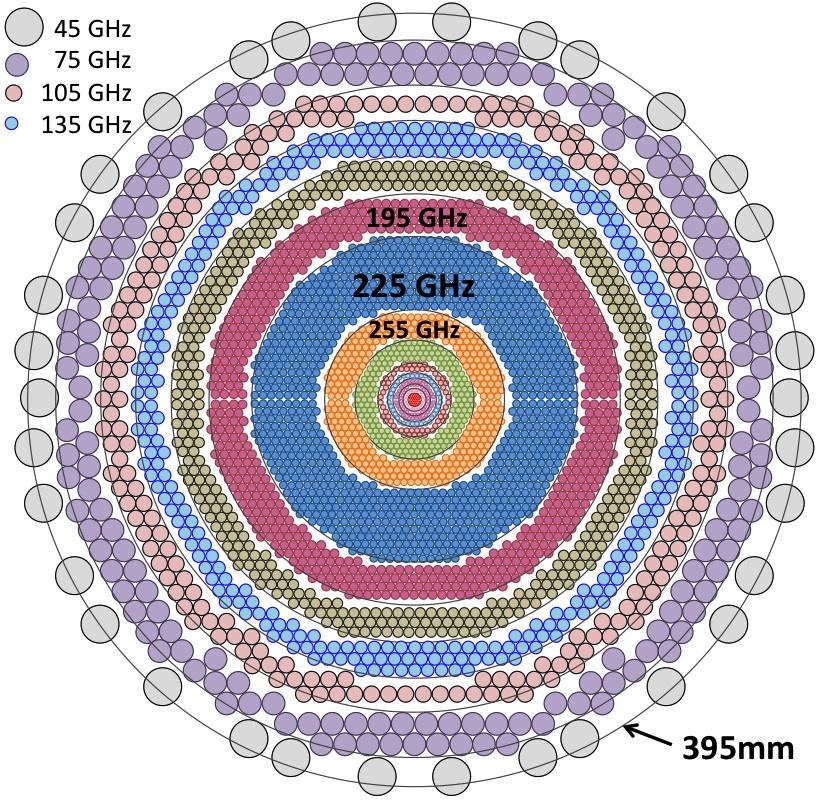}
 \hspace{1cm}
  \includegraphics[scale=0.5]{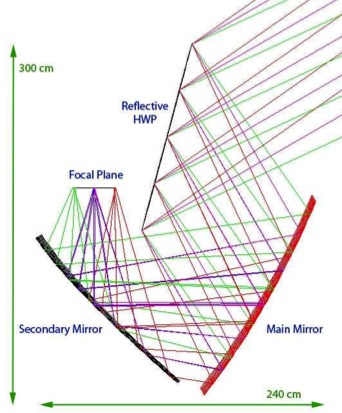}
\caption{\baselineskip=0pt
\small
Left: Focal Plane Unit. The highest frequency band (795~GHz) is at the center and the 45~GHz horns are at the periphery. Right: 
Ray tracing of the optical system showing the off-axis telescope and the RHWP.5}
\label{fig:FPU}
\end{center}
\end{figure}

The instrument (Fig.~\ref{fig:infairing}) uses an off-axis reflective telescope capable of providing
a large focal plane area (Fig.~\ref{fig:FPU} left) with limited aberration and cross-polarization. For reasons explained
in Sect.~\ref{sec:rhwp}, polarization modulation is achieved using of a reflective half-wave plate 
(RHWP) located at the aperture of the telescope. The RWHP modulates away or minimizes many of the systematic errors
that must be overcome for high precision polarization measurements. 
After reflecting off the RHWP, the incoming radiation passes through the telescope and is focused onto the 
detector arrays. While technology based on planar detector arrays has made tremendous progress 
\cite{Chattopadhyay} (used in BICEP2 for example), in order to reduce the risks associated to the lack of 
maturity, COrE uses feedhorn-coupled detectors. 
Horns have the advantage of defining a well-formed beam with rapidly decaying far sidelobes. 
These horns are coupled to ortho-mode transducers (OMTs) 
through a circular waveguide while each branch of the OMTs separating the two linear polarizations is coupled to 
a bolometric detector. The cold optics (feedhorns+OMT+detector pair) are cooled to 100~mK and surrounded by 
several successive temperature stages. The whole instrument is enclosed by a passively cooled 35~K shield. The 
telescope mirrors and the RHWP could reach a temperature as low as 30~K.
Table~\ref{tab:summary} gives the overall characteristics of the payload as a comparison to the Planck mission.

\begin{table}[htd]
\begin{center}
 \caption{\baselineskip=0pt
\small
COrE payload summary table compared to PLANCK. The full payload includes the service module and hydrazine. The data transfer rate of COrE assumes a compression factor of 4 and a 2 hours per day downlink.}
\begin{tabular}{c|c|c}
\hline
\hline
& COrE & PLANCK\\
\hline
Overall mass & 1832~kg&1974~kg\\
Power budget & 1279 W &1342~W\\
On-board data storage & 780 Gbit &32~Gbit\\
Data transfer rate & 54 Mbit/s &1.5 Mbit/s\\

\hline
\hline
\end{tabular}
\label{tab:summary}
\end{center}
\end{table}%

\subsection{Description of the measurement technique} 
\label{sec:meastech}

COrE will carry out a full-sky measurement of temperature anisotropies and polarization of the CMB. The choice of a 
far-Earth orbit, such as a halo or a quasi-periodic Lissajous orbit around the Sun-Earth Lagrangian point L2, is 
a requirement for COrE in order to provide sufficient control of straylight contamination from far sidelobe pick-up. 
Requirements for far-side lobe rejection are extremely stringent for COrE.
Because the gross features of the beam pattern at 
large off-axis angles are essentially the same for all detectors at a given frequency, the required rejection 
must be calculated against the total sensitivity per frequency channel. Using the PLANCK requirements as a guide, and 
assuming a factor of 10 better sensitivity for COrE in an L2 orbit, we need roughly -110 dB rejection for 
contamination from the Sun, -100 dB contamination from the Earth, and -80 dB contamination
from the Moon. Stringent, highly frequency-dependent rejection 
requirements will be imposed also by straylight from diffuse galactic emission entering via the intermediate side lobes. Although 
challenging, these straylight rejection requirements for COrE can be achieved with a very careful optical design and can be 
tested with a moderate extrapolation of the state-of-the-art technology developed by ESA for the PLANCK mission. 
Furthermore, the excellent thermal stability of the L2 environment is ideal to minimize thermal systematic 
effects on the instrument. The experiences of both WMAP and PLANCK have demonstrated exquisite thermal stability 
against small changes in the solar aspect angle, with very small and slow residual thermal drifts.

From the data analysis point of view, we need full sky coverage to optimize the
redundancy in each map pixel and the range of timescales for revisiting a given
pixel in order to reject ``naturally'' the statistical noise and most common
systematic errors. Moreover, polarization determination requires a wide variety
of measurement angles per pixel. In the first B-Pol proposal, all the
polarization modulation efforts involved payload motion. This led to a
sophisticated (but feasible) scanning strategy and set tight constraints on sun
shielding. For COrE, the addition of a HWP modulator relaxes these requirements
and allows for a scanning strategy similar to PLANCK, with only a precession of
the spin axis and possibly some level of nutation. If the COrE spacecraft is a
`spinner,' the spin rate $\dot{\mu}$ must be fast enough to ensure redundancy to
minimize $1/f$ noise and thermal fluctuations yet slow enough to be compatible
with the bolometer time constants. A `spinner' is a spacecraft with a
nonvanishing total angular momentum that is periodically re-oriented using
thrusters. The alternative is a three-axis stabilized spacecraft with
counter-rotating flywheels allowing for more complicated manoeuvring with
essentially no energy expenditure owing to the fact that the total angular
momentum is nearly zero at all times.  Based on the PLANCK experience, we
envisage a spin rate of order 0.5 rpm, and a step and point approach for the
spin axis like in PLANCK, with similar telescope de-pointing. This means that
the spin axis would be fixed for certain time allowing for a few rings, then
moved to another direction and kept fixed to allow for a few rings and so
on. The difference $\Delta\phi$ between two consecutive spin axis positions
would be of 1 to a few arcmin every few tens of circles and therefore maintain
the nearly anti-solar configuration ($\Delta\phi/\Delta t \sim
1^\circ$/day). This would yield sufficient redundancy in sky circles and,
simultaneously, good sampling of the sky for each single beam up to 220 GHz
(i.e., in all the main COrE cosmological channels). At higher frequencies, where
the angular resolution is $\theta_{fwhm}<4$~arcmin, adequate sampling would be
ensured by proper staggering of the feeds in the focal plane.

Tight constraints are imposed on the scanning parameters to avoid thermal and stray light artifacts and by 
data transmission requirements. The maximum excursion $\alpha$ of the spin axis from anti-solar direction must not exceed
$27.5^\circ$ to avoid thermal effects and straylight contamination from the Sun and 
Moon and to prevent power modulation due to spacecraft asymmetries or shadowing of the solar 
panels. Moreover the angle between the spin axis and the Earth direction must remain below $20^\circ$ to ensure adequate
visibility of a fixed medium gain antenna for telemetry activity.
While steerable antenna designs are possible, we 
do not favor their use since they would complicate the design and increase risk. Depending on the dynamical 
constraints and attitude control capabilities, modulation of the spin axis can be set either by precession (with 
the advantage of maintaining a constant solar aspect angle\footnote{The solar
aspect angles characterizes the orientation of the payload with respect to the
sun. Maintaining this orientation constant allows for simpler and more robust
design as far as thermal shielding and battery solar panels designs are concerned.}
or with cross-ecliptic excursions at an angle 
$\alpha = 90-\beta$. The optimal configuration will be determined via simulations that include polarization 
angle redundancy, de-striping efficiency, and figures of merit for straylight and thermal effects. 
Table~\ref{tab:scan_param_range} summarizes a possible baseline based on the PLANCK experience.

{\small
\begin{table}
 \caption{\baselineskip=0pt
\small
Scan parameter definitions and allowed variations}
\begin{tabular}{|l|l|c|l|}
\hline
Parameter & Definition & Range & Potential criticalities\\
\hline
$\beta$ & Telescope axis& $65-85^\circ$ & $>85^\circ$: stray light, polarization angle redundancy\\
&to spin axis angle &&$<65^\circ$: excessive excursion $\alpha$ angle required\\
$\dot{\mu}$ & Spin rate & $0.1-2$ rpm & $<0.1$ rpm : $1/f$ noise, thermal fluctuations\\
&&& $>2$ rpm : bolometer time constant\\
$\Delta\phi$ & Depointing angle & $1-$few arcmin & $<1$: limited redundancy in sky circles\\
&&&$>2$: insufficient sky sampling\\
$\alpha$ & Solar aspect angle & $5-27.5^\circ$ & $<5$: coupling with $\beta$ and full-sky requirement\\
&&&$>27.5$: thermal effects, stray lights\\
$\psi_{Earth}$ & Maximum Earth& $0-15^\circ$ & $>15$: TM/TC requirements\\
&aspect angle &&\\
$\dot{\phi}$ & Precession frequency & 1-24 weeks & polarization angle redundancy, destriping\\
&&&thermal effects\\
\hline
\end{tabular}
\label{tab:scan_param_range}
\end{table}
}

A three-axis stabilized satellite would allow improvement of the scanning
strategy concept above.  One could, for instance, imagine combining the best of
the PLANCK and WMAP scanning strategies by superimposing a daily nutation of
$20\dg$ about the antisolar direction, bringing the spin axis back to the
anti-earth position every 24 hours. This would distribute the basic ring
measurement\footnote{I.e., the measurements made with a stable pointing allowing
  revisiting the same pixel by the same detector but with several RHWP
  orientations} to fill an $\sim$ $40\dg$ annulus. The same pixels would be
revisited several times over a 40 day interval (constraining systematics of
differing periods) at the expense of somewhat uneven sky coverage. Optimization
is left as an option for further study.

In addition to the classical compromise between survey size and sensitivity, COrE has to account for the 
directivity of polarization. It is both an additional constraint on the system and a lever arm against 
systematic effects. Polarization is usually described in terms of the Stokes parameters $I$, $Q$ and $U$. In 
COrE we propose to implement a \emph{polarization modulator} made of a rotating half-wave plate (hereafter 
RHWP) which modulates simultaneously all the detectors in the focal plane, so that each detector measures
\begin{equation}
m = 0.5(I + Q\cos4\alpha + U\sin4\alpha) + n \label{eq:principle}
\end{equation}
where $n$ is the noise and $\alpha = \omega t$ the orientation of the RHWP. Each measurement is the 
linear combination of three independent parameters. Three independent measurements, each at a different angle 
$\alpha$, are therefore required (at least) to determine the polarization state. In the absence of a 
polarization modulator, these measurements would require rotating the detector. However that 
approach contaminates the sought polarization signal with beam shape and detector related artifacts.
Introducing the modulator very significantly simplifies the measurement and scan strategy. 
The satellite no longer needs to be rotated, and the sky scan can be very similar to that of PLANCK. 
The continuous rotation of the RHWP at a frequency incommensurate with the spin frequency 
provides the required change in $\alpha$ for each reobservation of the same sky pixel. After three
rotations, a ring of fully determined $I$, $Q$, $U$  has already been
obtained. Further rotations serve to improve signal-to-noise and to remove
systematics.
The 
idea behind this modulated measurement, as eqn.~\ref{eq:principle} shows, is that \emph{polarization 
is modulated at $4\alpha$} when the modulator is rotated by $\alpha$, \emph{whereas most systematic effects are 
fixed} in the instrument reference frame. Rotating the modulator with respect to the incoming polarization and 
providing sufficient angular coverage is therefore analogous to a lock-in detection that selects polarization 
and cancels out systematic effects. The rotation frequency of the modulator is around 0.1--0.5 Hz.

\begin{figure}[htbp]
  \begin{center}
  \includegraphics[scale=0.27]{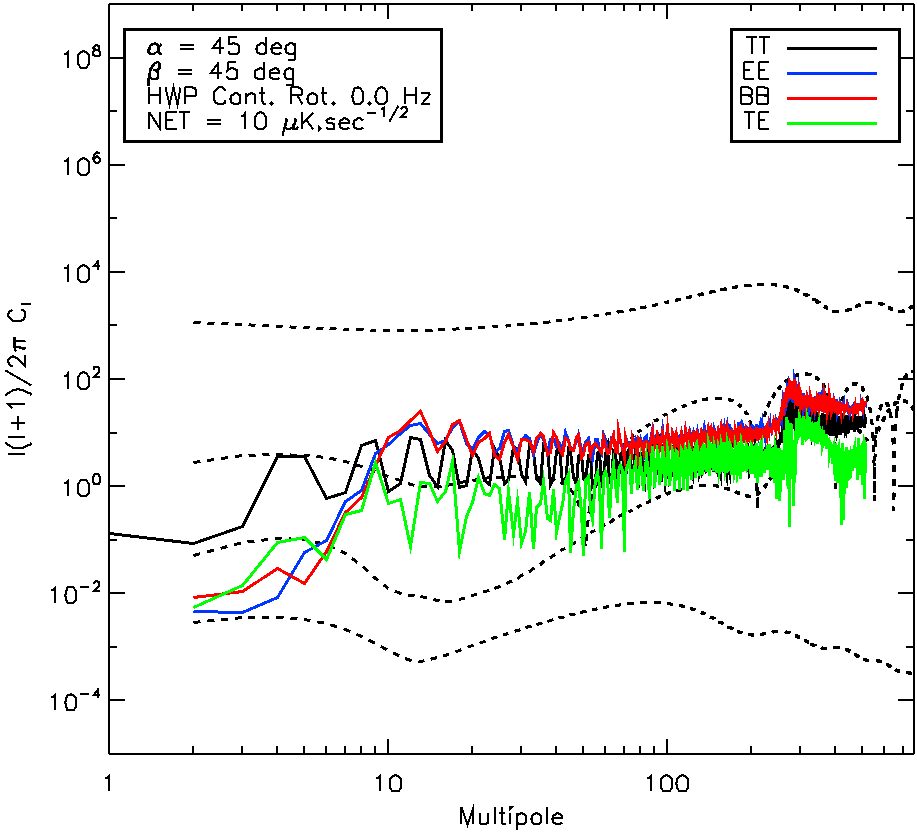}
  \includegraphics[scale=0.27]{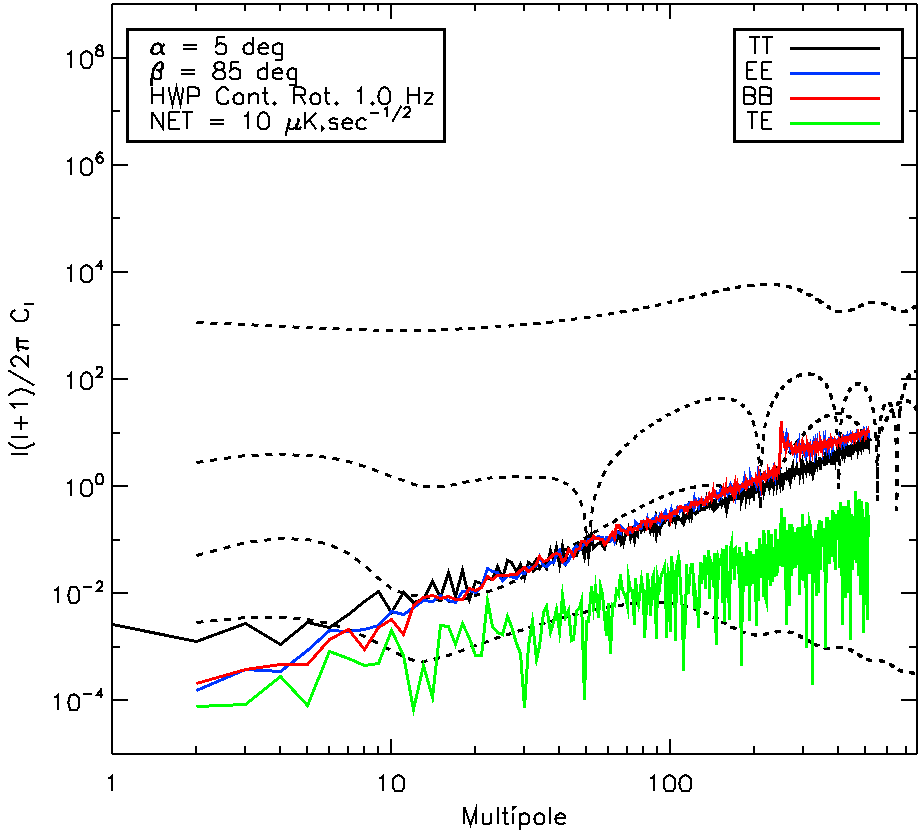}
\caption{\baselineskip=0pt
\small
{\bf Angular power spectra of the $1/f$ noise timelines.} Previous B-Pol proposal without HWP (left), PLANCK and 
continuously rotating HWP at 1Hz (right).}
\label{fig:cls_ipol_maps_planck}
\end{center}
\end{figure}

Results of detailed simulations of this scan and modulation strategy are summarized in 
Fig.~\ref{fig:cls_ipol_maps_planck}. The improvement compared to PLANCK (which was not 
optimized for polarization measurements) is substantial. To illustrate this point, we chose 
typical PLANCK parameters ($\alpha=5^\circ$, $\beta=85^\circ$, $\dot{\mu}=1$~rpm) with a continuous rotation
(1Hz). For these simulations, we assumed standard white noise and $1/f$ noise with $f_{knee} = 0.1$Hz and a $NET = 10 
\mu\rm{K\cdot sec}^{1/2}$.

A detailed optimization of scan parameters (angles and precession and nutation frequencies) deserves to be studied.
At present we conclude that the presence of the HWP enables COrE to relax previous 
requirements concerning sun-shielding and the scanning strategy. Our single detector simulations show that a 
continuously rotating HWP rejects $1/f$ noise better than a stepped HWP. However, we stress that
(1) $1/f$ rejection can be addressed using high-performance de-striping algorithms such as those 
developed for PLANCK, and (2) stepping the HWP enables fine tuning of the angular coverage and could provide 
a significant technical simplification.

Moreover, we have some estimates about some of the instrument parameters, more specifically about the satellite spin rate (0.5 rpm) and the RHWP rotation speed (0.1 to 0.5Hz) derived from either present experiments or preliminary simulation results. A more precise value of these parameters will require further study. However the sub-systems concerned are studied to accept a range of values which will be compatible with an improved scanning strategy.

\subsection{Instrument subsystems description and technology}
\subsubsection{Optical and RF configuration}
\label{sec:optics}

The specifications required for achieving the scientific goals of the mission relating 
to the optical design are the resolution, edge taper and spillover, beam ellipticity (requirement$<1\%$), and 
cross-polarization (requirement $<-30$~dB). To meet the requirements using high Technology Readiness Level (TRL)
technology within the size and weight budgets of a Soyuz launcher, we have chosen a payload with a single
reflective off-axis compact test range telescope configuration (such as for ground-based Clover \cite{Grimes} and QUIET 
\cite{Buder} projects) following the Dragone-Mitsugushi configuration. Used in conjunction with feedhorn-coupled 
detectors, this design allows for a large focal plane unit (FPU) with limited aberrations necessary to fulfil the 
scientific requirements. This technology is already rated TRL~9. However because of the large number of 
detectors, further development is required, mainly to reduce the mass, cost, and complexity and to ensure 
reliability. To limit the dimensions of the FPU, the horn apertures need to be small, 
resulting in a fast optical system. The resulting configuration has a 1.2~m projected diameter making use of large field of view horns (F2 aperture number), each having a 3$\lambda$ aperture diameter. Taking into account an array filling factor of 0.8 and 
leaving enough space for the spectral filtering, the overall FPU diameter is about 40~cm (Fig.~\ref{fig:FPU} 
right). The entire optical system is enclosed in a cavity formed by a shield at 35~K with the RHWP acting as a 
cold stop. The telescope could reach a temperature as low as 30~K with passive cooling 
(compared to 50~K on PLANCK).

While this first iteration optical concept is already producing good performance, improvement will be needed 
as an outcome of a Phase A study. While the cross-polarization of the whole system is predicted to be within the 
requirement, attention will have to be paid to the beam ellipticity which could be improved with a lengthy 
optimization of the telescope and RHWP designs as well as a reshaping of the FPU surface (curved instead of 
flat as assumed so far).

In the present unoptimized design, when the uniformizing effect of the RWHP (see 
\S~\ref{sec:calibration}) has not been taken into account, the beam ellipticity ranges from below 
0.5\%  to about 5\% for the edge pixels 
(depending on frequencies and location in the focal plane). Most of the pixels having an ellipticity of 
about 3\%. For comparison, PLANCK beam ellipticities range from 5\%  to about 15\% \cite{maffei}, 
\cite{sandri}. Also the possibility of having a colder RHWP will be investigated to reduce the spillover 
contribution and to limit the effects of its emissivity (\S~ \ref{sec:rhwp}).

The feedhorn transmits the full intensity of the incoming radiation through its circular 
waveguide onto an OMT that separates the two linear polarizations. The horn-OMT assembly determines
the minimum cross-polarization achievable between the two branches that guide each polarization to the 
detector. Waveguide OMTs are the best. Return loss and isolation of order -20~dB and -50~dB, respectively, 
have been already achieved at 100~GHz for a 30\% bandwidth \cite{pisano2}, and -40 and -70 dB, respectively, at 
30~GHz. Although further development is underway, this TRL~9 technology is potentially too heavy, and so far
allows similar performances only for frequencies below 150~GHz. For this reason the OMT baseline adopted
relies on planar technology.

\begin{figure}
\begin{center}

\includegraphics[scale=0.32]{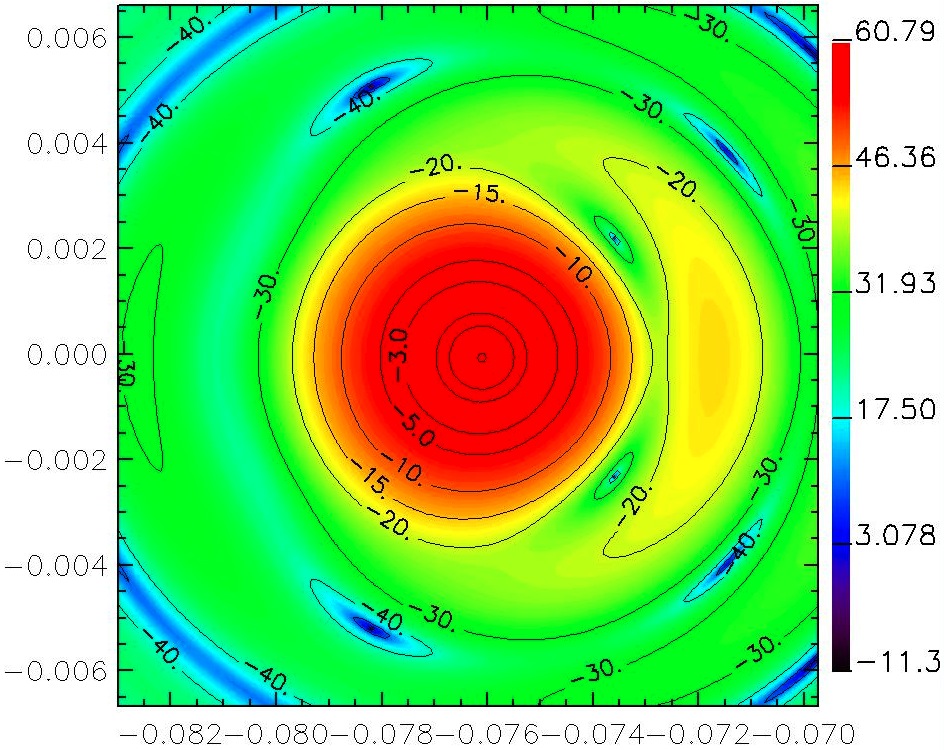}
  \includegraphics[scale=0.32]{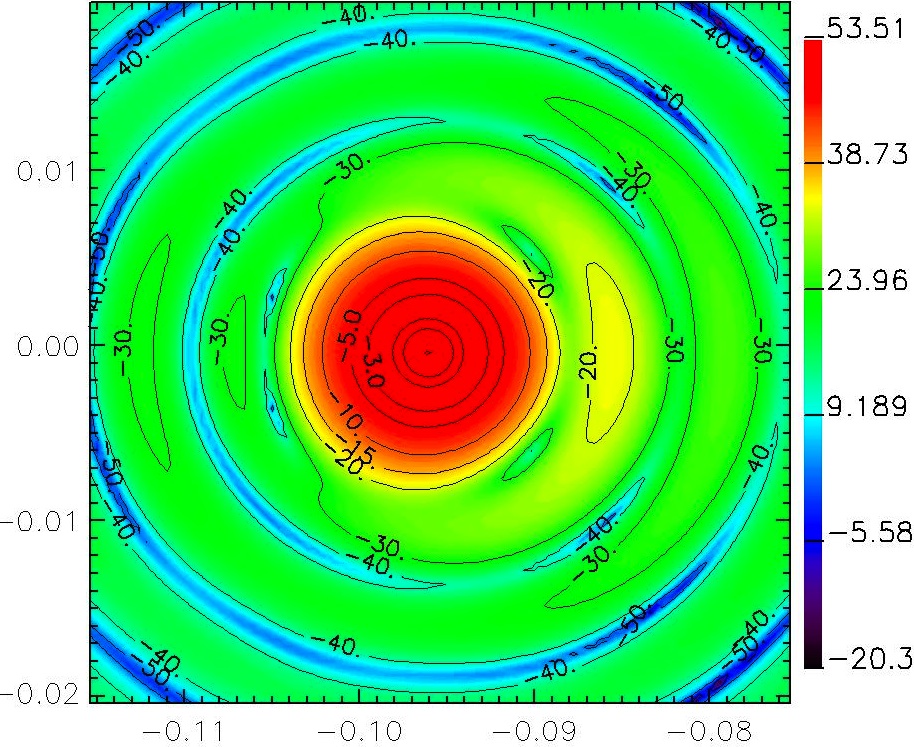}

\caption{\baselineskip=0pt
\small
{\bf GRASP beam simulations in the U-V plane for two extreme off-axis pixels.} (right) 45~GHz, (left) 105~GHz. Both have
5\% beam ellipticity. Optimization to reduce the beam ellipticities will be carried out during the Phase A study.}
\label{optics}
\end{center}
\end{figure}

\paragraph{Mirrors:}

The PLANCK telescope has demonstrated that the technology needed already exists \cite{tauber09}. The mirror
is fabricated using a carbon fiber reinforced plastic (CFRP) honeycomb sandwich technology and is coated 
with a thin reflective layer of aluminium. With dust contamination taken into account, the emissivity of the 
mirrors should not exceed 0.5\%, even for the highest frequency band (857~GHz). The surface quality is 
satisfactory; however, if even higher quality is required, other technologies such as the one used for the 
Herschel telescope are also rated at TRL~9.

\paragraph{Feed horns:}

Corrugated horns are extensively used in CMB experiments due to their very high performance. They exhibit  
very high polarization purity and low sidelobes levels, and the corrugation profile can be adjusted to 
meet electrical and mechanical requirements as well as to control the phase center location. However corrugated
horns are slightly heavier and more difficult to manufacture than smooth walled horns due to the definition and thickness 
of the corrugations. This leads to potential problems for large arrays of closely packed horns. 
Different solutions are in development at various institutes.

\begin{figure}[hbtp]
\begin{center}
\includegraphics[width=3.5cm]{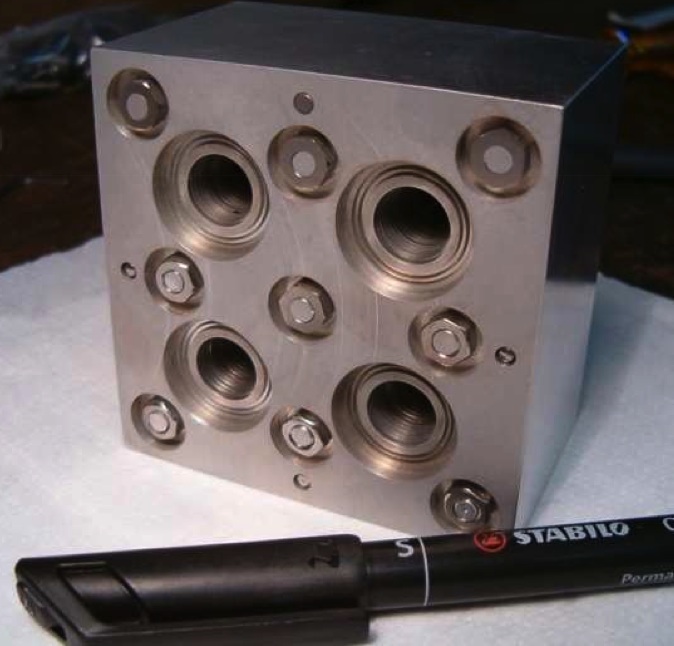}
\includegraphics[width=7cm]{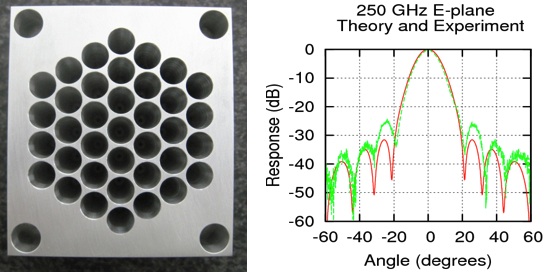}
\includegraphics[width=5cm]{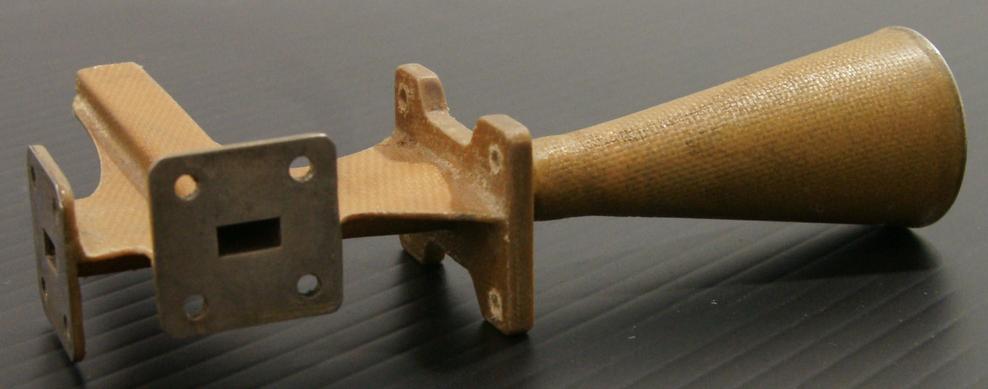}
\end{center}
\caption{\baselineskip=0pt
\small
Left: array of platelets horns (University of Milano, Italy). Center: array of drilled smooth walled horns 
together with a measured beam pattern (University of Oxford, UK). Right: combined horn and OMT made of Kevlar as 
a monolithic piece. The inside is coated with metal.}
\label{fig:horns}
\end{figure}

The University of Oxford is developing smooth walled horns (Fig~\ref{fig:horns} center) that are 
drilled to have a chosen profile \cite{kittara} giving performances nearly as good as corrugated horns. 
These performances might be good over a reduced bandwidth but large enough still to match COrE spectral 
bandwidths.
Most the development effort seeks inexpensive and reliable ways to manufacture lighter corrugated horns. As 
an example, we mention the platelets technique used by the QUIET experiment (US), also developed in Italy 
(University of Milano - Fig.~\ref{fig:horns} left) and France (APC - Paris). While the manufacturing process is 
simplified and less expensive, mass reduction still needs improvement. For that purpose, previous development efforts 
produced horns made of metal coated Kevlar (Fig.~\ref{fig:horns} right) which have been successfully flown 
(Italy/UK). Also, the University of Wisconsin has recently manufactured a 100~GHz prototype corrugated horn 
made of metal coated plastic with good RF characteristics for a first attempt through stereolithography techniques. 
Further development is needed in order to test its cryogenic performance.

Within a few years these techniques should reach the level of maturity required for incorporation in the final
instrument design.

\paragraph{Polarization separation:}
Several countries in Europe and the US are developing planar OMTs, some of which are now being used in experiments. 
Models shows that return loss of the order of -20~dB with an isolation of about -50~dB is  possible 
(measured for the C-BASS project at 5~GHz \cite{grimes2}). However similar results have not yet been confirmed 
experimentally at high frequencies (above 100~GHz). While planar OMT technology seems the most 
promising, it needs improvement, and other techniques will need to be investigated if 
planar technology subsequently proves to be inadequate. Parallel to horn fabrication R\&D, efforts are underway to
simplify the manufacture of waveguide components while retaining 
their exceptional RF performances. Platelets technology can also be applied to OMTs for which spark erosion 
manufacturing tolerances reache a few microns making possible the extension to high frequencies.

\begin{figure}[bthp]
\begin{center}
\includegraphics[scale=0.5]{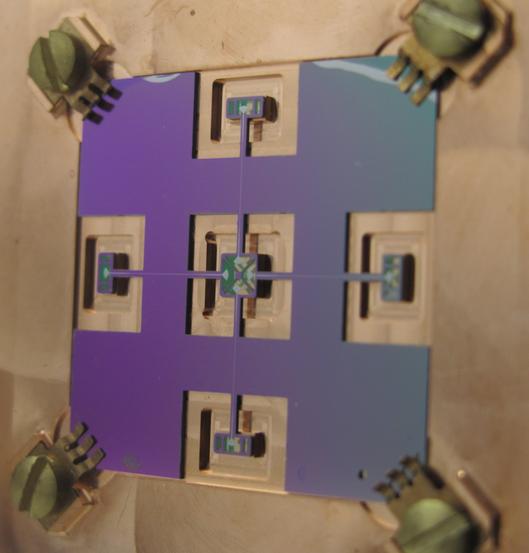}
    \includegraphics[scale=0.23]{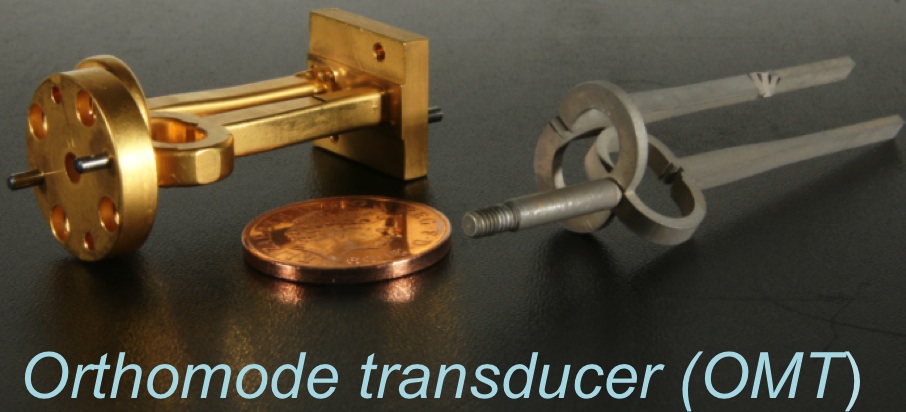}
\end{center}
\caption{\baselineskip=0pt
\small
Left: prototype of 100~GHz planar OMT (Fr). Right: Clover 97~GHz waveguide OMT (UK).}
\label{platelets}
\end{figure}
\subsubsection{Polarization modulator}
\label{sec:rhwp}

\begin{figure}[htbp]
\begin{center}
 \includegraphics[scale=0.2]{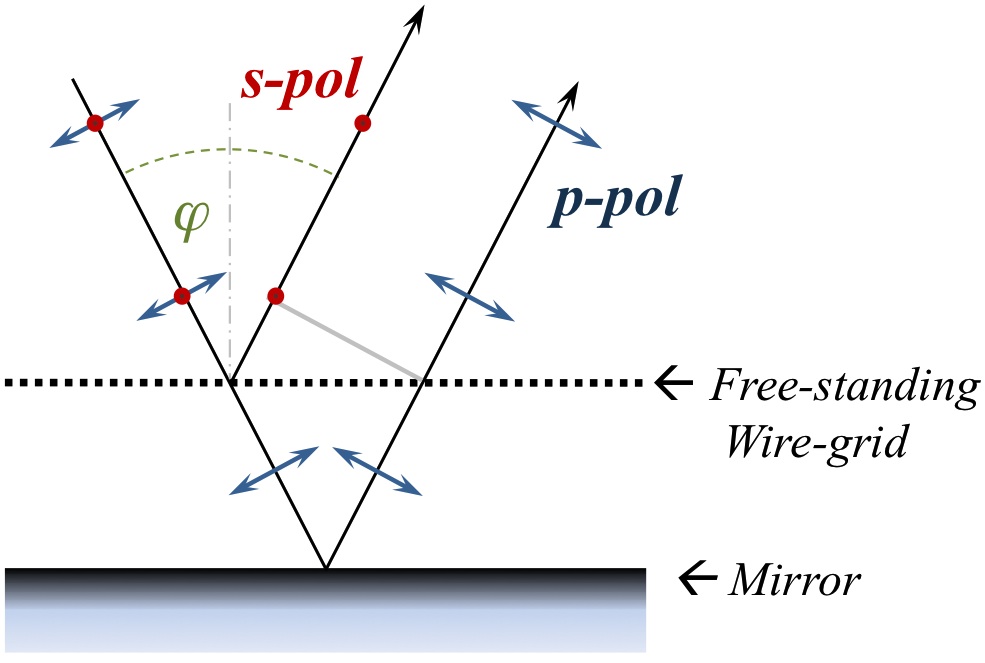}
  \includegraphics[scale=0.2]{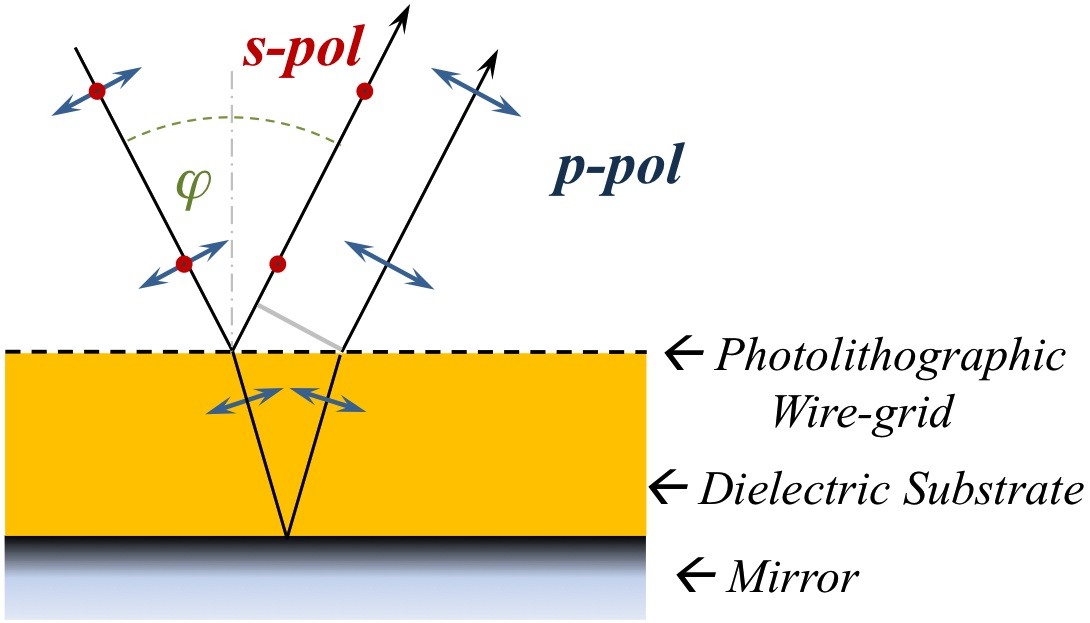}
\caption{\baselineskip=0pt
\small
Left: Free-standing RHWP. Right: Dielectric substrate RHWP}
\label{FreeRHWP}
\end{center}
\end{figure}

A traditional transmitting half-wave plate \cite{pisano} cannot cover the full COrE spectral range. 
It is important that the polarization modulator be placed as the first element of the optical chain, 
closest to the sky. Otherwise systematic errors remain that are not modulated away. This consideration
excludes a single telescope with many polarization modulators, each tailored to a specific frequency range. 
The alternative would be to use several telescopes, but since size is a prime cost driver for an instrument
of fixed physical size this option would entail unacceptable degradation of the angular resolution. 
We have therefore chosen to use a reflecting half-wave plate (RHWP) consisting of a free-standing wire grid 
polarizer held at a fixed distance $d$ from a flat mirror (Fig.~\ref{FreeRHWP}). 
The RHWP rotates aibout an axis 
orthogonal to its surface and passing through its center. With the RHWP oriented as in Fig.~\ref{FreeRHWP}, 
the RHWP introduces a phase shift between the two orthogonal `s' and `p' polarizations (orthogonal 
and parallel to the plane of incidence, respectively) that 
depends on the angle of incidence $\phi .$ 
When the phase shift is $180^{\circ}$ and the RHWP rotates with angular velocity $\omega ,$ 
the overall effect is a rotation of the reflected linear polarization at a frequency $2\omega$. This happens 
only when the path difference between the two waves is equal to $\pi$---that is, at the frequencies
\begin{equation}
\nu_n^{RHWP}=(2n+1)\nu_0^{RHWP} \qquad {\rm with} \qquad \nu_0^{RHWP}=\frac{c}{4\cos (\phi)d}
\label{eq:peaks}
\end{equation}
\noindent where $n$ is an integer and $c$ the speed of light. The modulation efficiency 
is a sinusoidal function of frequency with maxima at frequencies $\nu_n$. Without spectral 
filtering, the average modulation efficiency is 0.5. With spectral filtering (to be 
implemented, see \S~\ref{sec:filters}), higher efficiencies can be achieved in narrower bands. 
A window of 50\% bandwidth around one of the peaks gives an efficiency of 80\% . 
In our case, this means that the bandwidth of each channel would be 15~GHz. 
For the three highest frequency bands (555, 675, and 795GHz), we found more efficient to increase the bandwidth to 195~GHz.\footnote{
The optimization of the partition of the available focal plane area among the different frequency channels 
should be studied further. For the three highest frequency channels we have chosen broad bands containing 
several peaks (here 7) of the modulation efficiency so that $(\Delta \nu )/\nu \approx 1/3$ in order to increase 
sensitivity. An option for study is to further increase the sensitivity (in $\mu K \cdot {\rm arcmin}$) by approximately 40\% 
through introducing periodic sub-band filters to block the frequencies of low modulation efficiency. In both 
cases the polarization frequency bands have a comb-like structure with 7 teeth.}

{\it The absorption} is due to the power dissipated in the metal and along the wires. For a flat metal
absorption for the `s' and `p' polarizations are always slightly different, the latter being lossier. 
The losses increase with frequency but can be reduced by decreasing the device temperature. 
For a copper RHWP at 30~K used at $\phi = 45^{\circ}$ incidence and assuming an ideal wire-grid, the 
differential absorption coefficient between for the `s' and `p' polarizations is of order 
10$^{-4}$. However, while the RHWP is rotating, the s and p absorption coefficients fluctuate at a 
frequency 2$\omega$ with averaged fluctuation amplitudes of 6$\times10^{-5}$ and 1.5$\times10^{-5},$ 
respectively.

{\it A dielectric substrate RHWP} fabricated using photolithographic techniques could be used to solve the 
free-standing RHWP manufacturing problems. The air gap between the wire grid and the mirror is 
replaced by a dielectric substrate and the the wire grid is directly imprinted 
on one of the substrate surfaces 
(Fig.~\ref{FreeRHWP}). The presence of the dielectric substrate introduces frequency-dependent standing waves 
(multiple reflections). Their amplitude depends on the type of polarization `s' or `p' passing through.
Models show that modulation efficiency deteriorates with 
increasing substrate refractive index. For a substrate with $n=1.2,$ the efficiency drops to ~0.4,
which can be improved through spectral filtering. However, the absorption value then changes slightly with 
the RHWP orientation. These additional losses due to resonances appear at frequencies corresponding to 
the modulation efficiency peaks. For a RHWP at 30~K made out of copper, the values are within the following 
ranges: $A_s= 1.3-1.8\times10^{-4}$, $A_p=2.6-2.7\times10^{-4}$. (This means that the `s' and `p' components of an 
unpolarized background signal is modulated at 2$\omega$ with relative amplitudes 
of $5\times10^{-5}$ and $ 10^{-5}$ respectively). While presently rated only at TRL~4, we are choosing {\bf the 
dielectric substrate RHWP as the modulator baseline} for COrE as it offers the best potential 
performances and manufacturability. Indeed it is modelled on the dielectric embedded interference filter 
technology flown on PLANCK and Herschel.


Manufacturing the free standing RHWP is challenging because required diameter 
exceeds 1~meter. The mechanical structure of the free-standing wire-grid must guarantee 
a constant spacing between the wires and a constant distance from the mirror across the entire surface. To our 
knowledge, the biggest device of this type ever built has a diameter of 0.5~m \cite{Chuss}.

Commercially available dielectric laminates used in microwave engineering to build planar microstrip circuits 
could be used to build a dielectric substrate RHWP. These laminates come in different sizes, thicknesses, 
and refractive indices, and have copper deposited on both sides. One possibility is to use one metallic side as a 
mirror and photolithographically etch the other side of the substrate in order to produce metallic stripes 
in the place of a wire grid. The resolution required for the mask production in our frequency range is 
achievable with large commercial printers. The laminate can be joined to a thick metallic flat surface to 
increase the structure stiffness. Note that the main problems of the free-standing RHWP would thus be resolved. The 
constant spacing between wires is guaranteed by the high accuracy of the photolithographic techniques 
whereas the constant distance between the mirror and wire elements
comes directly from the accurate constant thickness of the laminates 
required in microstrip circuits.

\paragraph{Polarization modulator rotation mechanism}

The RHWP needs to be rotated. While a step and integrate rotation could be considered, a continuous rotation 
at 0.1 to a few rpm is preferable both for the observation strategy (\S~\ref{sec:meastech}) and to
avoid disturbing the rotation of the satellite. Such a strategy has been adopted on the 
Maxima and EBEX balloon-borne experiments 
for which a superconducting magnetic bearing has been developed. The HWP rotates at 
2~Hz with low frictional dissipation. This is a crucial parameter as the RHWP will have to operate at the lowest 
temperature possible (30K here) in order to reduce the systematic effects arising from thermal 
emissivity.

The rotation will be driven by a motor operated at a higher temperature within the service module where 
a larger cooling power is available (shown in Fig.\ref{fig:infairing}). The difficulty resides in having to 
rotate a HWP of 1.7~m diameter, weighting approximatively 25~kg including its support structure  
kept at 30~K. The technology is rated TRL~8, the associated risks being the mass and dissipated power.

The main problems associated with this subsystem are thermal and mechanical. The thermal dissipation of the 
bearings, thermal conduction through the shaft, and microphonic vibrations possibly coupling to the detectors
are areas of concern.

\paragraph{Shaft:} For transmitting the torque from the motor (located in the service module) to the RHWP, 
a long fiberglass shaft is a good solution but this option needs further development. 
The critical points are related to the thermal and dynamical aspects of the shaft. It will have to operate at low temperature (down to 35~K), be a good insulator, resist launch and should not create vibrations while rotating.
The effect of the differing thermal expansion coefficients at the 
interface between this composite material and the metal needs to be studied more carefully.

The shaft diameter will be studied 
with the stress criteria (torque transmission) and the dynamic criteria (modes for launch and no torsional modes 
to not interfere with the motor) potentially leading to higher thermal conduction between the temperature stage 
of the motor and the 30~K of the RHWP. If so, magnetic contact-less coupling devices could be investigated in 
order to reduce the thermo-mechanical aspects of the shaft.

\paragraph{Bearings:} Conventional bearing mechanisms for 
the RHWP will need to be studied for lubrications problems when operated at 30~K. A dry lubrication solution 
might be adopted. However, in order to decrease the thermal dissipation, a high temperature superconducting 
magnetic bearing, such as the one developed for EBEX \cite{Hanany}, would be a better solution completely 
avoiding the problem of lubrication, but this option requires further development.
Moreover, it has to be noted that NASA for instance is designing successfully flywheels for spacecrafts based on magnetic bearings. With a top speed of 40,000~rpm and a potential lifetime of 20 years the requirements are largely above ours. This is a parallel research on which we could also base our design for a superconductive magnetic bearing.

\subsubsection{Detectors and readout electronics}
\label{sec:tes}

In the frequency range of COrE, bolometers cooled to very low temperature ($T\le 100$mK) currently offer the 
best performances for broad bandwidth detection. With a careful instrument design, their sensitivity can approach
the photon shot noise of the astrophysical source. The understanding of bolometer performance in a space 
environment has dramatically advanced as a result of the PLANCK mission. In order to read out the 
large number of detectors required, transition edge superconducting bolometers (TESs) are used. Various 
materials can be combined to form the thermal sensor. Thin films such as NbSi or bilayer like MoCu are possible. The 
critical temperature of these sensors has to be matched to the power handling and cooling requirements of the 
instrument. TESs have been developed extensively for astronomical observations from millimeter through X-ray 
wavelengths and have a long history of use.

{\bf Microstrip-coupled TES technology is the detector baseline} for COrE, where thin-film waveguide antennas 
in the OMT are used to couple radiation from a horn antenna onto the TES through a superconducting microstrip 
transmission line terminated with a resistor. The detector would be located on the same wafer as the OMTs using 
microfabrication techniques demonstrated for superconducting bolometer arrays for ground based experiments like 
QUBIC, SCUBA2, or Clover.

Detailed models (assuming 50\% optical efficiency, a 30~K telescope with 1\% emissivity, and 15~GHz bandwidth) 
predict a photon NEP of about $4\times 10^{-18}W~Hz^{-1/2}$ in all wavebands except for the last 3 bands which have a larger bandwidth of 195~GHz. A sensitivity requirement for the 
detectors has been chosen at 70\% of the photon noise or $3\times 10^{-18}W~Hz^{-1/2}$ which increases the total NEP 
by about 22\%.\footnote{$NEP_{tot}=\sqrt{NEP_{det}^2+NEP_{phot}^2}$ where we have chosen $NEP_{det}={NEP_{phot}}/{\sqrt{2}}$} More details are given in table~\ref{tab:detectors}.

\begin{figure}[bhtp]
\begin{center}
\includegraphics[scale=0.33]{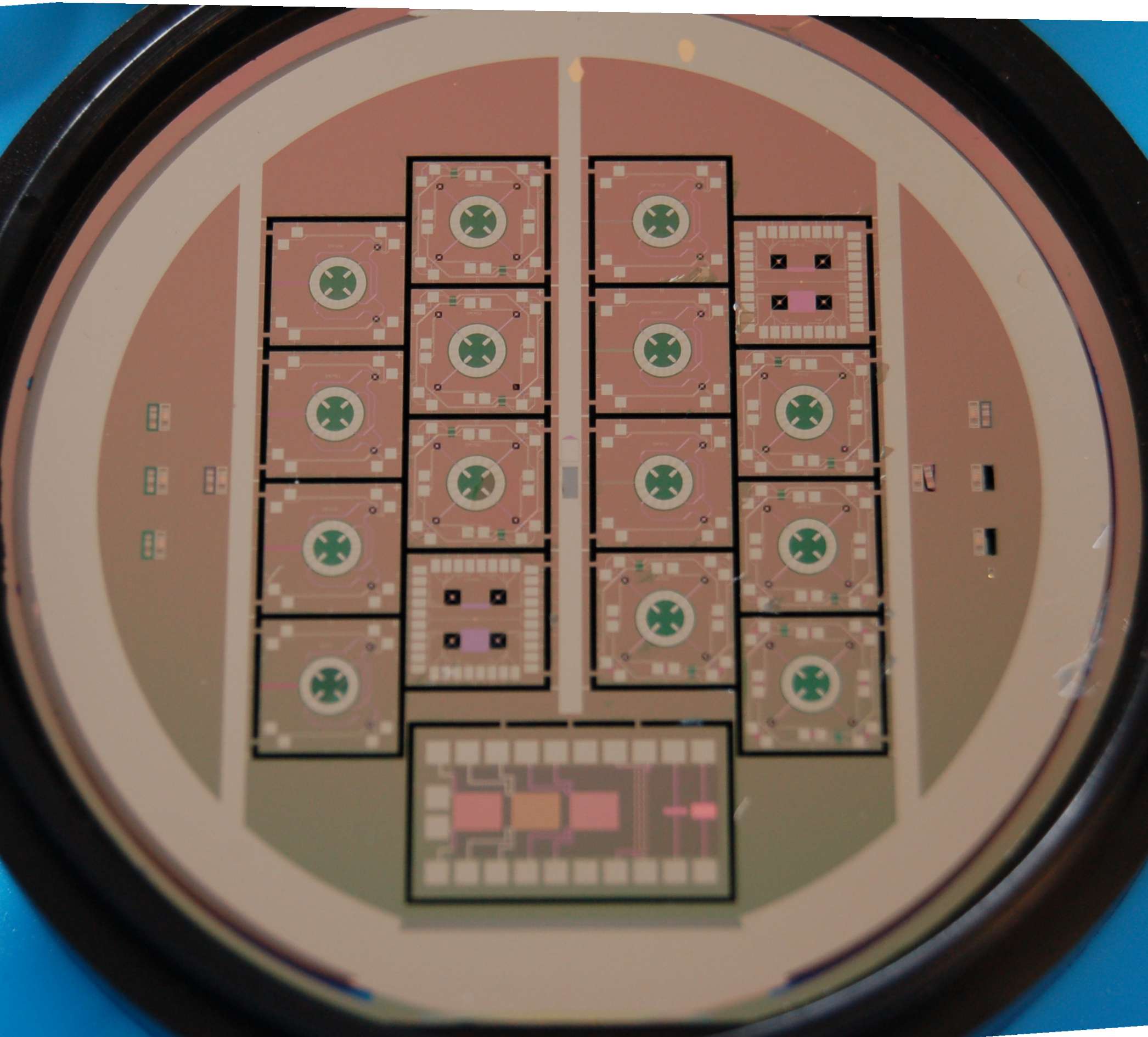}
\includegraphics[scale=1.00]{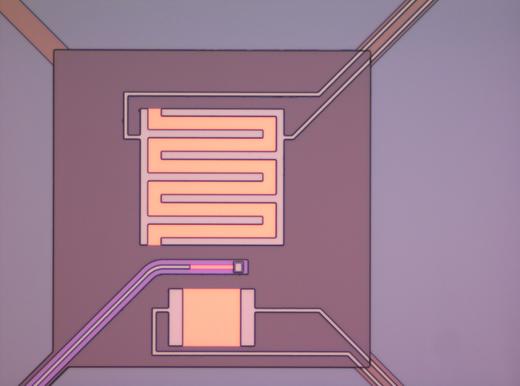}
\caption{\baselineskip=0pt
\small
Left: A wafer of polarization sensitive MoCu TESs. The 4, thin-film, SiN membrane suspended
waveguide probes can be seen (Cambridge Univ.). Right: A microstrip-coupled MoCu TES suitable for ground-based polarimetry (Cambridge Univ.).}
\label{fig:TES}
\end{center}
\end{figure}

In June 2010,
absorber-coupled TESs were down-selected by an international panel for the FIR imaging arrays
on the SAFARI instrument on the cooled-aperture telescope on SPICA. A European consortium now
routinely makes and measures MoCu, MoAu, and TiAu TES arrays having thermal conductances of 0.2~pW/K, and NEPs of
$5\times10^{-19}W.Hz^{-0.5}$. A substantial amount of work is currently taking place in the context of developing
space qualifiable, ultra-low-noise, SQUID multiplexed, TES arrays for SAFARI, and this work will be of
substantial benefit to COrE. COrE requires microstrip-coupled TESs having differential thermal conductances
of typically 10~pW/K, and NEPs of typically $<5\times10^{-18}W.Hz^{-0.5}$, which is an order of magnitude less
demanding than SAFARI. By combining the microstrip-coupled TES technology developed several years ago
for ground-based astronomy, having NEPs $<5\times10^{-17}W.Hz^{-0.5}$ (Fig~\ref{fig:TES}-right), together with the state-of-the-art
absorber-coupled TES technology being developed for SAFARI having NEPs $<5\times10^{-19}W.Hz^{-0.5}$, we can
produce TES detectors having the needed specifications for COrE. The time constant of the detectors should be
shorter than the time needed to scan a beam in order not to degrade the angular resolution. The stringent constraint is
therefore given by the high frequency channels where the beam size is smallest. Under the proposed design, the
required time constant ranges from 10~ms to 0.6~ms. TESs operate through a strong electro-thermal feedback
mechanism, which speeds up their effective time response to the sub-ms values required by COrE. Once again,
the work on SAFARI has shown that achieving the time constants needed for COrE is straightforward. Overall,
the main technological challenges will be associated with realizing the extremely large array, rather than any
fundamental problems associated with producing individual pixels having the needed performance.

The readout system of bolometer arrays requires multiplexing to decrease the thermal load on the 
final cryogenic stage and also to simplify the architecture. The COrE design uses time domain 
multiplexing with SQUIDs as the first amplifier stage. A standard SiGe BiCMOS ASIC (Application Specific Integrated Circuit)
cooled to 80K is used to 
control the multiplexing sequence and to amplify the signal from the SQUIDs. This technology has been tested in 
space by NASA during the MISSE-6 mission aboard the ISS. A multiplexing factor of 24 has been demonstrated 
with a heat load of 20~mW down to 4~K. Assuming the same power dissipation, we have finalized a new design to 
readout 128 detectors with one ASIC. The COrE architecture requires 50 of these new ASICs, leading to a 
power load of about 1~W. Further improvement in power dissipation and multiplexing factor is anticipated
during the phase A study.

\begin{figure}[bhtp]
\begin{center}
 \includegraphics[scale=0.4]{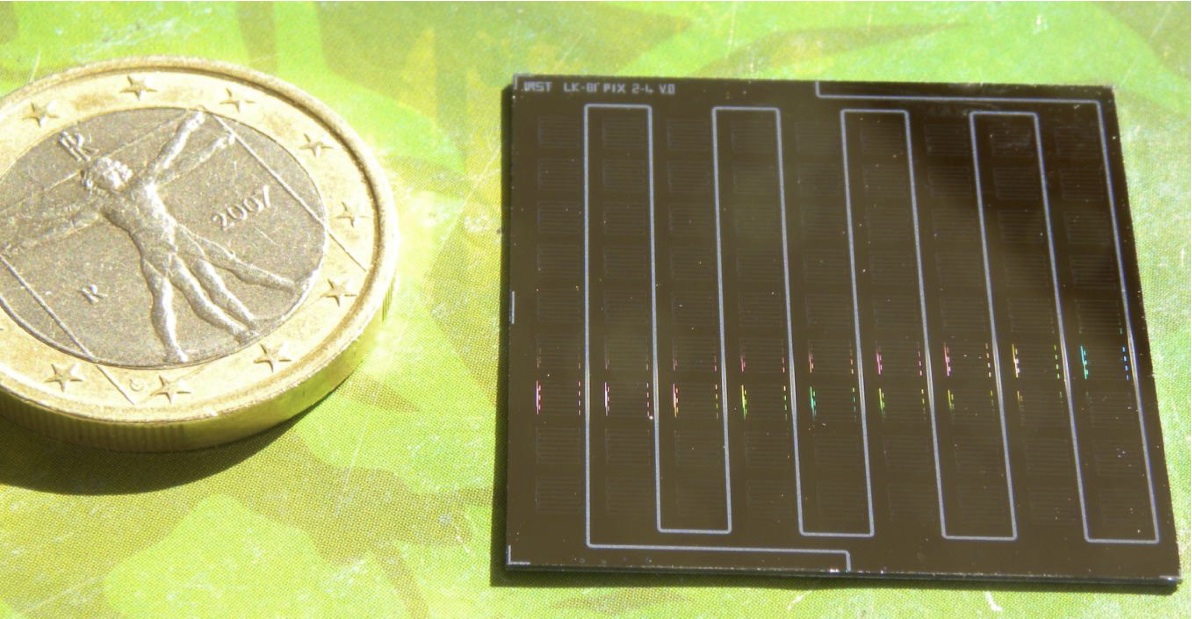}
\caption{\baselineskip=0pt
\small
An 81-pixels array of kinetic inductance detectors for the 140~GHz band,
developed in Italy under the RIC program of INFN and the ACDC program
of PNRA \cite{Calv10}.}
\label{fig:detector}
\end{center}
\end{figure}

We are also considering using superconducting microwave kinetic inductance detectors (MKIDs). These are new 
devices operating at 0.1K promising high-sensitivity large-format detector arrays. The MKIDs concept was 
proposed in 2002 \cite{Mazin}. In an MKID low energy photons (in the meV range) break 
Cooper pairs in a superconducting film, thus changing its surface impedance and in particular its kinetic 
inductance. The detector consists of a superconducting thin-film microwave resonator capacitively coupled to a probe 
transmission line. This is typically a quarter-wavelength of coplanar waveguide (CPW), capacitively coupled to 
a feedline (i.e., a distributed resonator). By exciting the electrical resonance with a microwave probe signal, 
the transmission phase of the resonator can be monitored, allowing the deposition of energy or power to be 
detected. The KIDs read each pixel at a separate frequency. Therefore only one pair of coaxial leads is needed to 
connect several thousand pixels to the readout electronics thus simplifying the thermal architecture. An example of some of these developments is shown in Fig.~\ref{fig:detector}. A different 
approach has been proposed \cite{Doyle} in which a lumped RLC series resonant circuit is inductively coupled to 
the feedline (LEKID). These devices have been demonstrated in a laboratory environment and have already been 
implemented in various demonstrator instruments in the millimeter and submillimeter bands 
\cite{Schlaerth}.

Cold electron bolometers are a third technology option worth mentioning. These detectors developed in Sweden 
(Chalmers) are starting to give promising laboratory results \cite{tarasov}.

Over the last few years many national funding agencies have invested heavily in establishing superconducting 
detector fabrication facilities in Europe. Major facilities exist in the UK, France, the Netherlands, Sweden, 
Germany, and Italy. These facilities not only provide clean rooms with lithography, processing, and test 
equipment, but also provide professional, traceable fabrication routes, producing high-quality high-yield 
devices. A major part of the technology development, fabrication, and test program will be to form a single 
network coordinated by one lead organization, with all of the partner organizations working towards the 
refinement of a single design across all of the chosen wavebands. Most of these laboratories are working on the 
development of TESs and KIDs as shown in Table~\ref{tab:manuf}.

\begin{table}[htd]
\begin{center}
 \caption{\baselineskip=0pt
\small
List of different detectors technologies which could be adapted to COrE specifications.}
\begin{tabular}{c|c|c|c}
\hline
\hline
Technology&Partnership&Achievements&Papers\\
\hline
Single and dual&UK: Cambridge and Cardiff&ClOVER, and study for a COrE&\cite{Audley}-\cite{Mauskopf} \\
 polarization TESs &&type mission - ESA TRP contract.& \\
with feedhorns&US: NIST and GSFC&& \cite{Henning}\\
 and waveguide&It - ASI technology contract&& \cite{Bagliani}-\cite{Vacc}\\
  probes coupled&Fr: APC, CSNSM, IAS&NbSi bolometers& \\
  to microstrip&IEF, Institut N\'eel, IRAP&with low temperature SiGe&\cite{Pajot}-\cite{Sou}\\
 transmission lines&L2E, LERMA, LPSC& ASIC for TDM readout&\\
 \hline
 TES detectors with&US&Detectors assembled&\cite{Calstr}\\
  feedhorns \& distributed&& in a large array (966 pixels)&\\
 absorber (unpolarized)&&Scientific breakthroughs at the SPT&\\
 \hline
 Planar antenna coupled&US: UC Berkeley &POLARBEAR&\cite{Hubmayr}\\
 TESs - dual polarization&&EBEX balloon experiment&\cite{Reich}\\
 broadband + multifreq.&&TRL=6 with about 1300 detectors&\\
 capability - lens-coupled&&&\\
 \hline
 Planar antenna coupled& US: JPL & BICEP and the Keck Array&\\
 TESs with planar phased &&(2048 bolometers) at the South Pole&\\
 array antennas&&balloon-borne experiment SPIDER&\cite{Crill}\\
 dual polarization&&Close to 99\% yield&\cite{Orlando}\\
 single frequency&&&\\
 \hline
Free space coupled&Fr: LETI- CEA &Herschel-PACS: Array of 2560 pixels&\cite{Revret} \cite{Revret2}\\
 detectors single or& \& CEA-DSM-DAPNIA&$NEP=1.5.10^{-16}$,  BW=5Hz&\\
dual polarization&&can be improved for the high&\\
 absorbers&& frequency bands of COrE&\\
 \hline
 LEKID/KID arrays& UK, It, NL, Fr, Sp&camera (2x112 pixels) at IRAM 30m&\cite{Do2} \cite{Do2}\\
 &&81 pixels array made in Italy-Fig~\ref{fig:detector}&\cite{Calv10} \cite{Monfardini} \\
\hline
\hline

\end{tabular}
\label{tab:manuf}
\end{center}
\end{table}%

\subsubsection{Spectral filtering}
\label{sec:filters}

Because bolometric detectors are sensitive to a broad spectral range, their spectral band definition requires 
special attention. COrE requires a high in-band transmission in order to reach a 50\% overall optical 
efficiency while rejection of optical/NIR power must be better than $\approx10^{-12}$ although a rejection of 
$\approx10^{-6}$ suffices in the FIR. Low frequency rejection is easily achieved through the horn waveguide 
design. The PLANCK-HFI-like high frequency band definition based on dielectric embedded interference filters
requires several multilayer mesh filters in series located on the different temperature stages also limiting the 
thermal power reaching the ultra-cold stages. {\bf This technology, selected as the baseline} for COrE, allows 
for use with mixed frequency array by arranging their geometry to suit the shape of the sub-array. It is also 
conceivable to make a unique filter made of several regions. However special attention is required to avoid 
straylight in between the different regions of this filter potentially leading to cross-talk between channels.

In order to reduce the number of interference filters mentioned above, {\bf superconducting planar bandpass 
filters} could be lithographed onto the detector chips, which together with a planar OMT would lead to intrinsic 
on-chip polarimetry. While parts of this technique has been demonstrated, further development is necessary.

\begin{figure}[htbp]
\begin{center}
 \includegraphics[scale=0.3]{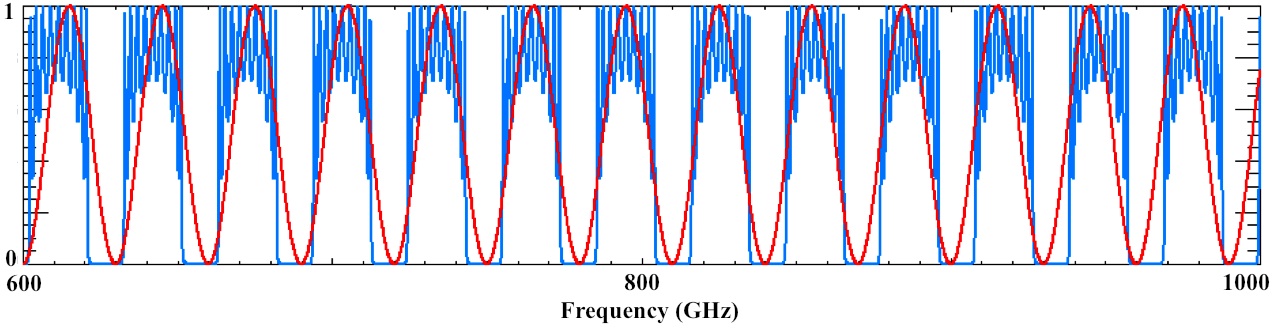}
\caption{\baselineskip=0pt
\small
Sub-band filtering: Filter transmission (blue) and RHWP efficiency (red).}
\label{filtering}
\end{center}
\end{figure}

{\bf Additional spectral filtering}: As mentioned in \S~\ref{sec:rhwp}, the modulation efficiency of the high 
frequency channels can be increased using broader bands in combination with sub-band selective filtering around 
the peaks. This can be achieved for example by using multilayer dielectrics that act as dielectric mirrors. 
However, while the RHWP modulation efficiency peaks are centered at an odd multiple number of its basic frequency 
as shown in eqn.~\ref{eq:peaks}, the filter transmission bands are centered at an even multiple number of its basic 
frequency following eqn.~\ref{eq:filter}. So a slightly different basic frequency $\nu_0^{Filter}$ will be used 
for the filter so that we have maximum coincidence between the two sets of peaks. The example in 
Fig.~\ref{filtering} shows a candidate filter around the 795~GHz peak using $\nu_0^{Filter}=15.29$ and $n=26$.
\begin{equation}
\nu_n^{Filter}=2n\times\nu_0^{Filter}
\label{eq:filter}
\end{equation}

\subsubsection{Cryogenic chain}
\label{sec:cryo}

Since the mass and power requirements and the temperature stages are similar to those of PLANCK, copying 
the PLANCK cryogenic chain would bring the whole system to TRL 9. However several modification would provide
superior performance. The main objective is to develop a system based on mechanical coolers and 
recyclable system to provide for a long duration mission. Below is a summary of a CNES study.

\paragraph{Passive cooling:}
The V-grooves used in PLANCK to radiate power into free space placed laterally to the spin axis 
have proven to be an efficient solution to a first passive cooling stage. They also provide cooling to 
intermediate stage electronics. An inner V-groove temperature of 50~K is currently achieved in orbit with an 
available power of the order of 1~W for PLANCK. With further optimization and the addition of a fourth 
V-groove, a temperature of about 35~K could be achieved for COrE inner V-groove.

\paragraph{Active cooling - Mechanical coolers} 

\paragraph{The 35K to 20K} stage could be provided by a H$_2$ JT (Joule-Thomson) cooler 
with metal-sorption compressor from JPL (PLANCK). However a pulse tube cooler (PTC) providing 15/20~K stage is 
under development at Air Liquide/CEA (with the goal to reach 15K). Note that a two-stage PTC with a 300~mW heat 
lift at 20~K (with intercept) has already been demonstrated on the GSTP program Ò20-50K Pulse TubeÓ (ESA 
contract 20497/06/NL/PA). Ongoing developments will allow more margin at the 20K stage. This 20~K PTC 
stage requires a pre-cooling at 80~K with a heat intercept cooling power of about 6~W. Note that pre-cooling at a higher 
temperature (e.g., 120~K) is possible but with a slight decrease of the performance at 20~K. This heat intercept 
can be provided either by a V-groove through thermal braids links for example (preferred solution), or by a 
dedicated 80~K PTC. Maturity level of the 20~K PTC is considered TRL 4.\\ {\bf The 15~K to 4~K} stage is 
provided by a Joule-Thomson (JT) cooler similar to the PLANCK cooler.

\paragraph{Active cooling - Sub-K coolers:} Because the open cycle dilution refrigerator (OCDR) providing both 
the 1.7~K and 100~mK stages from 4~K on PLANCK has a limited lifetime, a recyclable system is preferable. The 
0.1~K stage baseline could be provided by an {\bf ADR cryocooler} (e.g. from CEA SBT), requiring pre-cooling 
(e.g., at 1.7K and 4K on SAFARI/SPICA) or a continuous ADR 4-stage (NASA-GSFC) developed for MIRI (JWST). While 
ADRs are already at TRL~9 (XRS on ASTRO-E2), the strong associated magnetic field might affect the operation of 
the TES readout system (SQUIDs), although efforts are being dedicated to mitigate this difficulty.

For this reason, a {\bf closed cycle dilution refrigerator (CCDR)} based on the heritage of the flight-proven OCDR can offer 
a more convenient solution.
The re-circulation of the helium isotopes (He-3 and He-4) of this system allows 
for no theoretical limitation on lifetime. Contrary to the OCDR, the CCDR requires a pre-cooling stage 
around 1.7~K in order to evacuate the heat from thermalization of He-3 and He-4 gas flows and from condensation 
of He-3 (a fraction this heat is used for the evaporation of He-3 but not all of it). It is not possible to 
produce a 1.7~K stage with a He-3 Joule-Thomson expansion within this dilution system. As a consequence, the 
1.7~K stage is produced by a dedicated He-4 sorption cooler based on the concept of the Herschel He-3 sorption 
cooler. The CCDR system also provides extra cooling power between 200~mK and 600~mK if needed (e.g. for 
thermalization of mechanical supports or harnesses). The maturity level of CCDR is considered TRL 4, and should 
reach TRL 5 within 2 years with ongoing developments at Air Liquide/CNRS Neel institute.

\subsubsection{Data Rate}
\label{sec:datarate}

Every day the instrument will simply survey the sky for 24~hours according to the predefined program with a sampling rate matching the scan speed of 0.5~rpm and the FWHM of the beam for each frequency channel. Based on Planck experience, we request 16 bits per scientific data sample and 5 samples per beam. Taking into account a compression factor of 4, we reduce the resulting total raw data rate of 18~Mpbs (see Table~\ref{tab:detectors}) to 4.5~Mbps for transmission.

The on-board data storage system should be able to save a few days of data ( i.e. $4.5\times10^{6}\times86400\times2$~=~780~Gbit, instead of the 32~Gbit memory on board of Planck) in order to cope with downlink problems due to contingencies in the telecommunication system. The data will be dumped to the Earth via a high-gain antenna aligned to the spin axis for about 2 hours per day in the Ka band, during which the spin axis will be directed to the Earth. In this case we shall need a data dump rate of $4.5\times12$~=~54~Mbit/s.

\subsection{Expected performance}
\label{sec:performances}

The sensitivity of a perfect instrument with a uniform coverage of the celestial sphere, expressed in 
terms of noise angular amplitude spectrum $c_{noise}$, is given by eqn.~\ref{eq:cls} assuming a 
polarization modulation efficiency of 80\%, where $t_{miss}$ is the total mission duration, $N_{det}$ is 
the number of detectors and $s_{det}$ is the sensitivity in intensity of one detector
\begin{equation}
c^2_{noise}=2.4\times \frac{4\pi}{t_{miss}N_{det}}\times s^2_{det}=
\left(4.7\mu K\cdot arcmin\right) ^2\times \left( \frac{4~years}{t_{miss}} \right)
\times \left( \frac{400}{N_{det}} \right)\times \left( \frac{s_{det}}{50\mu K\cdot s^{\frac{1}{2}}} \right)^2. 
\label{eq:cls}
\end{equation}
To reach such sensitivity, the proposed instrument architecture is designed in order to have background 
limited performances. The detector number is furthermore optimized to reach the required sensitivity in 
each channel and to fill the focal plane area. With the current instrument design,
\footnote{Instrument efficiency: 50\%, telescope temperature: 30K with 1\% emissivity, detector noise: 
70\% of the photon noise}
this process leads to the numbers given in Table \ref{tab:CorePerf}. The 
proposed configuration has a total of 6384 detectors in 15 frequency bands. The greatest sensitivity is 
concentrated in the 
CMB channels (75GHz-225GHz) with 4950 detectors.

\begin{table}[htd]
\begin{center}
 \caption{\baselineskip=0pt
\small
COrE detector specifications with assumptions from \S~\ref{sec:tes} for NEPs calculations. Data rates based on \S~\ref{sec:datarate} assumptions give a total of 18~Mbit/s.}
\begin{tabular}{c|c|c|c|c|c}
\hline
\hline
Central Freq.&$NEP_{phot}$&$NEP_{det}$&$NEP_{tot}$&Data rate / waveband&Det. time constant\\
(GHz)&$(W.Hz^{-0.5})$&$(W.Hz^{-0.5})$&$(W.Hz^{-0.5})$&(kbit/s)&(ms)\\
\hline
45&$4.4\times10^{-18}$&$3.1\times10^{-18}$&$5.4\times10^{-18}$&38&10.7\\
75&$4.5\times10^{-18}$&$3.2\times10^{-18}$&$5.5\times10^{-18}$&298&6.4\\
105&$4.4\times10^{-18}$&$3.1\times10^{-18}$&$5.4\times10^{-18}$&557&4.6\\
135&$4.3\times10^{-18}$&$3.0\times10^{-18}$&$5.3\times10^{-18}$&985&3.6\\
165&$4.1\times10^{-18}$&$2.9\times10^{-18}$&$5.1\times10^{-18}$&1642&2.9\\
195&$4.0\times10^{-18}$&$2.8\times10^{-18}$&$4.8\times10^{-18}$&2975&2.5\\
225&$3.8\times10^{-18}$&$2.7\times10^{-18}$&$4.7\times10^{-18}$&5373&2.1\\
255&$3.7\times10^{-18}$&$2.6\times10^{-18}$&$4.5\times10^{-18}$&1945&1.9\\
285&$3.6\times10^{-18}$&$2.5\times10^{-18}$&$4.4\times10^{-18}$&1418&1.7\\
315&$3.6\times10^{-18}$&$2.5\times10^{-18}$&$4.4\times10^{-18}$&418&1.5\\
375&$3.6\times10^{-18}$&$2.5\times10^{-18}$&$4.4\times10^{-18}$&318&1.3\\
435&$3.6\times10^{-18}$&$2.5\times10^{-18}$&$4.4\times10^{-18}$&369&1.1\\
555&$13.7\times10^{-18}$&$9.7\times10^{-18}$&$1.7\times10^{-17}$&471&0.9\\
675&$14.2\times10^{-18}$&$1.0\times10^{-17}$&$1.7\times10^{-17}$&573&0.7\\
795&$14.5\times10^{-18}$&$1.0\times10^{-17}$&$1.8\times10^{-17}$&675&0.6\\
\hline
\hline
\end{tabular}
\label{tab:detectors}
\end{center}
\end{table}%

\subsubsection{Configuration evolution}

Tables~\ref{tab:CorePerf} and \ref{tab:detectors} give the present configuration. However in order to increase significantly the 
sensitivity at high frequency and strengthen the galactic science output, an option under study is to 
increase the number of detectors of the 795~GHz spectral band by a factor 5. As shown in 
figure~\ref{fig:FPU} this will have little impact on the overall focal plane dimension (all 64 detectors - 
32 pixels - are represented by the red circle at the center of the FPU). While the sensitivity of this band will be increased by a factor $\sqrt{5}$, the total data rate will increase by 15\%.

\subsection{Calibration}
\label{sec:calibration}

To keep the instrument simple and without a possible single point failure, no onboard calibration system is
included. 
COrE calibration will rely on ground calibration sequences and astrophysical sources that will 
have been thoroughly studied by PLANCK and follow-up observations. Systematic effects affecting the polarization 
purity of the incoming signal will be mitigated by the RHWP in front of the telescope. 
However the RHWP might introduce small beam shape distortions synchronous with its orientation that should
be detected and characterized in the calibration process to ensure a negligible impact on the measurements.
The ground calibration will consist of the following complementary tests:

{\bf Sub-system level}: (detectors, reflective optics, cold optics, RHWP, cooling system)

{\bf FPU}: spectral response of each pixel, dark and illuminated I-V curves for various FPU temperatures.

\begin{figure}[htbp]
\begin{center}
 \includegraphics[scale=0.35]{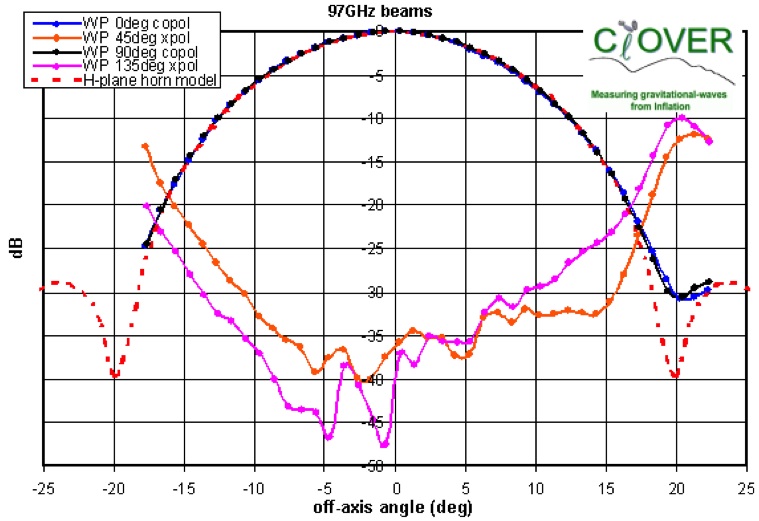}
\caption{\baselineskip=0pt
\small
Systematic effect caused by a birefringent HWP on a corrugated horn beam. Measurements performed with a vector 
network analyzer (75-110GHz) at the University of Manchester for the Clover project. The red dashed 
line is the predicted horn co-polarization beam, the solid lines the beams with different orientations of the 
HWP in front of the horn aperture.}
\label{fig:effect}
\end{center}
\end{figure}

{\bf Warm RF test in Compact Antenna Test Range}: RF test conducted during the PLANCK ground calibration campaigns have been extremely beneficial in order to characterize the main and far-side lobes of the telescope beams. RF performance have been measured with a few representative horns associated with OMTs and warm detectors (heterodyne 
instead of bolometers) in conjunction with the telescope and the shields.
Similar test including the RHWP will be performed 
in order to measure and model the systematic effects introduced by the RHWP. Such effects 
are already under study at pixel level. Several CMB polarization experiment are making use of a half-wave plate. 
Most of these are using birefringent HWP located in front of the focal plane, at the entrance aperture of the 
horns. Figure~\ref{fig:effect} is showing such a measurement taken in the frame of the Clover project. It shows 
the variation of the beam shape that can be expected when a birefringent HWP is rotated in front of a horn. In 
this instance the impact is minimal down to -25~dB and the cross-polarization integrated over the horn main beam 
remains below -35~dB. Preliminary studies are being carried out in the framework of R\&D programs with higher 
precision measurements allowing to develop and validate theoretical models predicting such systematic effects. 
Similar measurements on the effects of a RHWP in front of a telescope like COrE could be performed in the 
framework of a M3 mission.

{\bf Dedicated calibration facility} reproducing the cryogenic and radiative environment expected during the 
mission is envisaged. A large vacuum chamber fitted with a 2K cryogenic chain and a full beam source will be 
prepared. The chamber will allow operation of the entire instrument, observing a 2.7~K load and a polarized 
source modulated and oriented as needed. This setup will allow the precise determination of the polarimetric 
parameters of the instrument (polarization efficiency and main axis angles) and of the responsivity and noise of 
all the detectors for different radiative loading conditions. The proposers inherit considerable experience from 
similar calibration facilities built for Archeops and BOOMERanG as well as from the calibration campaign 
on PLANCK at CSL-Li\`ege.

%
%
%

\newpage
\section{Epilogue}

The COrE proposal was submitted to ESA on 13 December 2010 in response to a call for an M3 
`medium-sized' space mission in the framework of Cosmic Vision 2015-2025. The COrE collaboration 
includes the European CMB community (a full list of endorsers can be found on the website 
www.core-mission.org) and the US CMB community has expressed a strong interest in a NASA 
participation should the ESA proposal be successful. ESA is currently in the process of evaluating 
the COrE proposal.

Compared to the previous B-Pol proposal submitted to ESA three years ago, COrE has a higher angular 
resolution and broader frequency coverage, allowing COrE to broaden its science program 
substantially. Roughly speaking the size of the instrument is more or less fixed, being determined 
by the budget and physical dimensions of the launch vehicle. While B-Pol proposed several 
refractive telescopes, each tailored to a particular frequency range, and relied on a complex scan strategy 
to control systematic effects, COrE will have a single reflective telescope, suitable over the 
entire frequency range, and use a rotating reflective half-wave plate as the first optical element 
in front of the sky to modulate away most of the parasitic systematics. It is this feature that 
makes COrE unique and allows numerous requirements to be relaxed relative to what would be needed 
in the absence of such modulation. The enhanced resolution of COrE, owing to the large size of the 
mirror, is diffraction limited at all frequencies, unlike PLANCK, reaching a resolution of 1.3 arcmin in 
the highest 795 GHz channel. In the CMB channels the resolution ranges from
10 arcmin (105 GHz) to 4.7 arcmin (225 GHz) and the polarization
sensitivity (for a four-year mission) is around $4.5\, \mu\mathrm{K}\,
\mathrm{arcmin}$.

The broad frequency coverage in 15 narrow bands, and the high sensitivity,
will allow reliable subtraction of foregrounds in the search for primordial
B modes due to the gravitational wave background from inflation.
Our simulations indicate that a tensor-to-scalar ratio 
$r=10^{-3}$ should be detectable at $3\sigma$, corresponding to
an energy scale of inflation (few)$\times 10^{15}$ GeV. The high resolution in the CMB channels
will allow COrE to carry out virtually the 
ultimate CMB lensing experiment, with the deflection field reconstructed
with high signal-to-noise on all scales where linear theory is reliable.
In addition, COrE will have great discovery potential in the area
of non-Gaussianity being able to search for a variety of ``shapes'' of
the primordial $n$-point functions. This ability to fingerprint the
physics of inflation sets the CMB apart from other probes of non-Gaussianity.
COrE will also be able to do breakthrough galactic and extragalactic science in the far 
infrared, mapping the dust polarization at high resolution in the regions of diffuse emission, and 
hence map the galactic magnetic field, thus answering key questions concerning the mechanisms of star 
formation. Thanks to its exquisite resolution and sensitivity, COrE will also discover a myriad of 
new IR compact sources, both polarized and unpolarized. At the other end of the spectrum, at low 
frequencies COrE will be 30 times more sensitive than PLANCK LFI, providing exquisite maps of the galactic 
synchrotron emission free of Faraday rotation.

\newpage

\section{Selected references}

\noindent
We provide here an abridged set of essential references concerning
the polarization, CMB, and sub-millimeter science and instrumentation. 
\vskip 7pt 

\noindent
The COrE team has benefited from experience in a variety 
of suborbital and space CMB experiments, including most
notably Boomerang, CLOVER, and PLANCK (http://www.esa.int/SPECIALS/Planck/index.html).
We have also benefited from several B-mode polarization satellite
studies starting with the Frech CNES-sponsored SAMPAN study
and the Italian COFIS study of ASI,
leading to European B-Pol proposal three years ago. 
(www.b-pol.org) A more detailed American CMBPol study followed on this 
heritage, producing a number of useful and detailed white papers,
which broadly confirm the conclusions of the present study.
In particular, the following papers provide useful details 
and extensive references:
\begin{quotation}

\vskip -7 pt
{\parindent -10pt

\hskip-25pt J Dunkley et al.,
``CMBPol Mission Concept Study: Prospects for polarized foreground removal,''
(arXiv:0811.3915[astro-ph]) (2008)

A Fraisse et al., ``CMBPol Mission Concept Study: 
Foreground Science Knowledge and Prospects,''
(arXiv:0811.3920[astro-ph]) (2008)

D Baumann et al.,
``CMBPol Mission Concept Study: Probing Inflation with CMB Polarization,''
(arXiv:0811.3919[astro-ph]) (2008)

M Zaldarriaga et al.,
``CMBPol Mission Concept Study: Reionization Science with the Cosmic Microwave Background,''
(arXiv:0811.3918[astro-ph]) (2008)

K Smith et al.,
``CMBPol Mission Concept Study: Gravitational Lensing,''
(arXiv:0811.3916[astro-ph]) (2008)

}
\vskip -15 pt
\end{quotation}
\noindent
More information is at 
http://cmbpol.uchicago.edu/.
We following publications are particularly relevant:
\begin{quotation}
\vskip -7 pt
{\parindent -10pt

\hskip-25pt Reichborn-Kjennerud B. et al., ``EBEX: A balloon-borne CMB polarization experiment," 
Proceedings of the SPIE, Volume 7741, pp. 77411C-77411C-12 (2010) (arXiv:1007.3672)

D Chuss,
``Quasioptical Reflective Polarization Modulation for the Beyond Einstein
Inflation Probe" in {\it Polarization Modulators for CMBPol}, 
J of Phys: Conf. Series 155 (2009).

L Verde, H Peiris \& R Jimenez, 
``Optimizing CMB polarization experiments to constrain inflationary physics,''
JCAP 0601, 019 (2006) (astro-ph/0506036)

de Zotti, G., et al., ``Radio and millimeter continuum surveys and their
astrophysical implications'', (A\&ARv 18, 1) (2010). 

Lagache, G., et al., ``Dusty Infrared Galaxies: Sources of the Cosmic
Infrared Background'', (ARA\&A Annual 43, 727) (2005)

J Fergusson, M Liguori, P Shellard, 
``General CMB and Primordial Bispectrum Estimation I: 
Mode Expansion, Map-Making and Measures of $f_{NL},$''
Phys. Rev.D82, 023502 (2010)
(0912.5516/astro-ph.CO)

A Lewis \& A Challinor, 
``Weak gravitational lensing of the CMB,''
Phys. Rept. 429, 1 (2006) (astro-ph/0601594)

C McKee \& E Ostriker,
``Theory of Star Formation,''
ARAA 45, 565 (2007)

J Delabrouille \&  J Cardoso, 
``Diffuse Source Separation in CMB Observations,''
(astro-ph/0702198) 

S Leach et al.,
``Component separation methods for the PLANCK mission,''
A\& A 491, 597 (2008)
(arXiv:0805.0269) 

}
\vskip -15 pt
\end{quotation}
\noindent
A more detailed bibliography arranged section by section appears on the COrE website cited above.

\end{document}